\documentclass[LATO1COL,dvipsnames,table]{elsarticle}

\usepackage[dvipsnames]{xcolor}
\usepackage{amssymb}
\usepackage{graphicx} % Allows including images
\graphicspath{ {00_Figuras/} }
\usepackage{booktabs} % Allows the use of \toprule, \midrule and \bottomrule in tables
\usepackage{amsmath,amsfonts,mathrsfs}
\usepackage{subfigure}
\usepackage{epstopdf}
\DeclareGraphicsExtensions{.eps}
\usepackage{amsthm}
\usepackage{lscape}
 \usepackage{tabularx} 
\usepackage{color}
\usepackage{lineno}
\usepackage{caption}
\DeclareCaptionFormat{cont}{#1 (cont.)#2#3\par}
\emergencystretch=\maxdimen
\newcommand{\sout}[1]{} %Este redefine el comando \sout y elimina todo lo tachado

\newcommand{\NL}[1]{{\color{black}{#1}}}

\newcommand{\MS}[1]{{\color{black}{#1}}}

\begin{document}

\begin{frontmatter}

\title{Screening of seismic records \sout{for performing}\NL{to perform} time-history dynamic analyses of tailings dams: a power-spectral based approach}

\author[SRK,CURTIN,*]{Nicolás A. Labanda}
\author[SRK,FIUBA]{Mauro G. Sottile}
\author[SRK,FIUBA]{Ignacio A. Cueto}
\author[SRK,FIUBA]{Alejo O. Sfriso}
\address[SRK]{SRK Consulting, Argentina.}
\address[FIUBA]{Universidad de Buenos Aires. Facultad de Ingenier\'ia. Buenos Aires, Argentina.}
\address[CURTIN]{Curtin University, School of Earth and Planetary Science, Bentley, Western Australia, Australia.}
\cortext[*]{Corresponding Author at: Curtin University, Bentley, Western Australia, Australia (nlabanda@fi.uba.ar).}

\begin{abstract}
Time-history deformation analyses of upstream-raised tailings dams use seismic records as input data. Such records must be representative of the in-situ seismicity in terms of a wide range of intensity measures (IMs) including peak ground acceleration \NL{(PGA)}, Arias intensity \NL{(AI)}, cumulative absolute velocity \NL{(CAV)}, source-to-site distance, duration, \sout{and}\NL{among} others. No single IM is a sufficient descriptor of a given seismic demand (e.g. crest settlement) because different records, all of them compliant with any IM, can produce a very wide range of results from \sout{negligible}\NL{insignificant} damage to global failure. The use of brute force, where hundreds of seismic records compliant with a set of IMs are employed, has proven to be a reasonable workaround of this limitation, at least \NL{as it is} able to produce a probabilistic density function of demand indicators. This procedure, however, requires a large number of runs\sout{ of the numerical models}, and is therefore expensive and time-consuming.\sout{ Those} Analyses can be optimized if an \sout{a-priori}\NL{a priori} simple tool is used to predict which seismic records would yield a given demand, thus obtaining \sout{an estimate of demand indicator}\NL{estimations} with \sout{many less runs}\NL{much fewer runs}. \sout{In this study, a}\NL{In order to perform a more precise \sout{selections}\NL{selection}, a} semi-analytical screening procedure \sout{ for estimating the seismic demand of tailings dams} is \sout{proposed}\NL{presented} \NL{in this paper}. The procedure \sout{employs}\NL{makes use of} the spectral properties of \sout{a}\NL{the} seismic record\NL{, considering only the intensity  of the frequency content which is not} filtered by \sout{those of} the dam \sout{and is able to produce}\NL{to obtain} an \sout{a-priori}\NL{a priori} estimate of \sout{a given} demand\sout{ indicator}, expressed in this case in terms of displacements. \sout{The tool is \sout{applied to a large tailings dam subjected to strong earthquakes and is validated} \NL{validated} using \NL{analytical and} numerical models that show the robustness of the tool by proving \NL{its} insensitivity to \sout{constitutive models}\NL{the constitutive models used}}\NL{The tool is validated using analytical and numerical models that prove insensitivity to the constitutive model used in the analysis, and is applied to a large tailings dam subjected to strong earthquakes}. 

\end{abstract}

\begin{keyword}
\sout{Dynamic, }Liquefaction, Tailings dams, \NL{Earthquake}, \NL{Deformation analysis}\sout{Correlations, Time-history analysis}
\end{keyword}

\end{frontmatter}

%\linenumbers

\section{Introduction}
\label{section:Introduction}

Tailings are mine residues in the form of rock flours. They are generally deposited hydraulically as a viscous mixture into storage facilities (TSFs). The lack of post-deposition compaction \sout{entail}\NL{results in} loose states, which can be locked as such by chemo-electrical interactions among the particles and early diagenesis \citep{Bachus2019}. The storage facility construction method can be downstream, centerline or upstream, named after the crest movement direction during the raise. Upstream-raised dams are highly attractive from \sout{the economical}\NL{an economic} point of view as they minimize construction volumes; however, they are the least safe, as stability largely relies on the strength of the tailings themselves. \sout{Recent upstream-raised TSFs massive failures}\NL{Recent massive failures of upstream-raised TSFs} (such as Merriespruit, \sout{Mount Polley,}Samarco and Brumadinho \citep{Santamarina526}) have depicted their vulnerability against liquefaction. Upstream-raised \NL{TSFs} have an annual probability of failure five to ten times larger than hydroelectric dams \citep{Davies2002}. 

According to Rico et al. \cite{Rico2008}, liquefaction is one of the main causes of tailings dams failures, \NL{with} dynamic liquefaction representing around 15$\%$ of the cases. Liquefaction occurs when loose water-saturated tailings undergo an increase of pore pressure and loss of strength due to undrained shearing or \sout{by} internal fabric collapse. In the context of \sout{the} static liquefaction, due to the difficulty of \sout{preventing events triggering failure}\NL{preventing and predicting events that might trigger the failure},  guidelines used as international best practice (e.g. \citep{Ancold2019}) recommend to assume that \sout{an event}\NL{triggering} will occur during the lifetime of the TSF and \sout{to}\NL{then} perform limit equilibrium (LE) analyses \sout{while} adopting a fully-softened shear strength. The \sout{factor of safety}\NL{safety factor} required for such analysis is set close to 1.0 \citep{Ancold2019}. While safe, this approach makes no allowance for the amount of strain required for a given dam to reach a point of uncontrolled progressive failure \cite{Sottile2021}. \sout{This is reasonable for designing}\NL{This approach is reasonable to design} new TSFs but \sout{fails}\NL{it falls} short in assessing the risk posed \sout{by}\NL{on} existing TSFs, both operating and abandoned. 

Liquefaction imposed by seismic loading is generally not analyzed as a process \cite{Bray2011}. Instead, the risk of liquefaction is estimated based on the (static) brittleness of the tailings. If the material is deemed liquefiable, static stability analyses considering post-seismic undrained shear strength are performed, \sout{with no inertia effects considered at all}\NL{without considering inertia effects at all}. This is a very limited approach, producing essentially the same design regardless of the seismicity of a given site. In this sense, numerical time-history deformation analyses become imperative to better understand the dynamic liquefaction vulnerability of a given dam and to warrant that the designed freeboard is adequate.

One of the crucial aspects of dynamic liquefaction assessments is the selection of a proper and  representative set of seismic records as input for the time-history deformation analyses. Several ground motion intensity measures (IMs) have been proposed to characterize the destructive potential of a record: peak ground velocity (PGV) and peak ground acceleration (PGA) are the most widespread but limited IMs \citep{Ebrahimian2012}; Arias intensity (AI), proportional to the total energy content of the signal \citep{Arias1990}; modified cumulative absolute velocity (CAV and CAV5), the latter being the integral of the acceleration after application of a 5 $cm/s^2$ acceleration threshold \citep{Kramer2006}; a normalized hysteretic energy, an empirical relation between dissipated shear energy and residual excess pore pressure ratio \citep{Green2000}, among many others.

In the context of liquefiable soils, new approaches have been proposed recently. Kramer et al. \cite{Kramer2016} \sout{reviews}\NL{review} procedures to detect the time of liquefaction triggering, comparing their performance with empirical methods. The research is focused on signal analysis using \sout{short term}\NL{short-term} Fourier transform (STFT), spectrograms, wavelets transforms and Stockwell spectrum procedures, showing that the mean frequency content tends to reduce in signals recorded above a liquefied stratum \citep{Kramer2018384,MEZAFAJARDO2019292,ozener2020}.

Motivated by its simplistic and computationally efficient model, other researchers have used sliding-block Newmark-type models \citep{Newmark65} to estimate settlements of slopes under shaking, and adopting empirical modifications of the Arias intensity to characterize the excitation \citep{CHOUSIANITIS2014}. Combinations \sout{between}\NL{of} Newmark-type models\NL{,}  as a displacement estimator\NL{,} and some selected IMs like PGA, CAV and AI\NL{,}  are popular in hazard analysis in geotechnics \citep{Jibson1993521, JIBSON2007209,Armstrong2013,Roy2016,DEYANOVA2016210}.

Due to the improvement of \sout{computers}\NL{computational} capacity, numerical methods are progressively replacing analytical models. Naeini et al. \cite{NAEINI2018179} addressed the problem of tailings dams subjected to dynamic loads \NL{by} focusing \sout{in}\NL{on} detecting resonance points by means of transfer functions \citep{Severn99,Hwang2007}. Jin et al. \cite{Jin2018} proposed a theoretical framework \sout{of}\NL{for} mudslides based on experimental and numerical models, pointing out that instability mechanisms are somehow similar, relating \sout{a proper}\NL{an appropriate} design of a tailings dam to the site seismicity rather than \NL{to} its \sout{constructive}\NL{construction} procedure.  \sout{Its}\NL{It} is worth noting that computational models have become the standard procedure to perform forensic studies over real dam failures, and some remarkable works are recommended as a reference \citep{ISHIHARA20153,BAOTIAN14,Wenlian12,swidqs2016}. Despite advances in this subject matter, tailings engineering keeps using correlations of engineering demand parameters with classical IMs, usually leading to highly scattered results \citep{HARIRIARDEBILI2019761}. 

This paper presents a generalization of Arias Intensity (AI) based on the spectral properties of the input signal\sout{,} and \sout{applies}\NL{it uses} it as a tool to produce \sout{an a-priori}\NL{a preliminar} estimate of the seismic demand of any \sout{seismic record}\NL{given earthquake}. The use of the tool is validated \sout{against}\NL{using} classical/analytical methods and dynamic numerical models of a large TSF, and the correlation between this new IM and the maximum displacement during time-history analyses is shown \sout{for}\NL{in} this case \sout{and compared to that of AI}\NL{, comparing its performance with the \NL{results} obtained with classical IMs}. 

\sout{The paper is organized as follows:} \NL{The paper starts describing} an example of a \sout{routine }time-deformation analysis of a TSF in Section \ref{section:tailing}, \NL{that will be used to show the strength and consistency of the proposal}. In Section \ref{section:Const} the constitutive model PM4Sand, used for \NL{subsequent numerical}\sout{ the} liquefaction assessment, is calibrated \sout{with}\NL{using} laboratory and field tests. In Section \ref{section:apriori}, the proposed IM based on spectral power is presented in detail. Numerical results related to deformation induced by pore pressure buildup are presented and discussed in Section \ref{section:Num_res}, and the predictive capability and efficiency of the tool is tested. Finally\NL{,} some conclusions and outlooks are drawn in Section \ref{section:Conc}.

\section{Deformation analysis of a TSF}
\label{section:tailing}

\subsection{Description}

\begin{figure}[h!]
\begin{center}
{\includegraphics[height=5.4cm]{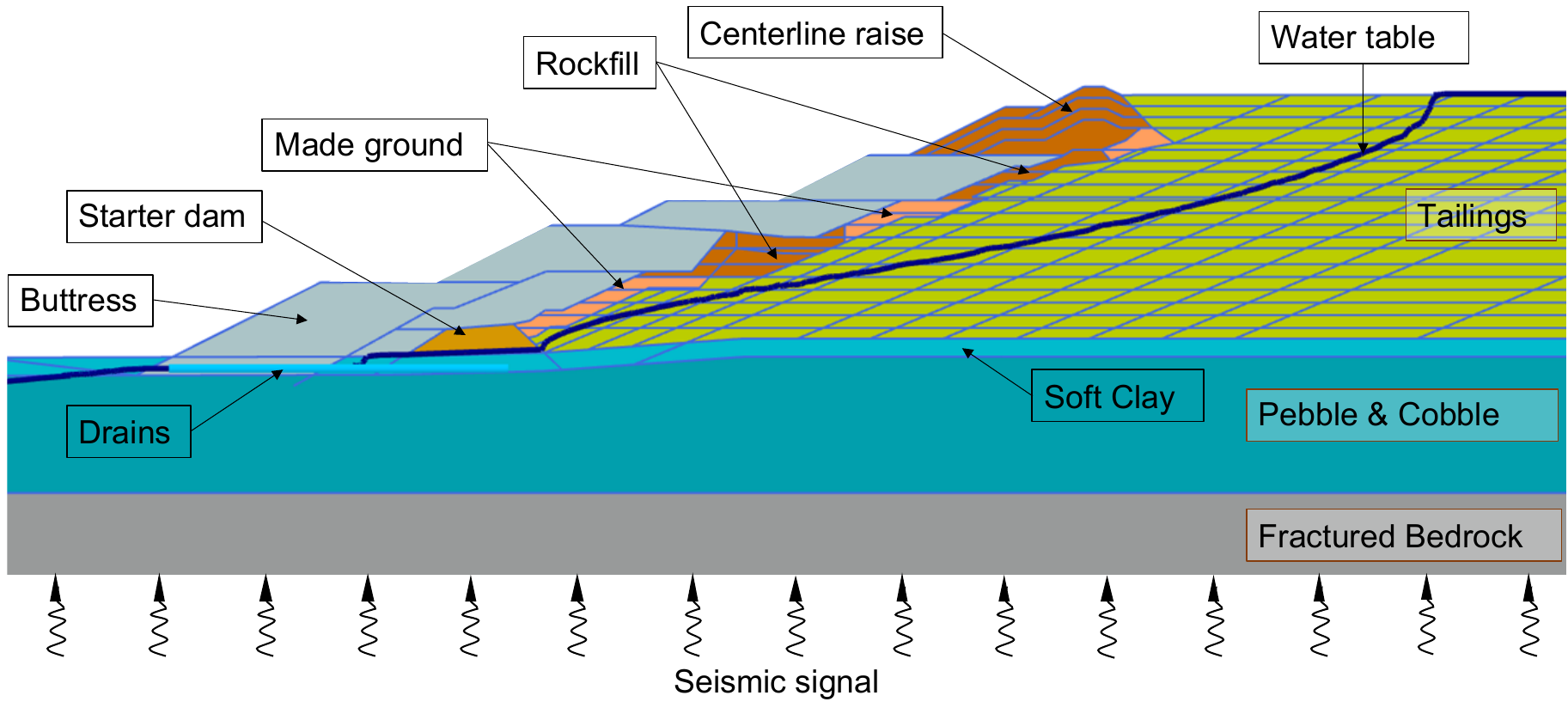}}
\end{center}
\caption{TSF \sout{shown as example. Geotechnical units.}\NL{considered in the paper and its geotechnical units.}}
\label{fig:TailingMaterials}
\end{figure}

\sout{The process for the deformation \sout{modelling} of a TSF is shown by example}\NL{As an example, the process for the deformation modeling of a TSF is shown}. The TSF considered \sout{here }is an upstream-raised facility located in a region with high seismicity. It has a current approximate height of 70 $m$; an additional raise of 10-15 \NL{$m$} is expected to be deposited in the next few years. The facility consists of a starter dam, embankment raises forming a 3H:1V slope; a sandy silt/silty sand loose tailings body; and a reinforcement buttress built at the toe. The combination of loose tailings and high seismicity results in a high risk of liquefaction, should \NL{the} tailings be near saturated or saturated. \sout{An}\NL{A} schematic \sout{cross section}\NL{cross-section} of the TSF in its final configuration is presented in Figure \ref{fig:TailingMaterials}. This is a fairly large installation where the value of numerical deformation \sout{modelling}\NL{modeling} is apparent.

\subsection{Numerical model}

A numerical model must be set-up and many runs must be carried out in order to estimate the seismic demand of the TSF or, in other words, the required freeboard and \sout{-maybe-}the need of a reinforcement buttress. In this section, a brief description of the numerical model used in this example is presented.

\subsubsection{Mesh and boundary conditions}

The \sout{models was}\NL{model is} \sout{made}\NL{generated} in Plaxis 2019 \NL{finite element software}. The mesh has 12281 \sout{6-nodes}\NL{6-node} triangular elements with an average size of 2.8 $m$,  a maximum size of 27.27 $m$ and a minimum size of 0.18 $m$.

Far field boundary conditions \sout{were}\NL{are} used at the left and right boundaries. A compliant base \sout{was}\NL{is} employed, where horizontal acceleration time signals \sout{were}\NL{are} introduced as input \NL{at the bottom} for the dynamic modeling. \NL{For simplicity,} no vertical accelerations \sout{were}\NL{are} considered in this study.

\subsubsection{Constitutive models}

HS-Small \NL{model} \sout{was}\NL{is} used \sout{for simulating}\NL{to simulate} the construction sequence \sout{and static loading }of all materials other than \NL{the} bedrock, where linear elasticity \sout{was}\NL{is} employed. \NL{Once the final configuration is reached,} HS-Small \NL{model} \sout{was}\NL{is} changed \sout{for}\NL{to} PM4Sand \NL{model only} for the tailings. Immediately before the application of the time-history ground motions, \sout{and }a nil-step \sout{was}\NL{is} \sout{inserted}\NL{performed} to assure a fully converged \sout{starting point for the }\NL{state prior to the} dynamic analyses. 

It is worth noting that HS-Small \NL{model} \sout{was}\NL{is} preferred \sout{for simulating}\NL{to simulate} the construction stages \NL{in order} to overcome \NL{the} limitations of the PM4Sand model \NL{when dealing with} monotonic compression stress paths and to produce a realistic stress-field in the dam body, relevant \sout{for}\NL{to} the dynamic response of the dam. An insight of the model calibration is provided in the next section. 

\subsubsection{Construction sequence}

\begin{figure} [h!]
\begin{center}
{\includegraphics[height=14.0cm]{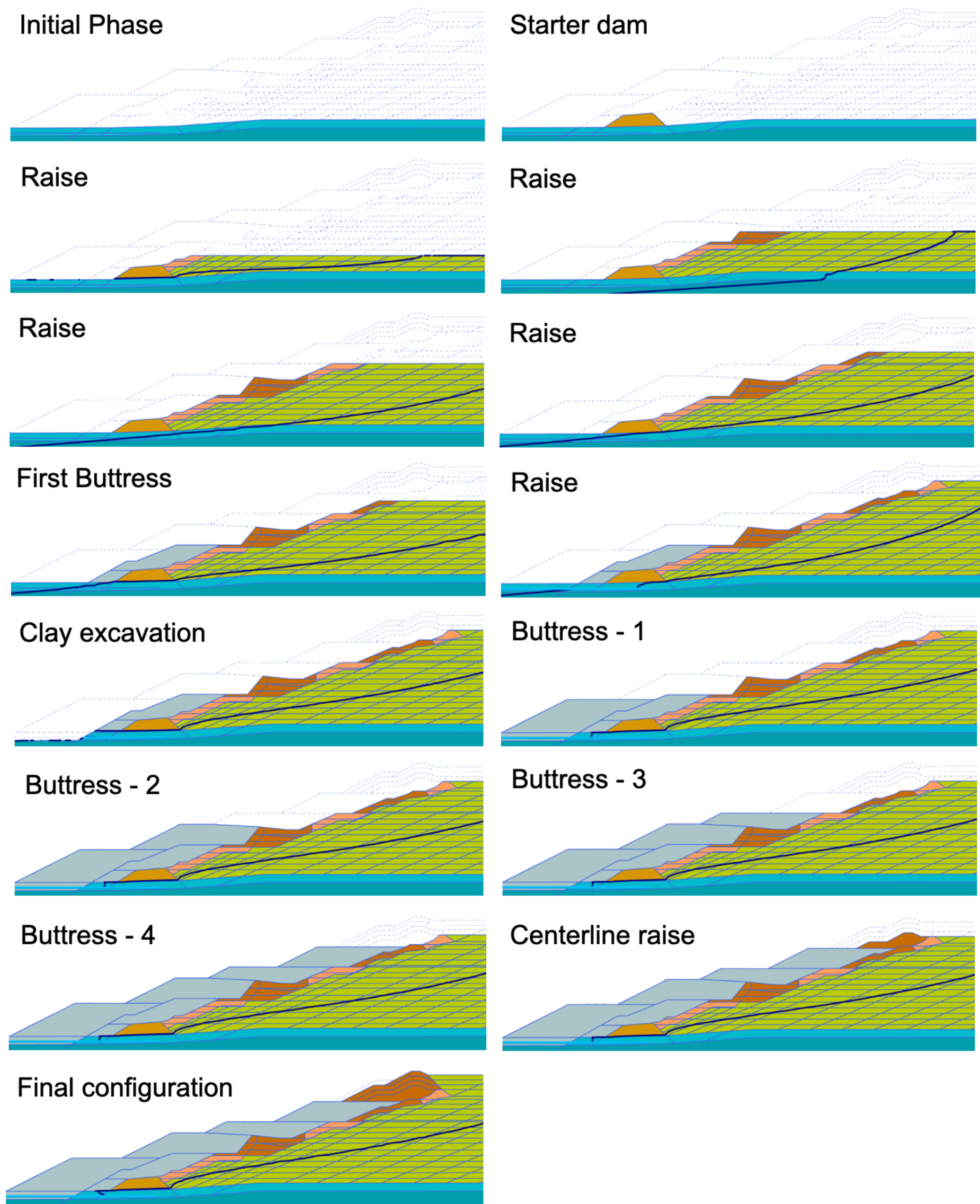}}
\end{center}
\caption{Simulation of dam construction and raises.}
\label{Fig:RaiseDam}
\end{figure}

Figure \ref{Fig:RaiseDam} shows the construction sequence \NL{that is used to simulate the staged construction prior to the dynamic deformation analyses}. \NL{During the dam rise, the deposition of each layer modeled using subsequent consolidation-type stages, allows excess pore pressure generation and dissipation}. After reaching the final configuration, a full consolidation of excess pore pressures \sout{was permitted}\NL{is allowed}, \NL{since the design earthquake will take place after a recurrence period of 7500 years, following ANCOLD guidelines \cite{Ancold2019}}. \sout{Then, the constitutive model \sout{of}\NL{used for the} tailings was changed from \sout{HSSmall}\NL{HS-Small} \NL{model} to PM4Sand \NL{model} and a plastic nil step was performed to ensure a full converged state as a starting point for the dynamic runs.}

\subsubsection{Rayleigh damping}

Rayleigh damping is required to account for the small-strain damping\NL{,} which is not captured adequately by both, HS-Small and PM4Sand  \NL{models}.  To determine  Rayleigh damping, the fundamental frequency of the tailings body is computed as:

\begin{equation}
f_{1} = \frac{V_{s,av}}{4H} \sout{,}
\label{eq:freq}
\end{equation}
where \sout{$v_{s,av}$}\NL{$V_{s,av}$} is the average shear wave velocity in the tailings body, \NL{$f_{1}$ is the fundamental frequency of the system} and $H$ is the \NL{maximum} height of the TSF. Far behind the crest of the slope, the mean pressure at mid-height of the TSF is approximately 200 $kPa$. Using the parameters introduced in Section \ref{section:Const}, the average shear modulus is:
\begin{equation}
 G_0= {G^{ref}_0} \left(\frac{p}{p_{ref}} \right)^{m} =45 \left[MPa\right] \left(  \frac{200 \left[kPa\right] }{100 \left[kPa\right]}  \right)^{0.75}  =75 \left[MPa\right] \sout{,}
 \label{eq:G0}
\end{equation}
which translates into an average shear wave velocity of \sout{$v_ {s,av} =200 \ m/sec$}\NL{$V_ {s,av} =200 \ m/sec$} and a natural frequency of about $0.70 \ Hz$. If the same calculation is done at the crest of the final elevation, the fundamental frequency is approximately $0.90 \ Hz$. Therefore, the expected range of the tailings \sout{first mode} natural frequency is \NL{from} $0.70$ to $0.90 \ Hz$.
At the buttress, the first natural frequency mode is estimated to be approximately $2.5 Hz$.  \sout{Following}\NL{In addition} \cite{Hudson94}, an upper bound target frequency \NL{$f_2$} should be:

\begin{equation}
f_2 = \frac{f_{fund}}{f_1} \sout{,}
\label{eq:freq2}
\end{equation}
where $f_{fund}$ is the fundamental frequency of the signal,\sout{ a representative value of $4 \ Hz$ is adopted for practical purposes,} \NL{being $4 \ Hz$ the most representative value for the considered set of seismic records}. Then, the upper bound target frequency is about $5 \ Hz$ for an average tailings first natural frequency mode of $0.80 \ Hz$.
In this case, values adopted for the Rayleigh coefficients $\alpha$ and $\beta$ are $0.04189$ and $0.00212$ respectively, \sout{such}\NL{so} that: the damping is 1$\%$ \sout{between}\NL{from} 0.4 \sout{and}\NL{to} 1.0 $Hz$, \sout{between} 1.0 to 2.0 $\%$ for frequencies \sout{between}\NL{from} 1.0 \sout{and}\NL{to} 3.0 $Hz$, and 2.0 to 3.0 $\%$ for frequencies \sout{between}\NL{from} 3.0 \sout{and}\NL{to} 5.0 $Hz$, \NL{which is} the estimated upper bound. \NL{This is aligned with Ref.} \cite{Boulanger2017}.

\section{Constitutive models calibration for tailings}
\label{section:Const}

\MS{The mechanical behavior of the tailings material has been characterized by means of laboratory and field testing. Full details of isotropically consolidated drained and undrained triaxial compression tests (CIDC /CIUC) results can be found in Ref. \cite{Sottile2019}. In addition, the authors provide the calibration of NorSand model, along with numerical cavity expansion analyses and inverse interpretation of the state parameter along a similar cross-section of the TSF using CPTu data (after Ref. \citep{Shuttle2016}). The calibration of the NorSand model parameters is presented for a fine and coarse tailings sample, named Sample A and B, respectively. Table \ref{t:CSL} shows a summary of the parameters that define the Critical State Line (CSL) according to the expression shown in Eq. \eqref{eq:CSL}. Given the tailings complex deposition sequence, an average CSL is used in this study.

\begin{equation}
e_{c} = \Gamma - \lambda_{10} \cdot \log \ p'  \sout{,}
\label{eq:CSL}
\end{equation}

\begin{center}

\begin{table}[h!]
\caption{Critical state line parameters.}
\label{t:CSL}
%\tabletext{
\centering
\begin{tabular}{llllll}\hline
    Sample   & $\Gamma$ & $\lambda_{10}$ & $e_{min}$ & $e_{max}$ \\\hline
\sout{ CS Data $N^{\circ}1$}\NL{Sample A }  & 1.02 & 0.170 & 0.285 & 1.218  \\
\sout{ CS Data   $N^{\circ}2$}\NL{Sample B} & 1.12 & 0.175  &0.229 & 1.114 \\
 Average      & 1.07 & 0.172  & 0.257 & 1.166 \\
\hline
\end{tabular}%}
\end{table}

\end{center}

The PM4Sand model calibration strategy is summarized as follows: i) the average NorSand CSL is converted from $e - \log p'$  to  $D_r - \log p'$ space using the average minimum and maximum void ratios, $e_{min}$ and $e_{max}$ respectively; ii) the PM4Sand parameters that define the CSL, $Q$ and $R$, are adjusted to approximate to the NorSand CSL within a representative stress range; iii) the contraction rate parameter ($h_{p0}$) is calibrated using two available cyclic direct simple shear tests (CDSS); iv) a representative relative density ($D_r$) is computed at different sub-regions of the TSF to reflect the in-situ state of the material, as a function of the mean effective stress ($p'$) before the earthquake and the already interpreted state parameter ($\psi$). The remaining PM4Sand parameters are adopted with default values proposed by Ref. \cite{ZIOTOPOULOU2013268}.

A comparison between the averaged NorSand model and the calibrated PM4Sand model CSL is shown in Figure \ref{fig:CSL}. The calibration of PM4Sand $Q$ and $R$ parameters aims to approximate the NorSand model CSL to the range of stresses that are of interest for this study, which varies from $50$ to $600 \ kPa$. Values of $Q=11.8$ and  $R=3.9$ are obtained; as shown in Figure \ref{fig:CSL}, the matching is acceptable.

\begin{figure}[h!]
\begin{center}
\subfigure[Void Ratio $e$ versus mean effective pressure $p'$ plot.]{\includegraphics[height=5.75cm]{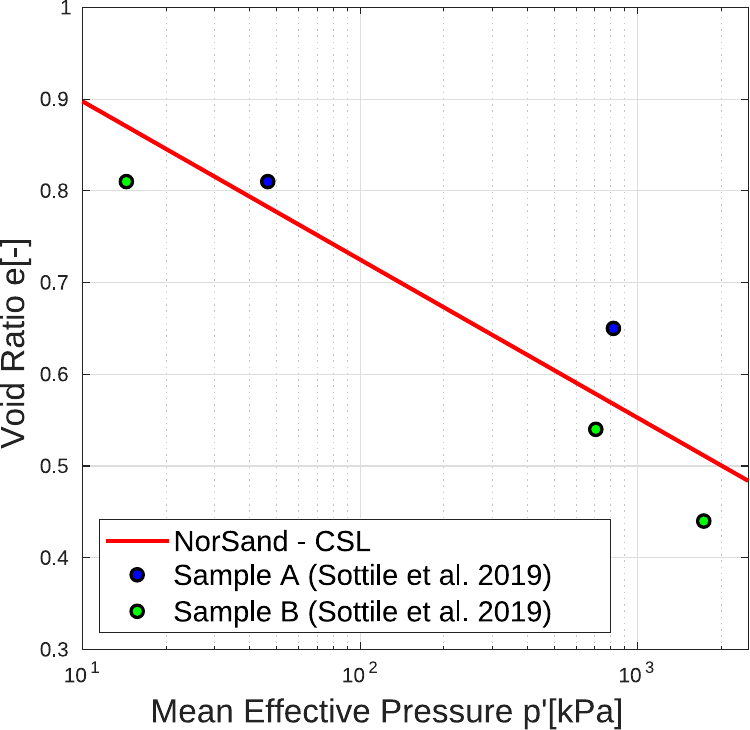}}
\subfigure[Relative density $D_r$ versus mean effective pressure $p'$ plot.]{\includegraphics[height=5.75cm]{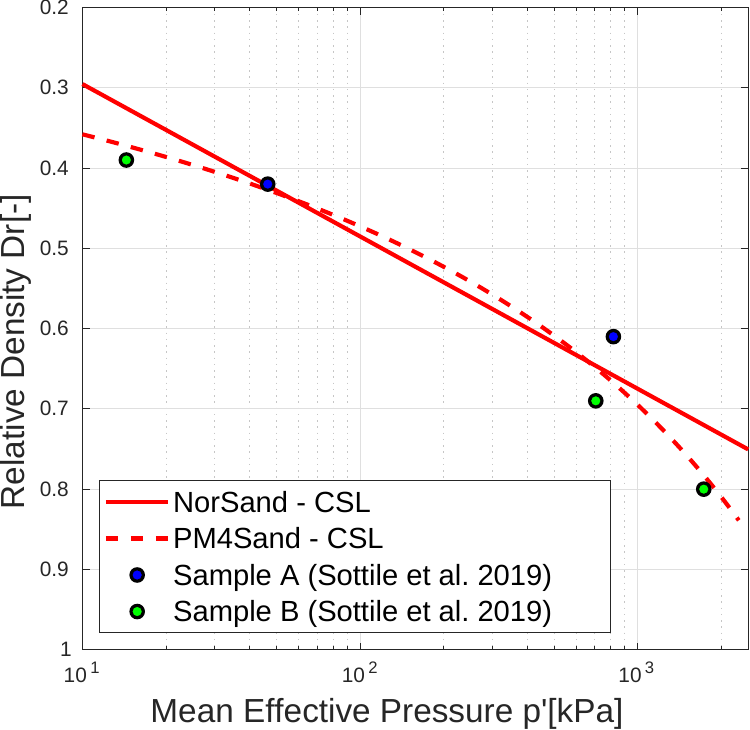}}
\end{center}
\caption{Critical State Line (CSL) calibration and comparisons with isotropically consolidated undrained and drained tests.}
\label{fig:CSL}
\end{figure}

Two Cyclic Direct Simple Shear tests (CDSS) were performed on the coarser tailing (Sample B). The results are shown in Figure \ref{fig:test1} and are summarized as follows: i) Test 1 entails an initial vertical effective stress of $\sigma'_{v}=200 \ kPa$  and an initial void ratio of $e_0=0.642$; for a Cyclic Stress Ratio $CSR = 0.12$, 30 cycles were needed to fail the sample (defined as shear strain of $3.75\%$); ii) Test 2 had $\sigma'_{v}=200 \ kPa$  and $e_0=0.523$, corresponding to a denser state than Test 1; for $CSR = 0.12$, 69 cycles were needed to results in failure. It must be noted that the initial configuration for both tests entails negative state parametes, ranging from $-0.10$ to $-0.20$ approximately; therefore, for looser configurations, a lower amount of cycles for a given $CSR$ is expected in order to reach failure.  Elemental CDSS tests were performed using PM4Sand model and compared with the available test, from which a conservative  value $h_{p0} = 0.75$ is determined. A summary of the PM4Sand parameters is shown in Table \ref{t:PM4sand}.

\begin{figure}[h!]
\begin{center}
\subfigure[PM4Sand  \NL{model} vs. laboratory tests (N$^{\circ}$1)]{\includegraphics[height=8.35cm]{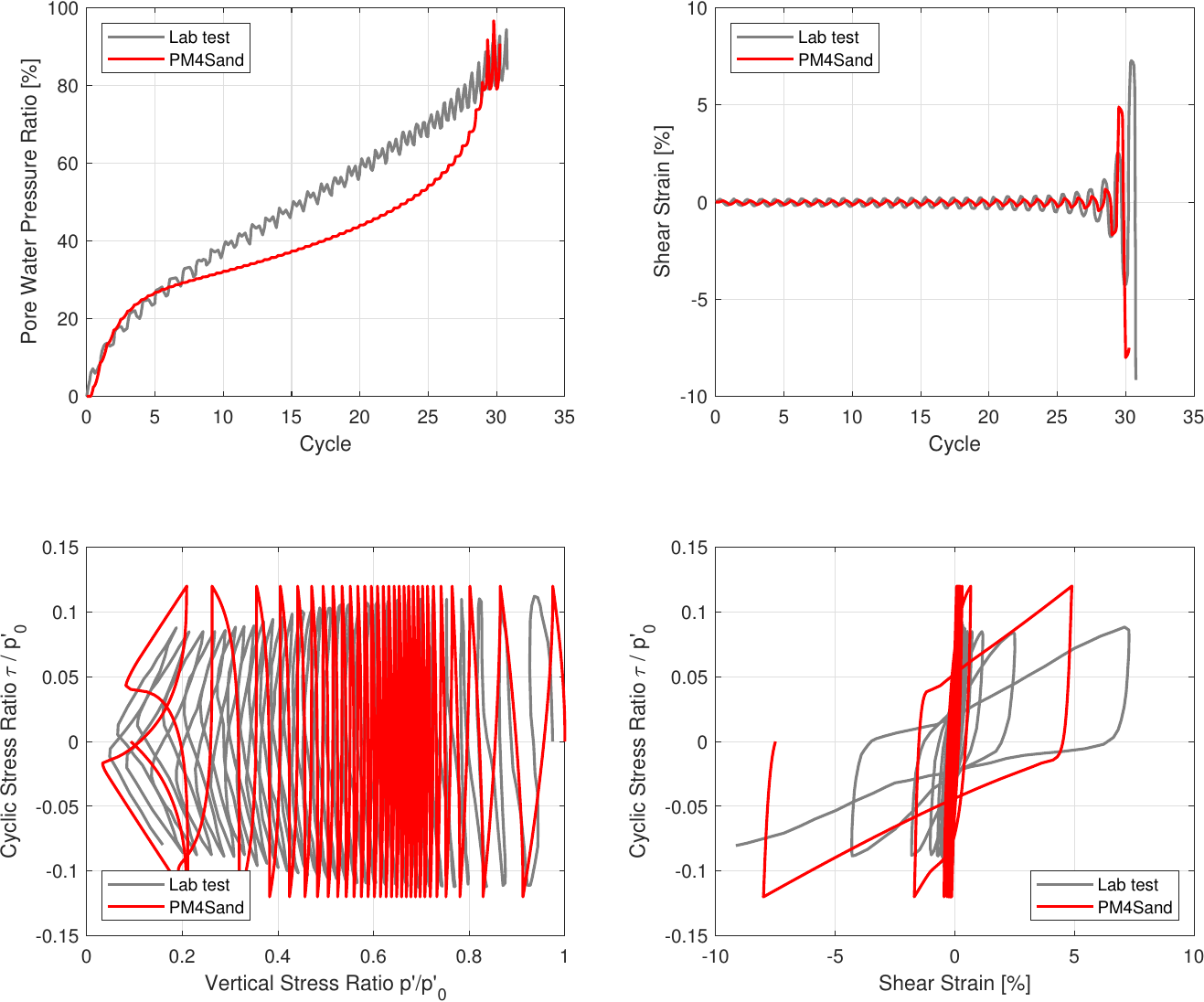}}
\subfigure[PM4Sand  \NL{model} vs. laboratory tests (N$^{\circ}$2)]{\includegraphics[height=8.35cm]{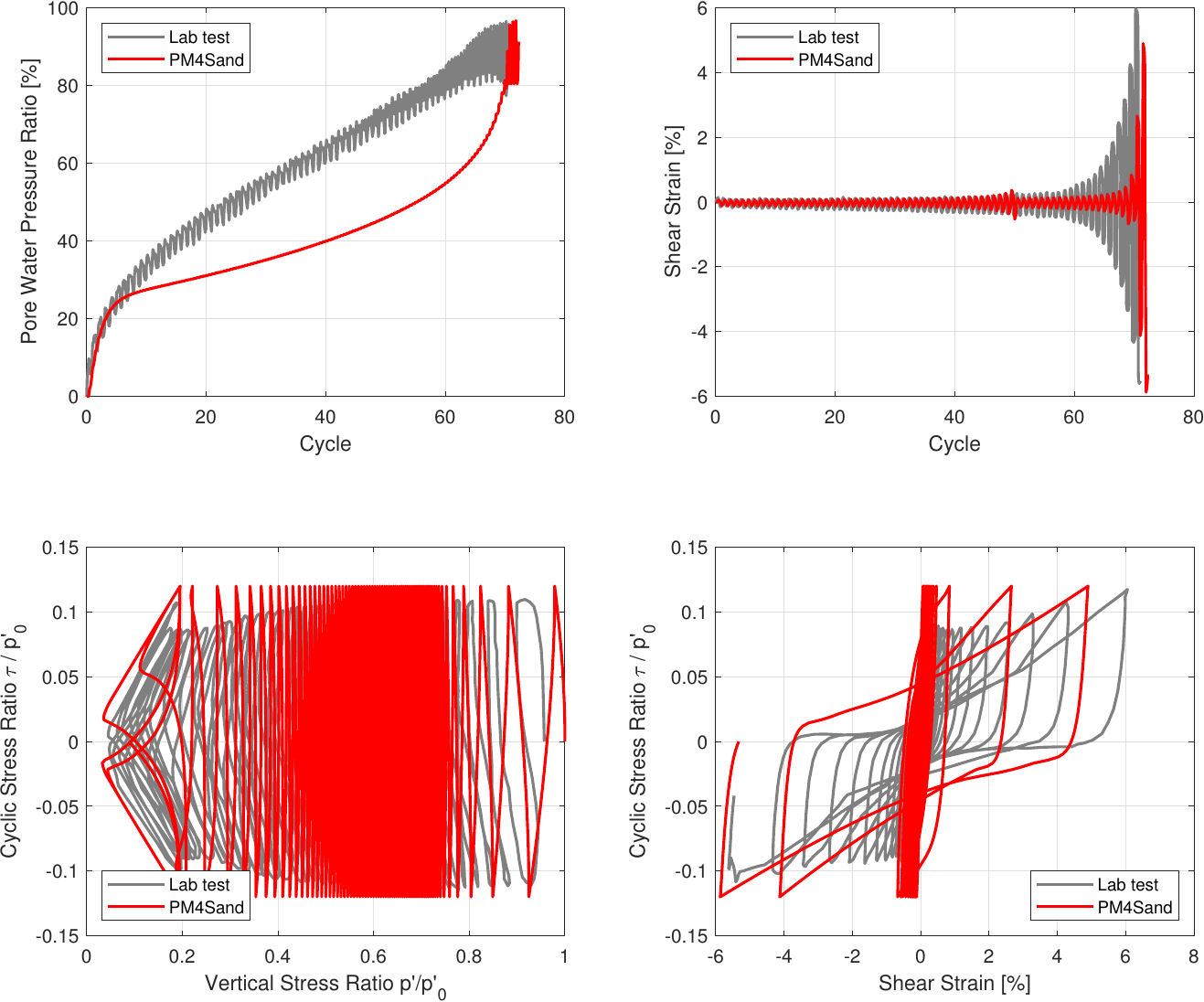}}
\end{center}
\caption{Comparison between PM4Sand \NL{model} and CDSS tests.}
\label{fig:test1}
\end{figure}

\begin{center}
\begin{table}[h!]
\caption{PM4Sand \NL{model} parameters for tailings, used in dynamic stages.}
\label{t:PM4sand}
%\tabletext{
\centering
\begin{tabular}{llll}\hline
       & Unit & Parameters\\\hline
 $\gamma_{sat}$  & $kN/m^3$ & 18.0 \\
$G_0$     &         -        &  450    \\
$h_{p0}$ &  -    &  0.75 \\
$p_{ref}$&  $kPa$    &  103 \\
$e_{max}$& -    & 1.166 \\
$e_{min}$&  -   & 0.257 \\
$n^b$  &  -     &  0.50 \\
$n^d$ & -  & 0.10 \\
$\phi'_{cv}$ & $^{\circ}$  & 32 \\
$\nu$  & -  & 0.20 \\
$Q$ & -  & 11.8 \\
$R$ & -  & 3.4 \\
 $k$          &  m/s  & $1x10^{-7}$ \\
\hline
\end{tabular}%}
\end{table}
\end{center}
\sout{To relate the state parameter with the in-situ relative density, the initial state parameter is defined as $\psi_0=e_0-e_c$, where $e_c$ is the void ratio at the critical state, is first computed using the methodology proposed by Ref. \cite{Shuttle2016}, which entails the calibration of NorSand model, numerical cavity expansion analyses and inverse interpretation of CPTu data. Details of this procedure for these tailings can be found in Ref. \cite{Sottile2019}. The CSL is calculated as}

After calibrating all the PM4Sand model parameters, a representative relative density is determined at each region of the TSF model. Figure \ref{Fig:CalibrationPM} shows the details of this procedure, which is summarized as follows: 1) the mean effective stress distribution at the TSF is obtained before the earthquake event, and an average value is computed at each soil cluster of the model; 2) the state parameter is interpreted using CPTu data (after Ref [31]) and four different regions of constant $\psi$ are defined: $0.000$, $0.025$, $0.050$ and $0.075$ ; 3) using the mean effective stress at each soil cluster and the assigned state parameter, a relative density is computed using the equation
\begin{equation}
D_r = \frac{e_{max} - \left( e_c + \psi \right)}{e_{max} - e_{min} } ,
\label{eq:relativedensity}
\end{equation}
where  $e_c$   is the void ratio at the critical state line for the mean effective stress $p'$  average over the soil cluster; 4) using a Python script, a PM4Sand material with a proper relative density is generated for each region of the model and assigned respectively. 
\begin{figure}
\begin{center}
{\includegraphics[width=11.85cm]{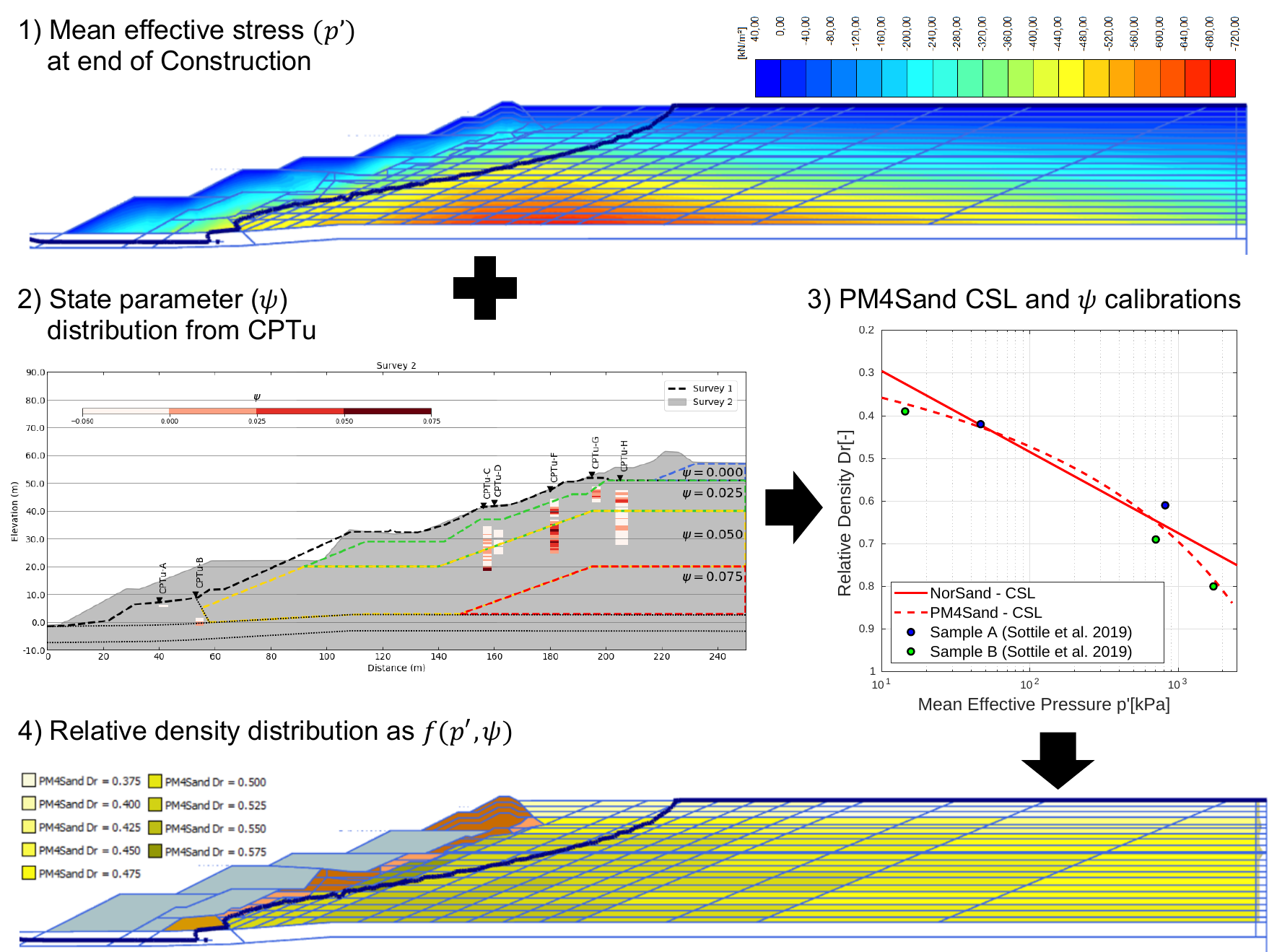}}
\end{center}
\caption{\NL{PM4Sand relative density definition strategy.}}
\label{Fig:CalibrationPM}
\end{figure}
}
\sout{PM4Sand model parameters were calibrated from CPTu data and remolded lab tests. Two Cyclic Direct Simple Shear tests (CDSS) were used: Test 1 had an initial vertical effective stress $\sigma'_{v}=200 \ kPa$ and an initial void ratio $e_0=0.642$. For a Cyclic Stress Ratio $CSR=0.12$, 30 cycles were needed to fail the sample. Test 2 had $\sigma'_{v}=200 \ kPa$ and $e_0=0.523$,corresponding to a denser state than Test 1. For $CSR=0.12$, 69 cycles were needed to achieve failure. 
PM4Sand  formulates the CSL in terms of relative density $D_r$ rather than void ratio $e$ or state parameter $\psi$ \cite{Been1985}. 
The resulting CSLs predicted by NorSand and PM4Sand  are presented in Figure \ref{fig:CSL} in terms of $D_r$ (a) and $e$ (b).
A comparison between the CDSS tests and the simultations are shown in Figure \ref{fig:test1}; a reasonably good fit is obtained.}
%\begin{figure} [h]
%\begin{center}
%{\includegraphics[height=2.75cm]{State_Parameter_distr2.png}}
%\end{center}
%\caption{Distribution of the state parameter $\psi$ within the tailings body.}
%\label{Fig:StateParam}
%\end{figure}
%
%\begin{figure} [h]
%\begin{center}
%{\includegraphics[height=3.8cm]{DRdistribution2.png}}
%\end{center}
%\caption{Distribution of relative density $D_r$ within the tailings body on the excess pore pressure contour.}
%\label{Fig:DRdistrib}
%\end{figure}
On the other hand, Table \ref{t:HSmall} presents the parameters of HS-Small \NL{model used for the dam raising and validation of the proposed seismic selection method}\sout{; Table \ref{t:PM4sand} those of PM4Sand {model}}. For the sake of brevity, the calibration of HS-small \NL{model} is not presented here.
\begin{center}
\begin{table}[h!]
\caption{\sout{HSsmall}\NL{HS-Small} parameters for tailings, used in static stages.}
\label{t:HSmall}
%\tabletext{
\centering
\begin{tabular}{llll}\hline
       & Unit & \sout{HSsmall}\NL{HS-Small}  \\\hline
 $\gamma_{sat}$  & $kN/m^3$ & 18.0 \\
 $\phi'$   & $^{\circ}$ & 32   \\
 $c'$       & $kPa$ & 0   \\
 $\psi$        & $^{\circ}$ & 0   \\
 $G^{ref}_0$          & $MPa$ & 45 \\
 $\gamma_{0.7}$          & - &  $10^{-4}$   \\
 $E^{ref}_{ur}$          & $MPa$ & 60 \\
 $E^{ref}_{50}$          & $MPa$ & 5   \\
 $E^{ref}_{oed}$          & $MPa$ & 9   \\
 $m$          & - & 0.75    \\
 $\nu_{ur}$          & - & 0.20   \\
 OCR          & - & 1.00   \\
 $K^{nc}_{0}$          & - & 0.50  \\
\hline
\end{tabular}%}
\end{table}
\end{center}

\section{\sout{A-priori}\NL{A priori} estimator for the demand of seismic records}
\label{section:apriori}

This section represents the core of the paper  where, first, an intensity measure which is a generalization of Arias Intensity based on spectral power is mathematically defined and then, calculations are performed to a set of seismic signals and the proposed IM is determined for \sout{each earthquake}\NL{each earthquake}.

\subsection{Definition of an intensity measure based on spectral power}

Calculations to obtain the spectral power content of a seismic signal are based on the Fast Fourier Transform (FFT), an efficient implementation of the Discrete Fourier transform (DFT). The formulation is stated in a discrete space, in order to explain the physical aspects more clearly. 

Let $\left\lbrace \boldsymbol{a}_n \right\rbrace = a_0, a_1, ..., a_{N-1}$ be a finite set of $N$ elements uniformly spaced of time-history accelerations, the DFT is defined by means of Euler's formula:
\begin{equation}
\begin{split}
\mathscr{F}\left\lbrace \boldsymbol{a}_n \right\rbrace \left( k \right) = \left\lbrace \boldsymbol{A}_k \right\rbrace &= \sum^{N-1}_{n=0} a_n \cdot e^{-i2 \pi k\frac{n}{N}} \\
&= \sum^{N-1}_{n=0} a_n \cdot \left[\cos\left(\frac{2 \pi}{N} k n\right) + i \cdot \sin\left(\frac{2 \pi}{N} k n\right) \right] \sout{,}
\end{split}
\label{eq:DFT}
\end{equation} 
where $ \left\lbrace  \boldsymbol{A}_k \right\rbrace$ is a set of complex vectors which represents  the amplitude and phase of a complex sinusoidal component and $k$  an integer representing the frequency domain. 

%As the goal is to analyze the power in each time step of the seismic signal, a generalization of the DFT is considered, the so-called Discrete Short-time Fourier transform (STDFT), defined as:
%\begin{equation}
%\mathscr{F}_{s}\left\lbrace \boldsymbol{a}_n \right\rbrace \left( \omega_n  , k \right)  =  \sum^{N-1}_{n=0} a_n \ \omega_n \cdot e^{-i2 \pi k\frac{n}{N}}. 
%\label{eq:STDFT}
%\end{equation} 

The power spectrum density in terms of the frequency is defined as 
\begin{equation}
S_{xx}\left( k \right) = \| \mathscr{F} \left\lbrace \boldsymbol{a}_n \right\rbrace \left( k \right) \|^2 \sout{,}
\label{eq:PSD}
\end{equation} 
while the total spectral power of the signal is expresed as
\begin{equation}
P_{0 - \infty} = \sum^{\infty}_{k=0} S_{xx}\left( k \right) \Delta k \sout{,}
\label{eq:PS}
\end{equation} 
with $\Delta k$ being the frequency sampling. 
In this paper, the seismic intensity measure used is a windowed version of the spectral power expressed in equation \eqref{eq:PS}, where the highest frequency considered for the calculations is the limit of power accumulation i.e., $P_{0 - X \ Hz}$ is the accumulated power between the frequencies 0 to  $X \ Hz$. 
It is worth noting that, due to Parseval's Theorem
\begin{equation}
\sum^{N-1}_{n=0} \|  a_n \|^2 = \sum^{\infty}_{k=0} \|   \mathscr{F} \left( k \right) \|^2 \sout{,}
\label{eq:parseval}
\end{equation} 
which means that, when the considered power spectrum is computed considering the full frequency domain, the intensity measure expressed in terms of the spectral decomposition tends to represent classical intensity measures based on the integration of the seismic signal such as the Arias intensity. In this way, the IM proposed here is stated as a generalization of these classical IMs. 

Finally, the spectrogram expressing a signal decomposition in terms of time, frequency and spectral power, is plotted in terms of the spectral power expressed in decibels $dB$, computed as
\begin{equation}
P_{dB} = 10 \log_{10} \left( \frac{P}{P_{r}}\right) \sout{,}
\label{eq:dbpower}
\end{equation}
where $P$ is the computed spectral power and $P_r = 10^{1.5}$ is a reference power. The reference power behaves like a simple shift in the accumulated power and does not modify the proposed correlation.

\subsection{Application to a set of seismic records}

In order to evaluate the liquefaction risk of the TSF described in  Section \ref{section:tailing}, a set of 25 seismic records \sout{were}\NL{are} selected based on the \sout{50th}\NL{85th} percentile of the 7500-year event, for which a  peak ground acceleration (PGA) of 0.78 $g$ is expected. Twenty seismic records correspond to seismographs located on dense soil/soft rock (NEHRP site class C); two (records 3 and 8) belong to firm/hard rock (NEHRP site class B), and three (records 4, 7 and 23) belong to stiff soils (NEHRP site class D). 
For each seismic record, the acceleration-time signal and spectrogram \sout{were}\NL{are} computed by using short-time fourier transform  \citep{Oppenheim99} and presented in \ref{section:Appendix}, where a Hamming window is used for the calculations. 
The main characteristics of the records are summarized in Table \ref{T:SeismicRec}. The spectral power is included together with arias intensity (AI), Cumulative Accelerate Value (CAV) and Cumulative Accelerate Value above 0.05 $g$ (CAV5). Spectral power is expressed in relative terms, i.e. the argument of the logarithm,  as explained in equation \eqref{eq:dbpower}.

\begin{figure} 
\begin{center}
%\subfigure[]{\includegraphics[height=6.5cm]{SpecPwrWdw10Hz(2).pdf}}
%\subfigure[]{\includegraphics[height=6.5cm]{SpecPwrWdw15Hz.pdf}}

\subfigure[Spectral power calculated using a Hamming window fixed from  $0 Hz$.]{\includegraphics[height=5.1cm]{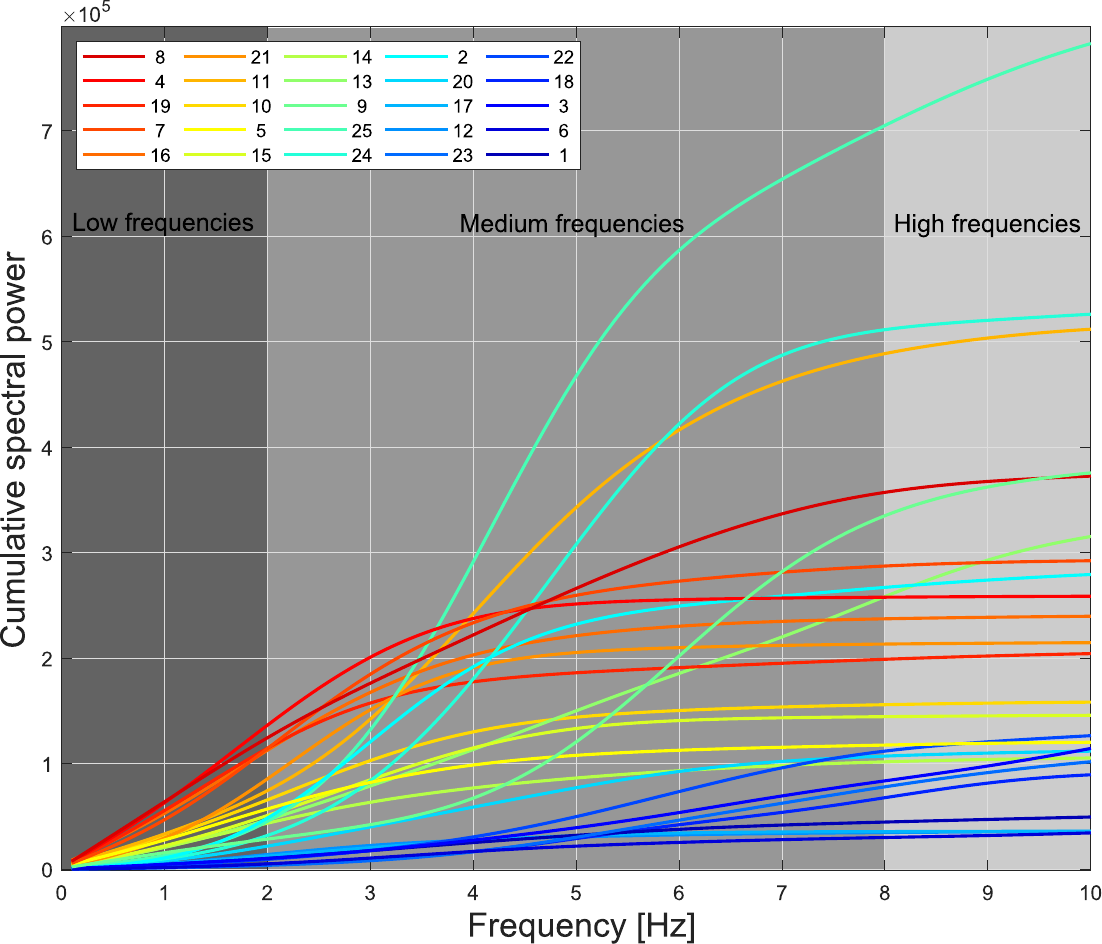}} 
\subfigure[Spectral power calculated using a Hamming window with $2 Hz$ width.]{\includegraphics[height=4.9cm]{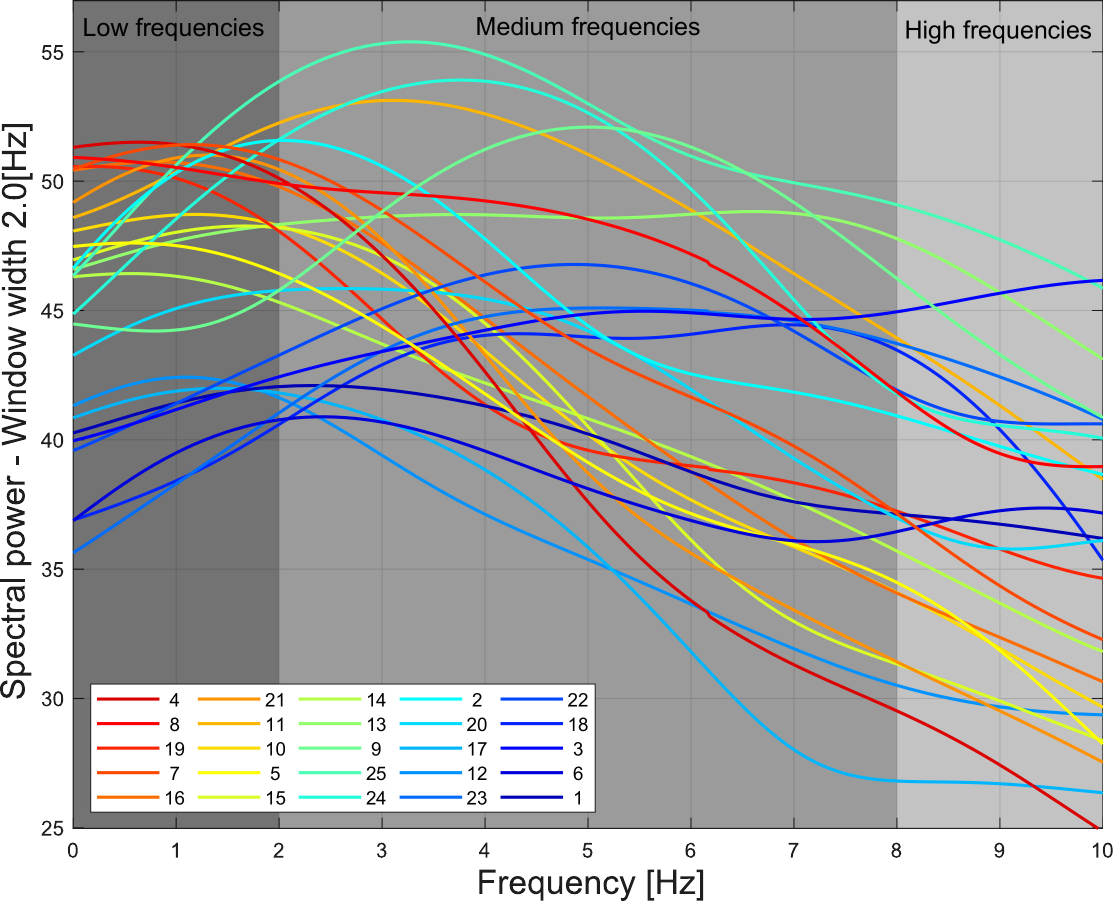}}
%\subfigure[]{\includegraphics[height=6.5cm]{CumulativeSpecPwrZOOM.pdf}}
\end{center}
\caption{Acumulated spectral power for the considered set of signals, obtained with sprectrograms presented in \ref{section:Appendix}.}
\label{fig:spectrogramsrecords_acumul}
\end{figure}

Figure \ref{fig:spectrogramsrecords_acumul} (a) plots the cumulative spectral power of the seismic records. Some of them have more power at low frequencies; as the frequency window widens, others increase their spectral power. The curves have been \sout{ordered such}\NL{organized so} that those with higher spectral power in low frequencies are plotted in red, while those with less power are plotted in blue. Seismic records 8, 4, 19 and 7 are those with higher spectral power in the window 0-2 $Hz$, while 18, 3, 6 and 1 are the ones with lower power in this range.  Another way to express these results is presented in Figure \ref{fig:spectrogramsrecords_acumul} (b) where a mobile Hamming window 2 $Hz$ wide is used. It can be seen that some signals like number 9 have power accumulated in medium frequency range, while their content for low and high \NL{frequency} is relatively low; others like number 8 have higher spectral power in the low band, while the medium and high bands are negligible.

In the following section, it will be demonstrated that the demand of a seismic record on the dam shown in Section \ref{section:tailing} is strongly correlated with the spectral power content at low frequencies.

\begin{landscape}
\begin{table}
	\centering
	\caption{Seismic records used for the seismic analysis, scaled to PGA of 0.78g.}
	\label{T:SeismicRec}
	\begin{small}
	\begin{tabular}[t]{lccccccccccc}
		\toprule
		ID & Event Name & Record Station & Duration & AI  & CAV   & CAV5 & $\frac{P_{0-1.5 \ Hz}}{P_r}$ & $\frac{P_{0-2.0 \ Hz}}{P_r}$  \\
		   &    &   & $\left[sec \right]$ & $\left [mm/sec\right]$ & $\left[mm/sec\right]$ & $\left[mm/sec\right]$ & $\left[-\right]$ & $\left[-\right]$ \\
		\midrule
		1 &	Nahanni Canada &	Site 1	& 10	&1931	& 7337 &	7340 & 7021\sout{.07} & 10627\sout{.58} \\
		2 &	Duzce Turkey &	Lamont 375 &	41	& 9445	& 25599	& 25956 & 24509\sout{.54} & 47292\sout{.67} \\
		3 &	Landers &	Lucerne &	48 &	7944 &	26919 &	26937 & 7116\sout{.28} & 9981\sout{.88} \\
		4 &	Kobe Japan &	KJMA &	150 &	9057	 & 20488 &	21410 & 98654\sout{.39 }& 135265\sout{.10 }\\
		5 &	San Salvador & Geotech Inv C. &	9 &	3764	 & 8832	& 8849 & 41613\sout{.12 }& 56137\sout{.66 }\\
		6 &	Parkfield-02 CA &	Parkfield - Cholame 3E &	21 &  1475 &	5274	 & 5566 & 2961\sout{.05} & 4877\sout{.42} \\
		7 &	Chuetsu-oki & Tamati Yone Izumozaki &	120 &	 9378 &	27126 &	28674 & 77178\sout{.63} & 112413\sout{.05} \\
		8 &	Tabas Iran & 	Tabas &	33 & 	11941 &	30548 &	31031 & 95086\sout{.15} & 124176\sout{.07} \\
		9 &	Iwate &	MYG004 &	300 &	12444 &	32509 &	34949& 22159\sout{.07} & 28378\sout{.75} \\
		10 & Northridge-01 &	Jensen Filter Plant Building &	29 & 	4947 &	13887 &	14166 & 46433\sout{.23} & 64496\sout{.35} \\
		11 &	 Chi-Chi Taiwan &	TCU071	& 90	 & 16834	 & 44434	 & 44986 & 49931\sout{.39} & 71563\sout{.90} \\
		12 & Mammoth Lakes-06	& Long Valley Dam  &	26 & 1289 & 6143	& 6566 & 9121\sout{.63} & 13607\sout{.46} \\
		13 & 	Loma Prieta &	WAHO	 & 25 &	11009 &	26645 &	26753 & 32168\sout{.89} & 45838\sout{.66} \\
		14 & 	Victoria Mexico &	Cerro Prieto &	25 & 	3579	 & 12868 &	13045 & 31633\sout{.45} & 42897\sout{.50} \\
		15 &	 Loma Prieta &	Corralitos &	40 & 	5868	 & 16206	 & 16543 & 34638\sout{.57} & 49401\sout{.24} \\
		16 &	 Chi-Chi Taiwan &	CHY028 & 	90 & 	7677 &	21532 &	22451 & 80261\sout{.05} & 110293\sout{.29} \\
		17 &	 Coalinga-05 & 	Transmitter Hill	& 22 & 	1205 & 	5174	 & 5558 & 8452\sout{.32} & 12217\sout{.04} \\
		18 & 	Chuetsu-oki	& Joetsu Oshimaku Oka & 	60 & 	3032	 & 10029	& 10865& 3412\sout{.88 }& 4918\sout{.03} \\
		19 & 	C. Mendocino &	Petrolia	 & 36	& 6556	& 18185	& 18899 & 87179\sout{.11} & 115902\sout{.93} \\
		20 & 	Coalinga-05 &	Oil City	& 21 &	 3760 &	8758 &	8955 & 13845\sout{.83} & 21345\sout{.23} \\
		21 & 	Chi-Chi Taiwan &	CHY041 & 	90 & 	6681 &	26188	& 26830 & 52214\sout{.76}  & 82736\sout{.82} \\
		22 &	 Christchurch NZ	& LPCC & 	24 & 	4737 & 	11037 &	11156 & 6221\sout{.59} & 9153\sout{.39} \\
		23 &	 N/A &	Ward Fire St &	164 &	3749	 & 9307 &	10112  & 2422\sout{.59}  & 3830\sout{.22} \\
		24 & N/A &	ANGOL &	180 &	17082 &	66227 &	68397 & 17876\sout{.66} & 30719\sout{.94} \\
		25 &	 N/A &	PICA	 & 252 &	27110 &	67414 &	69258 & 24022\sout{.93}  & 43339\sout{.76} \\
		\bottomrule
	\end{tabular}
	\end{small}
\end{table}
\end{landscape}

\begin{figure}
\begin{center}
%\subfigure[Arias intensity versus spectral power between 0 and 1.0 Hz]{\includegraphics[height=3.15cm]{AIvsPower0a1.pdf}}
\subfigure[Arias intensity versus spectral power between 0 and 2.0 $Hz$]{\includegraphics[height=4.75cm]{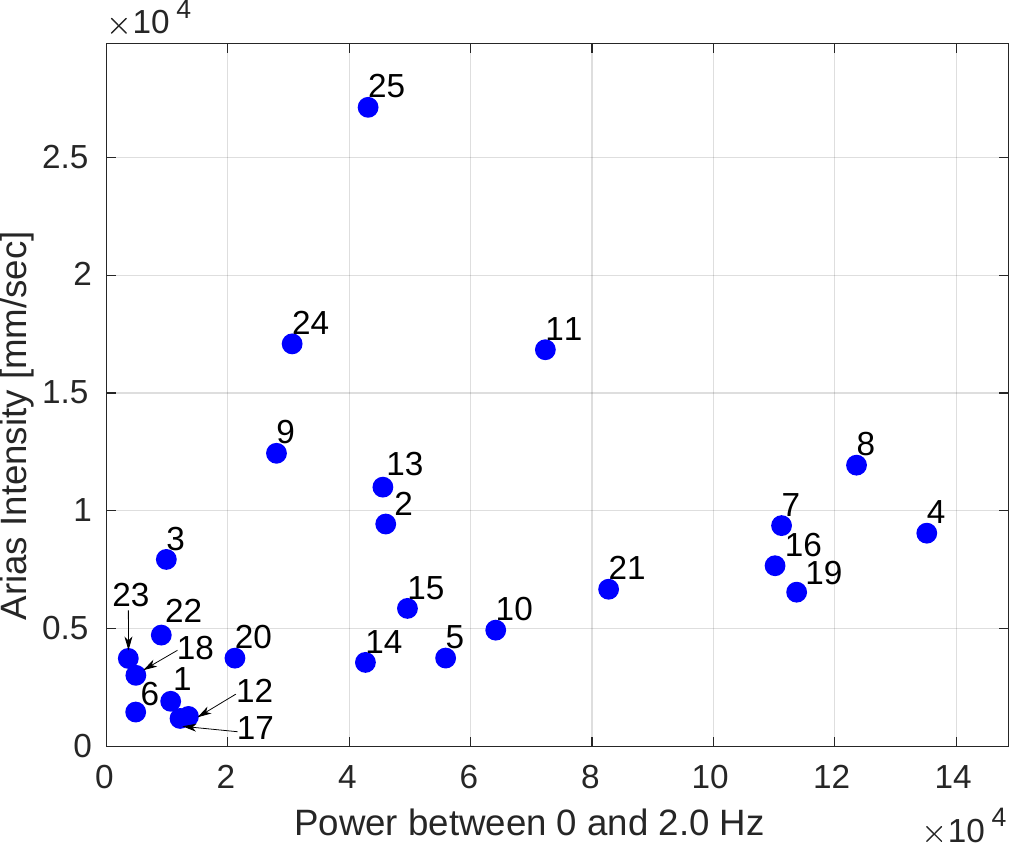}}
\subfigure[Arias intensity versus spectral power between 0 and 15.0 $Hz$]{\includegraphics[height=4.75cm]{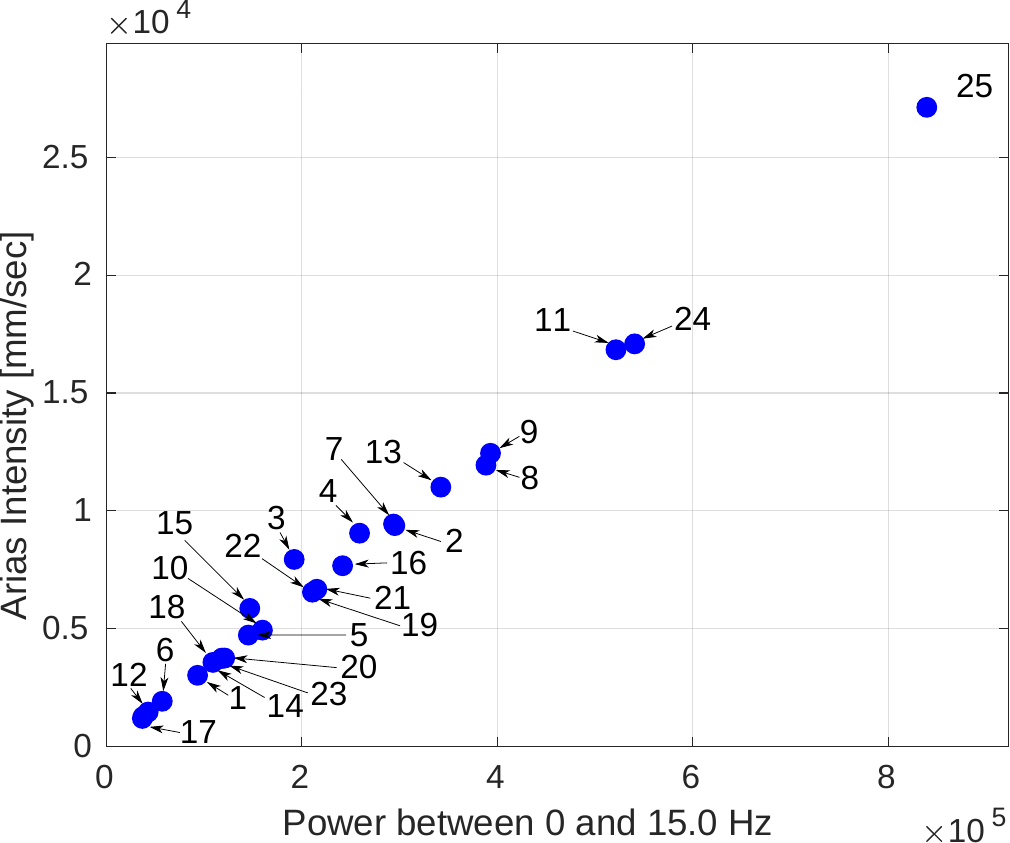}} \\
%\subfigure[Cumulative Absolute Velocity versus spectral power between 0 and 1.0 Hz]{\includegraphics[height=3.15cm]{CAVvsPower0a1.pdf}}
\subfigure[Cumulative Absolute Velocity versus spectral power between 0 and 2.0 $Hz$]{\includegraphics[height=4.75cm]{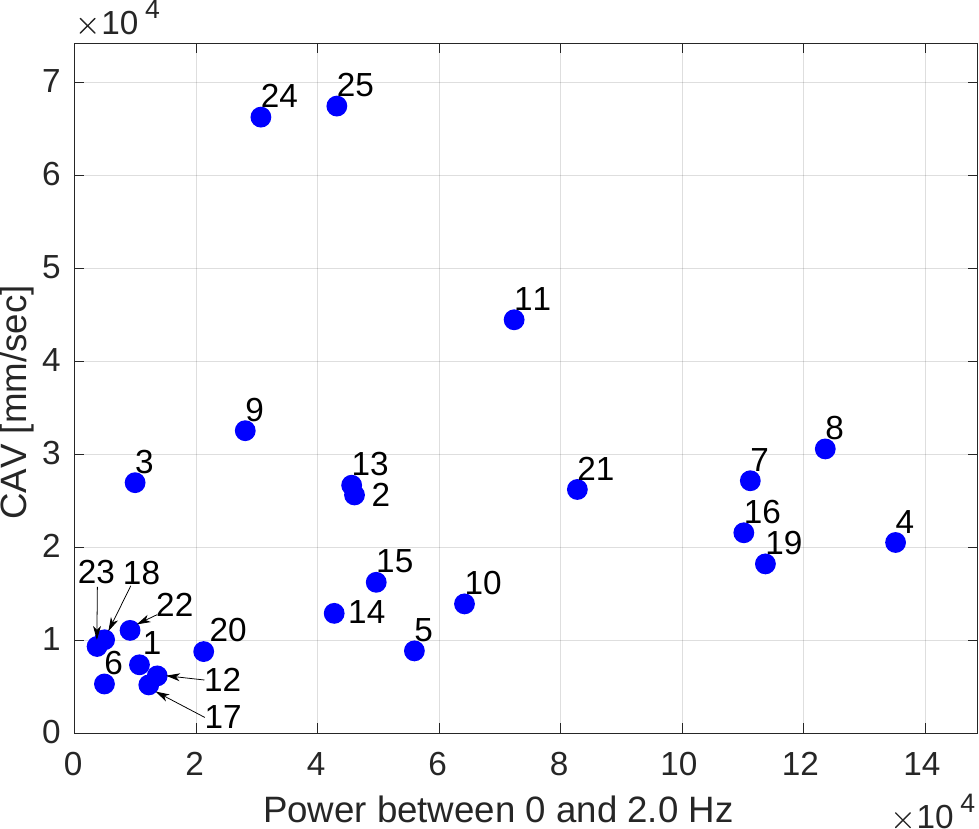}}
\subfigure[Cumulative Absolute Velocity versus spectral power between 0 and 15.0 $Hz$]{\includegraphics[height=4.75cm]{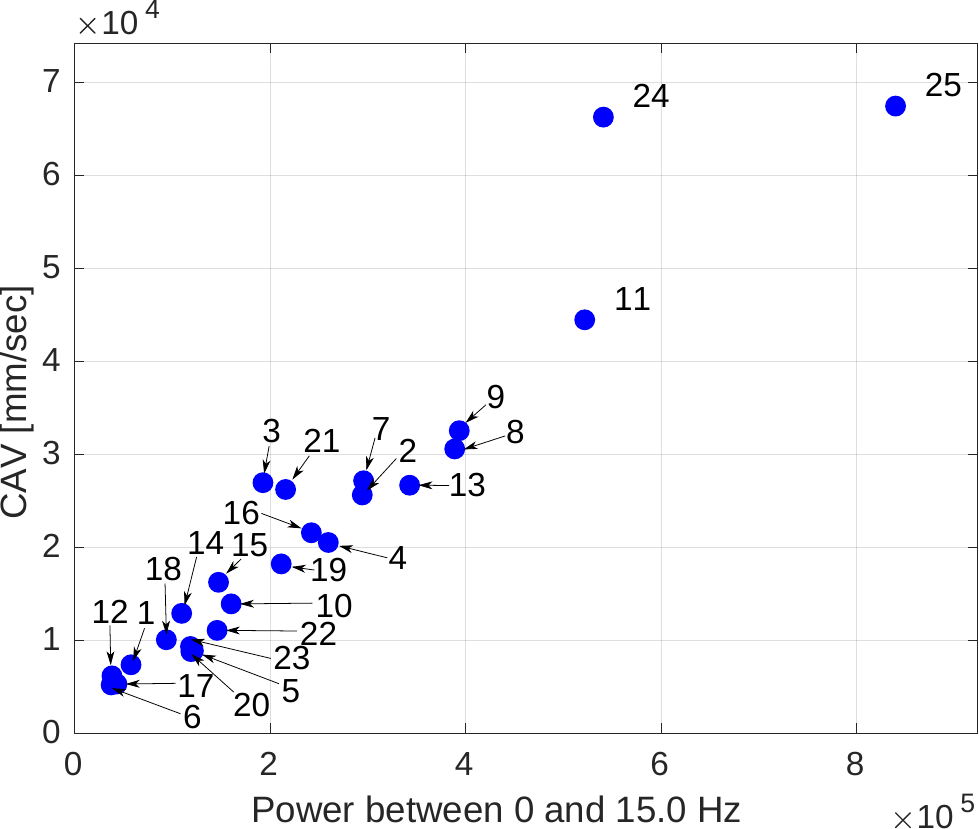}}\\
%\subfigure[Cumulative Absolute Velocity above 0.05g versus spectral power between 0 and 1.0 Hz]{\includegraphics[height=3.15cm]{CAV5vsPower0a1.pdf}}
\subfigure[Cumulative Absolute Velocity above 0.05g versus spectral power between 0 and 2.0 $Hz$]{\includegraphics[height=4.75cm]{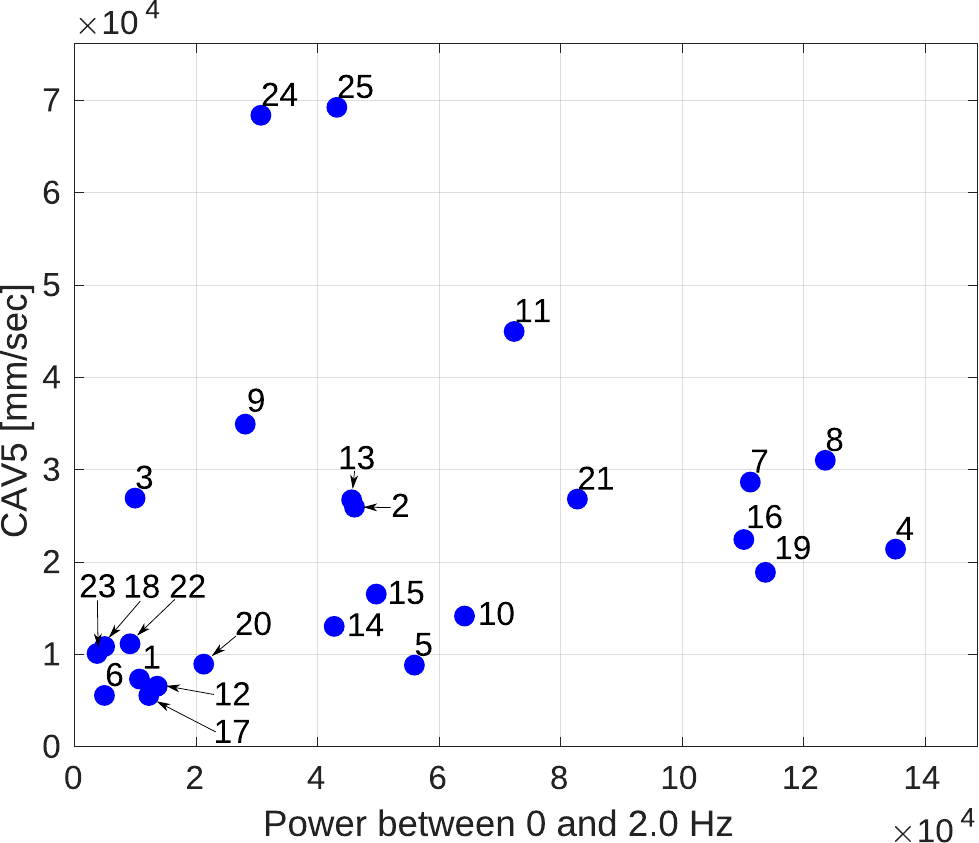}}
\subfigure[Cumulative Absolute Velocity above 0.05g versus spectral power between 0 and 15.0 $Hz$]{\includegraphics[height=4.75cm]{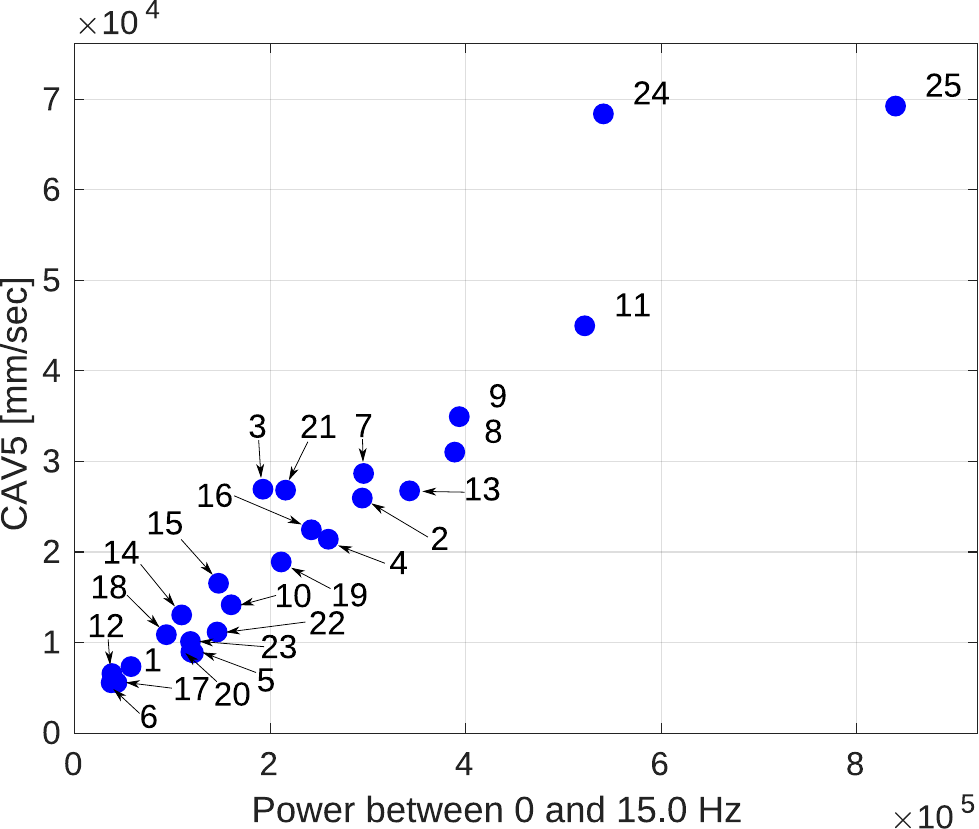}}
\end{center}
\caption{Comparison of the proposed intensity measure based on spectral power and classical IMs.}
\label{fig:correlations}
\end{figure}

Equation \eqref{eq:parseval} is a generalization of classical seismic intensity measures, filtering the power of frequencies that just brings scatter to correlations. It will be shown that \sout{these}\NL{this} spurious power in the case of the tailings dam shown as example \sout{are those at}\NL{is that at} medium-high to high frequencies (above 5  $Hz$). Figure \ref{fig:correlations} shows the comparisons AI in Fig. \ref{fig:correlations} (a) and (b), \sout{showing}\NL{proving} that Parseval's theorem is accomplished.  Fig. \ref{fig:correlations} (c) to (f) \sout{shows}\NL{show} CAV and CAV5  with cumulative spectral power within different ranges of frequencies. When the band frequency is narrow, i.e. up to 2 $Hz$, a considerable scatter is found \sout{between}\NL{among} the intensity measures in all cases but, as the frequency band becomes wider, the accumulated spectral power tends to represent AI, CAV and CAV5 obtaining a perfect fitting in all cases. It is worth noting that for some records like 21 or 24, the filtering produced by CAV and CAV5 does not represent a straightforward energy generalization. 

\section{Numerical validation}
\label{section:Num_res}

This section presents the validation of the spectral power intensity measure for three \sout{analysis types}\NL{types of analyses}: Newmark-type rigid block displacement, time-history deformation models using HS-Small (representative of industry-standard analyses) and time-history deformation models using PM4Sand \NL{model} (aiming at taking into account dynamic liquefaction \sout{problems}\NL{susceptibility of the tailing}).

\subsection{Correlation with a Newmark-type model}

First, the proposed intensity measure is contrasted with the Newmark displacement \cite{Newmark65}, a \sout{popular}\NL{widely-used} damage indicator \cite{DU2018286,VEYLON2017518,KOKUSHO2019121,CATTONI2019221}. Relevant parameters presented in Table \ref{t:HSmall} have been used in all cases.

\begin{figure}
\begin{center}
\subfigure[Newmark displacement vs AI]{\includegraphics[height=4.5cm]{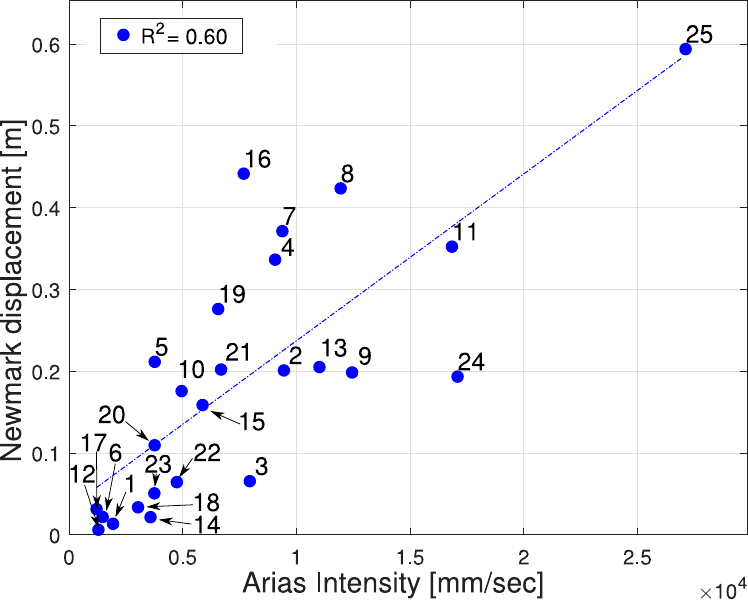}}
\subfigure[Newmark displacement vs CAV.]{\includegraphics[height=4.5cm]{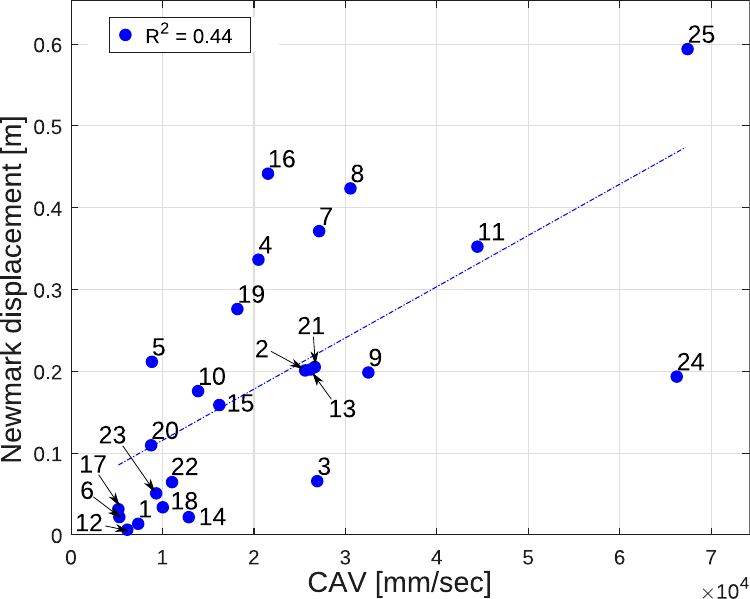}}
\subfigure[Newmark displacement vs CAV5.]{\includegraphics[height=4.5cm]{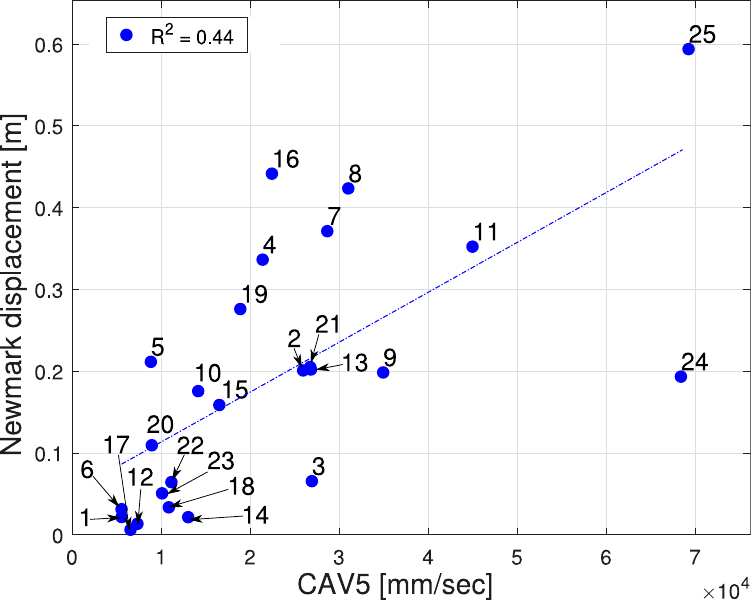}} 
\subfigure[Newmark displacement vs $P_{0-1.0 \ Hz}$.]{\includegraphics[height=4.5cm]{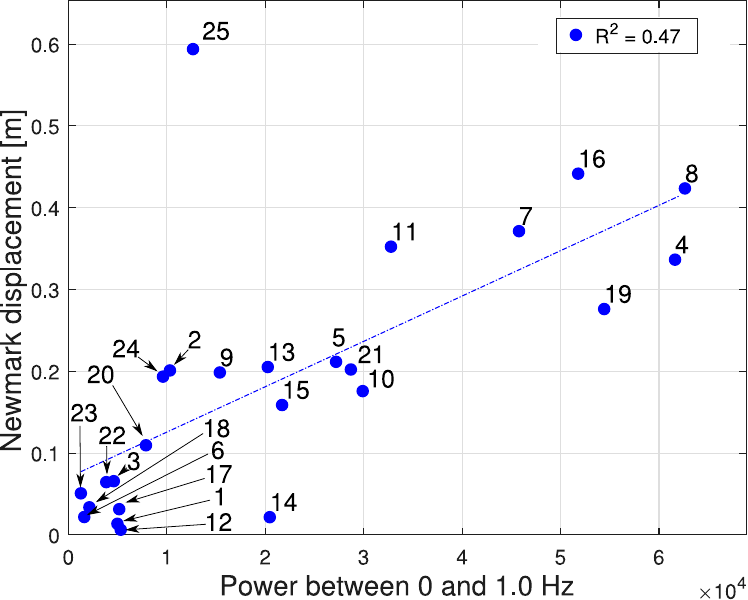}} 
\subfigure[Newmark displacement vs $P_{0-2.0 \ Hz}$.]{\includegraphics[height=4.5cm]{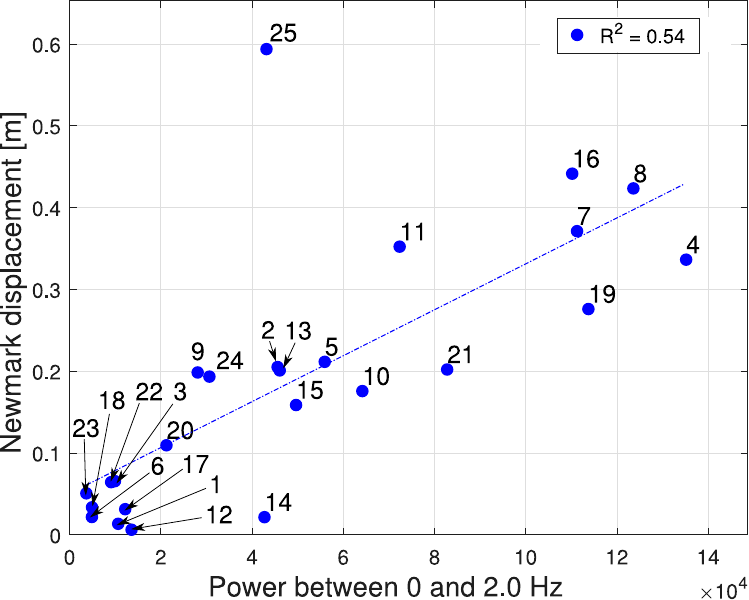}}
\subfigure[Newmark displacement vs $P_{0-5.0 \ Hz}$.]{\includegraphics[height=4.5cm]{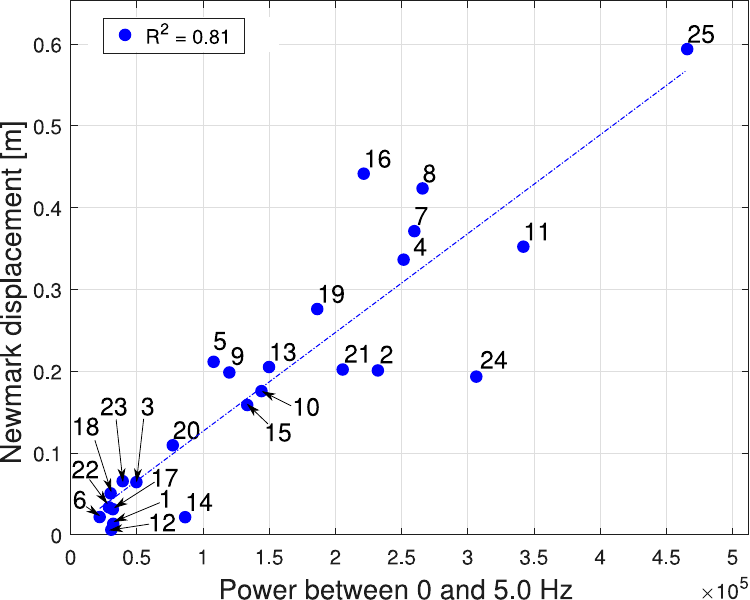}}
\end{center}
\caption{Newmark displacement compared with classical and the proposed intensity \sout{measure}\NL{measures}.}
\label{fig:correlationsND}
\end{figure}

Correlation between Newmark displacement and some IMs are presented: Arias Intensity in Fig. \ref{fig:correlationsND}  (a), CAV in Fig. \ref{fig:correlationsND} (b), CAV5 in Fig. \ref{fig:correlationsND} (c), cumulative spectral power $P_{0-1.0 \ \textit{Hz}}$ in  Fig.\ref{fig:correlationsND}  (d), $P_{0-2.0 \ Hz}$ in Fig.  \ref{fig:correlationsND}  (e) and finally $P_{0-5.0 \ Hz}$ in Fig. \ref{fig:correlationsND}  (f). 

When classical intensity measures are used, an $R^2 = 0.60$ is obtained for the best estimator (AI). Results improve considerably for the proposed intensity measure, where the fitting arises to  $R^2 = 0.81$ when the spectral power is accumulated in the window $0-5.0 \ Hz$, confirming that, at least in this case, the higher demand correlates with higher power content in low and medium to low frequency ranges.

\subsection{Comparison with time-history deformation models using HS-Small}

\begin{figure}
\begin{center}
\includegraphics[height=3.5cm]{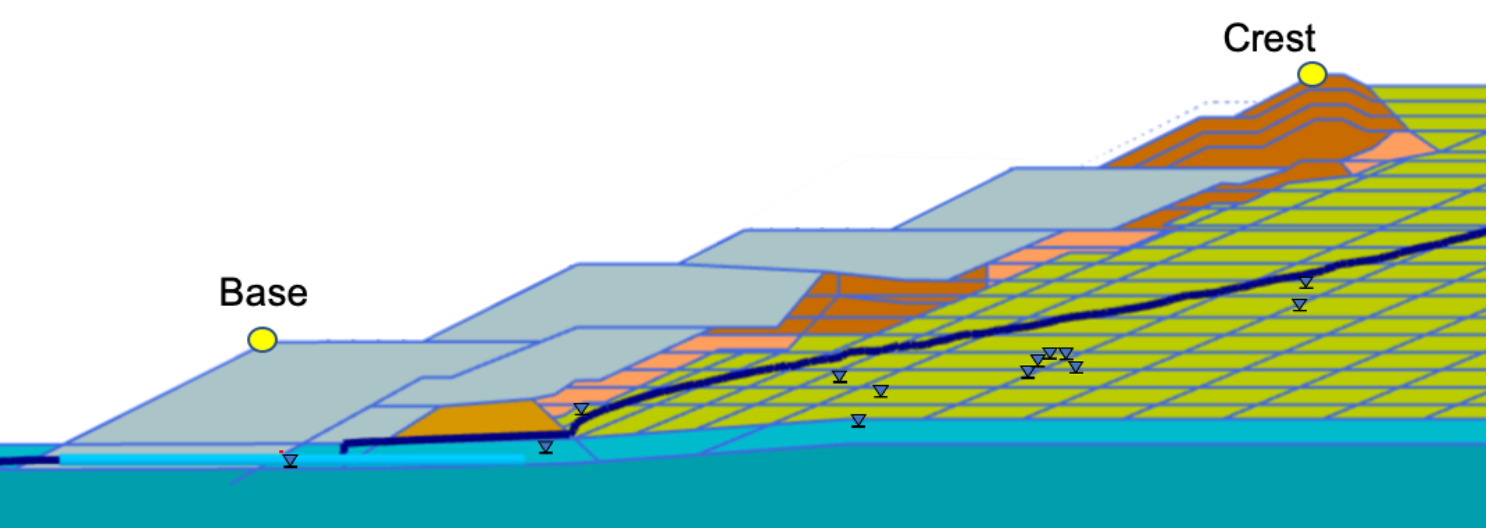}
\end{center}
\caption{Selected point to measure the displacements in the buttress.}
\label{Fig:Points}
\end{figure}

In this section, the configuration and materials of the model are those used for simulating the construction sequence. \sout{HSsmall}\NL{HS-Small}  is unable to generate excess pore pressure due to cyclic loading, and therefore this exercise is representative of many problems where seismic liquefaction is not relevant.
Two points were selected to quantify the demand induced by each seismic record in terms of displacement: one in the crest and one in the base \sout{or toe} of the buttress (see Figure \ref{Fig:Points}).

\begin{figure}
\begin{center}
\subfigure[Crest displacements vs AI]{\includegraphics[height=4.5cm]{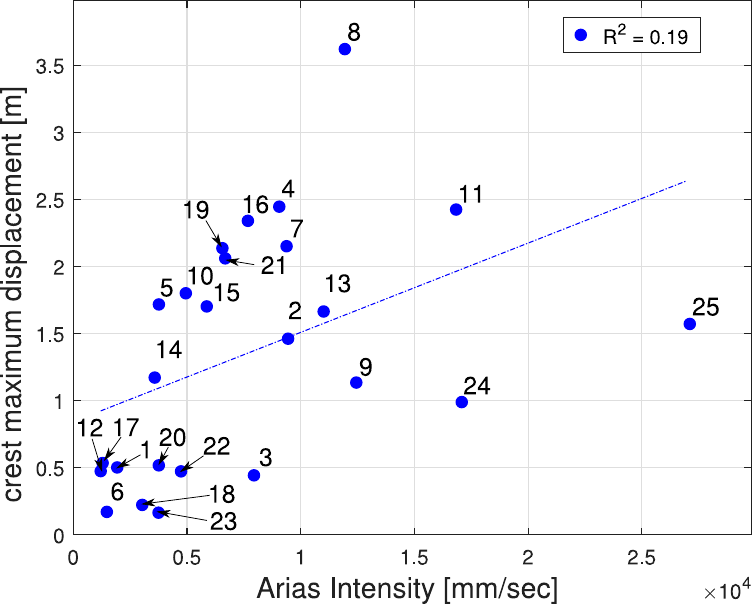}}
\subfigure[Crest displacements vs CAV.]{\includegraphics[height=4.5cm]{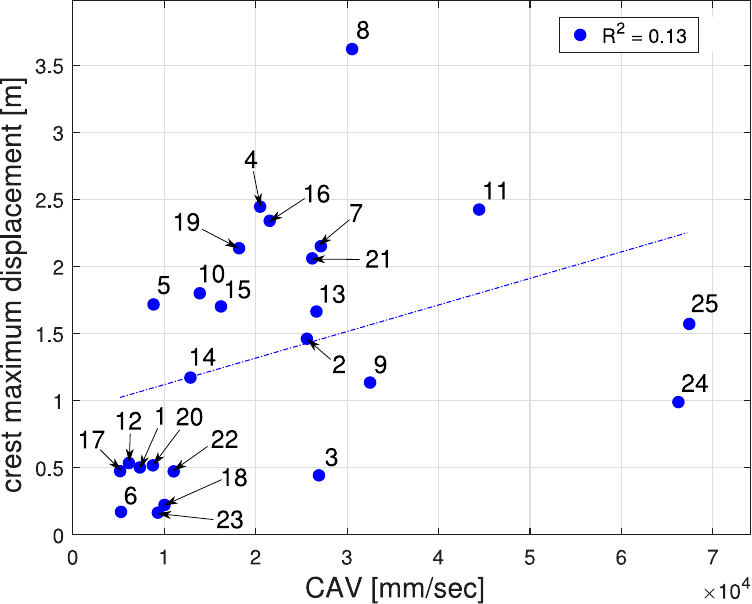}}
\subfigure[Crest displacements vs CAV5.]{\includegraphics[height=4.5cm]{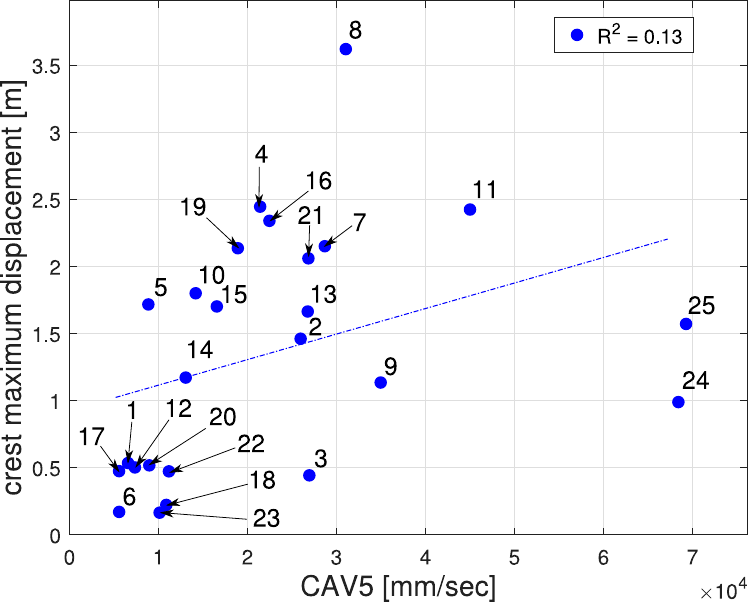}}
\subfigure[Crest displacements vs $P_{0-1.0 \ Hz}$.]{\includegraphics[height=4.5cm]{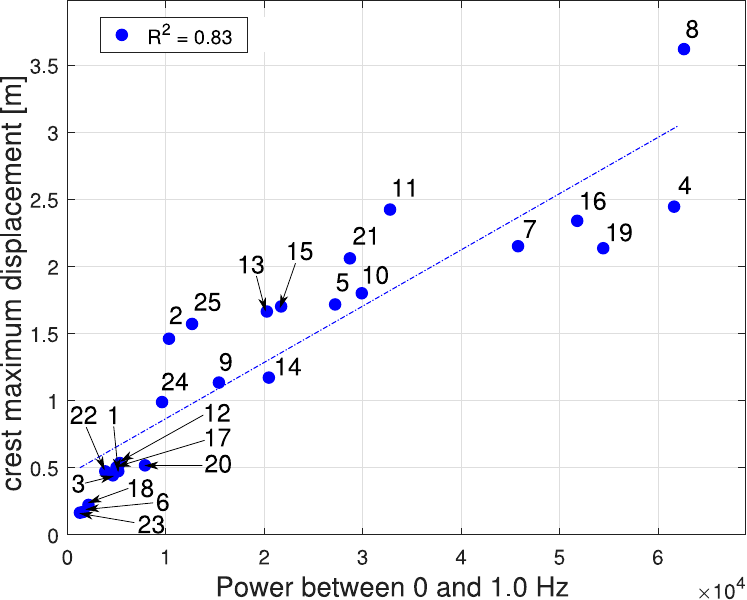}}
%\subfigure[HSsmall]{\includegraphics[height=3cm]{Energy0a15HSsmall.pdf}}
%\subfigure[HSsmall]{\includegraphics[height=3cm]{Energy0a2HSsmall.pdf}}
\subfigure[Crest displacements vs $P_{0-2.5 \ Hz}$.]{\includegraphics[height=4.5cm]{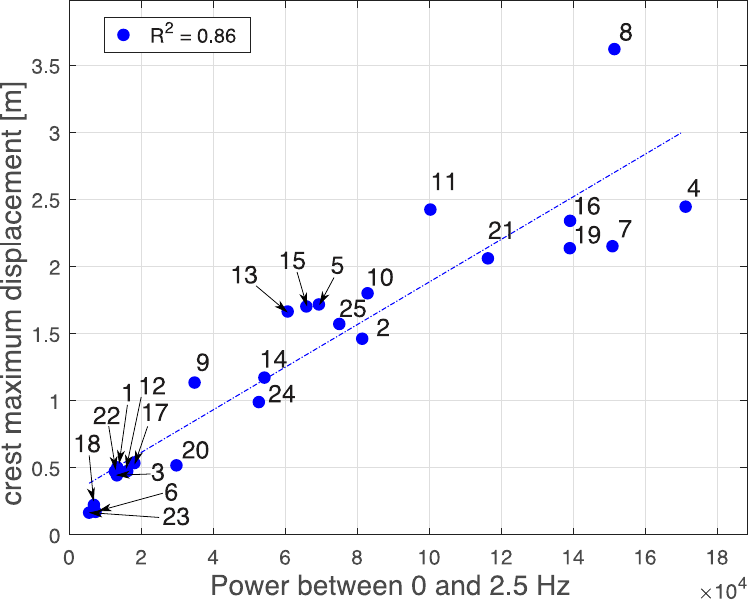}}
%\subfigure[HSsmall]{\includegraphics[height=3cm]{Energy0a3HSsmall.pdf}}
%\subfigure[HSsmall]{\includegraphics[height=3cm]{Energy0a5HSsmall.pdf}}
%\subfigure[HSsmall]{\includegraphics[height=3cm]{Energy0a10HSsmall.pdf}}
\subfigure[Crest displacements vs $P_{0-15 \ Hz}$.]{\includegraphics[height=4.5cm]{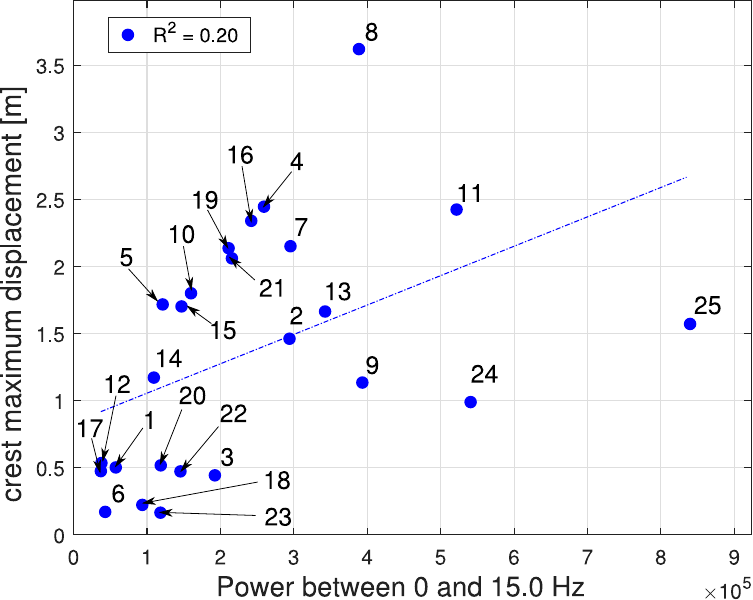}}
\end{center}
\caption{Crest displacements obtained with the Hardening Soil Small (\sout{HSsmall}\NL{HS-Small}) model compared with classical and the proposed intensity measures.}
\label{fig:DisplacementbaseHSsmall}
\end{figure}

Figure \ref{fig:DisplacementbaseHSsmall} shows the results of the \sout{verious}\NL{various} IMs, compared with the crest displacement. Correlations for classical IMs are rather poor. The best fit is again obtained by AI with a $R^2 = 0.19$ in Fig. \ref{fig:DisplacementbaseHSsmall} (a). Results for CAV and CAV5 are presented in Fig. \ref{fig:DisplacementbaseHSsmall}(b) and (c) respectively. When the spectral power is used, the correlation becomes much better. In this case, the best fit is obtained when the spectral power is accumulated between $0$ and $2.5 \ Hz$, i.e. $P_{0-2.5 \ Hz}$ where the $R^2 = 0.86$ as shown in Fig. \ref{fig:DisplacementbaseHSsmall} (e). It is again observed that seismic records with high power in the low and medium-low frequency range produce higher demands.

\begin{figure}[h]
\begin{center}
\subfigure[Base displacements vs AI.]{\includegraphics[height=4.5cm]{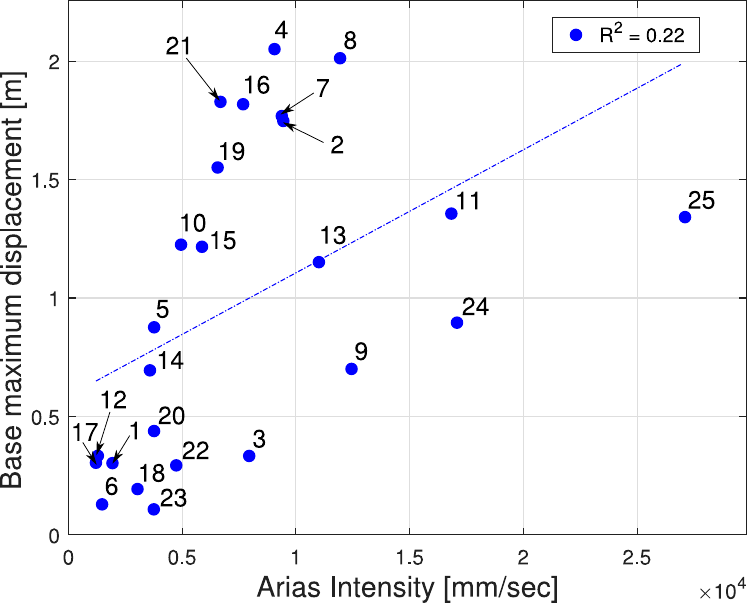}}
\subfigure[Base displacements vs CAV.]{\includegraphics[height=4.5cm]{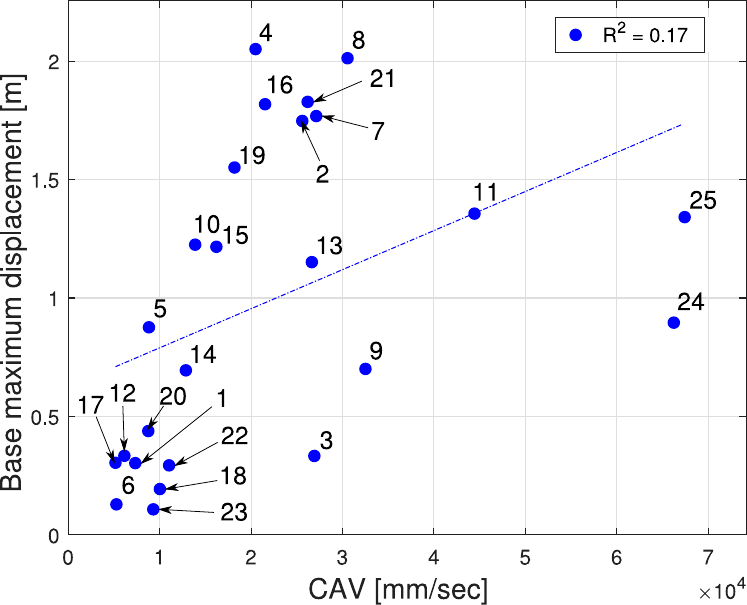}}
\subfigure[Base displacements vs CAV5.]{\includegraphics[height=4.5cm]{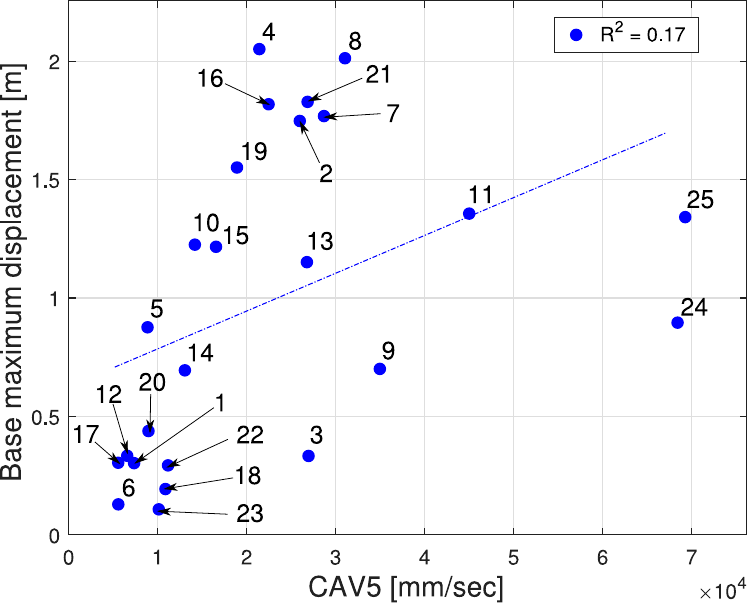}}
%\subfigure[HSsmall]{\includegraphics[height=3cm]{Energy0a05Base.pdf}}
%\subfigure[Base displacements vs $P_{0-1.0Hz}$.]{\includegraphics[height=4.5cm]{Energy0a1Base.pdf}}
%\subfigure[HSsmall]{\includegraphics[height=3cm]{Energy0a2Base.pdf}}
\subfigure[Base displacements vs $P_{0-2.5 \ Hz}$.]{\includegraphics[height=4.5cm]{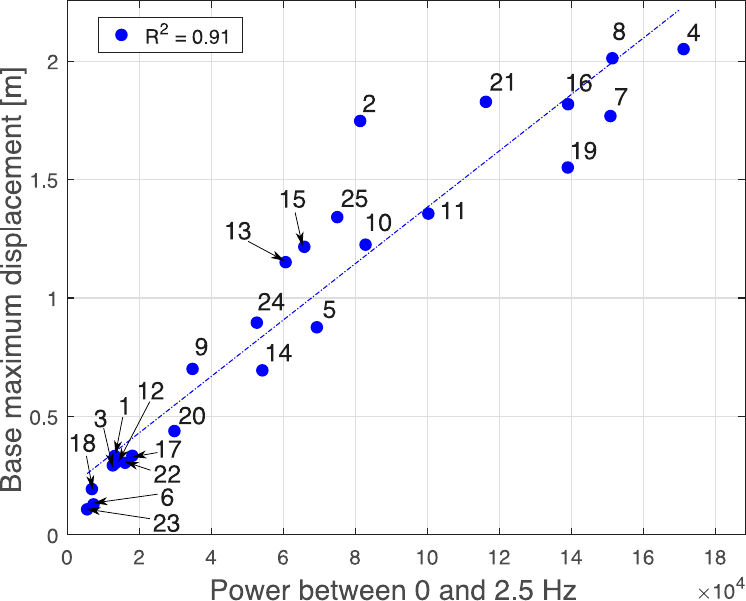}}
%\subfigure[HSsmall]{\includegraphics[height=3cm]{Energy0a3HSsmallBase.pdf}}
%\subfigure[HSsmall]{\includegraphics[height=3cm]{Energy0a5Base.pdf}}
%\subfigure[HSsmall]{\includegraphics[height=3cm]{Energy0a10Base.pdf}}
%subfigure[Base displacements vs $P_{0-15Hz}$.]{\includegraphics[height=4.5cm]{Energy0a150Base.pdf}}
\end{center}
\caption{Base displacements obtained with the Hardening Soil Small model compared with classical and the proposed intensity measures.}
\label{fig:Displacementbase}
\end{figure}

Similar results are obtained for base displacements. Figures \ref{fig:Displacementbase} (a) to (c) show base displacements compared with AI, CAV and CAV5 respectively, showing a best fit for AI with an $R^2 = 0.22$. Figure \ref{fig:Displacementbase} (d) shows the correlation with $P_{0-2.5 \ Hz}$, obtaining a better fit.


\subsection{Comparison with time-history deformation models using PM4Sand \NL{model}}

\begin{figure}[ht]
\begin{center}
{\includegraphics[height=9.5cm,trim={0 0.0cm 0 0.27cm},clip]{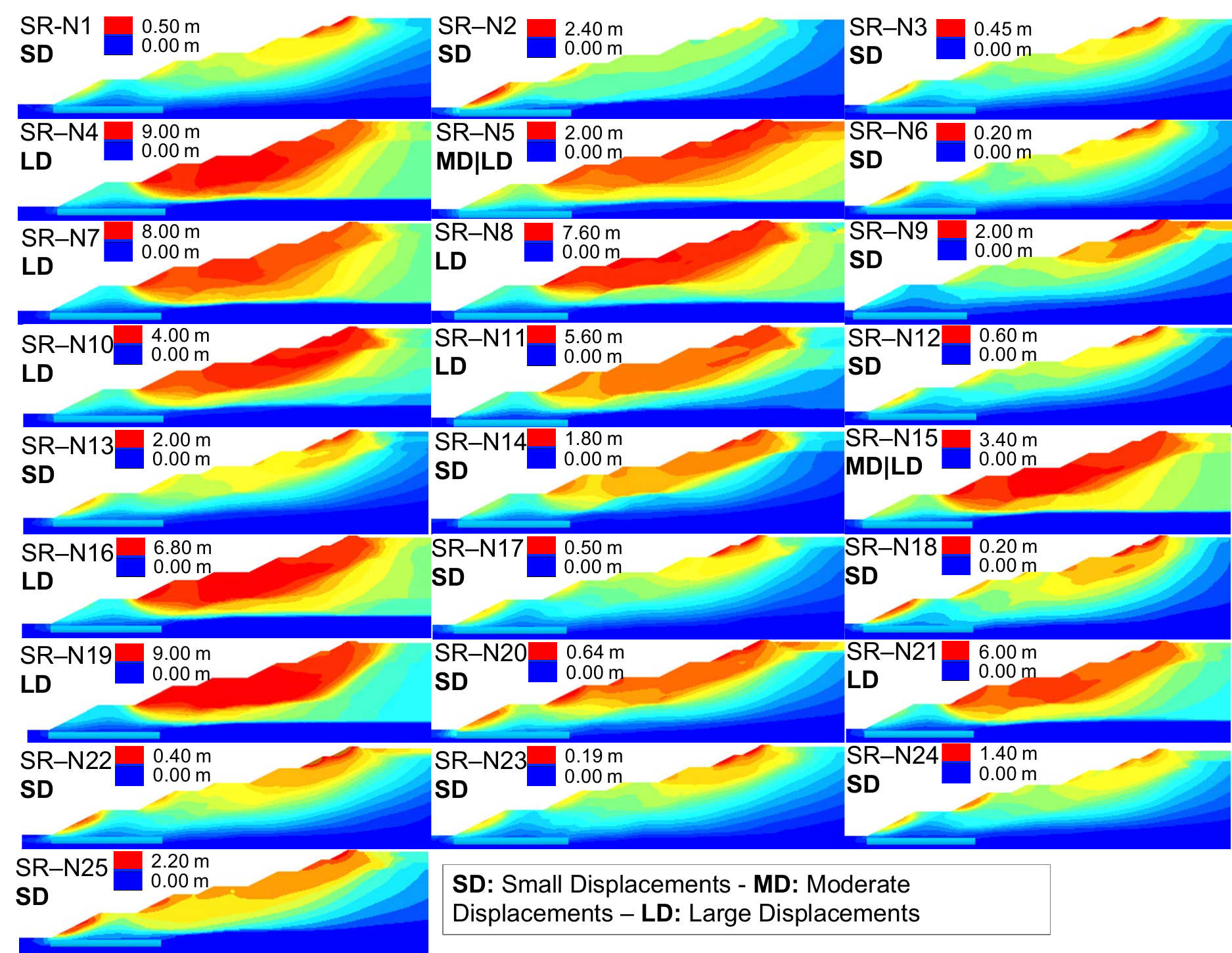}}
%\subfigure[]{\includegraphics[height=6.5cm]{ButressN1.png}}
\end{center}
\caption{Map of displacements obtained for each seismic record \NL{using PM4Sand model}.}
\label{fig:Butress}
\end{figure}

The exercise is repeated using PM4Sand \NL{model}, and can be considered representative of problems where seismic liquefaction must be studied \NL{for contractive and saturated materials}. Figure \ref{fig:Butress} shows the displacement map produced by each seismic record. Results have been \sout{ordered by those with}\NL{organized into groups of} small (SD), moderate (MD) and large displacements (LD). In some cases, like records 1 to 3, damage is negligible with displacements concentrated on the edge of the buttress. Records 4, 8, 16, 19, 21, among others, produce a considerable damage with a huge portion of tailings sliding down due to liquefaction.

\begin{figure}[ht]
\begin{center}
%\subfigure[Small displacements - Base point]{\includegraphics[height=3cm]{Norm_U_Base_Green_Gold.pdf}}
%\subfigure[Moderate displacements - Base point]{\includegraphics[height=3cm]{Norm_U_Base_Yellow_Gold.pdf}}
%\subfigure[Large displacements  - Base point ]{\includegraphics[height=3cm]{Norm_U_Base_Red_Gold.pdf}}
\subfigure[Small displacements - Base point]{\includegraphics[height=3.25cm]{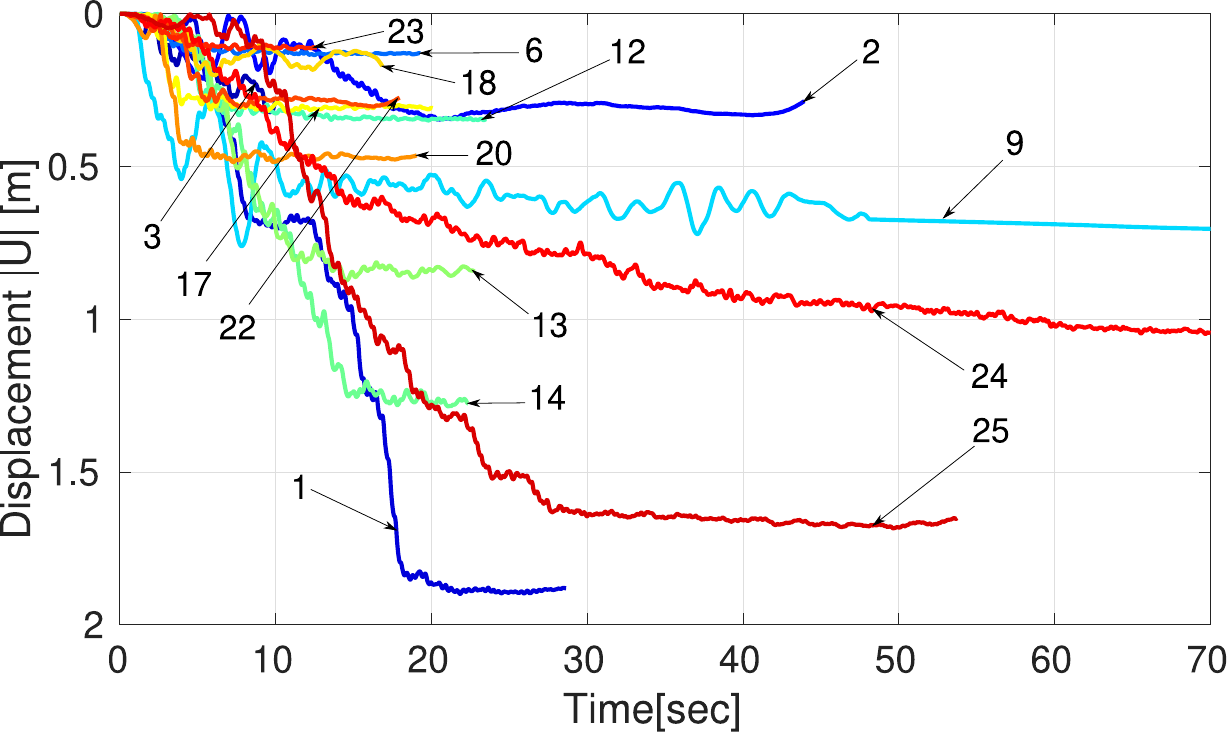}}
\subfigure[Moderate displacements - Base point]{\includegraphics[height=3.25cm]{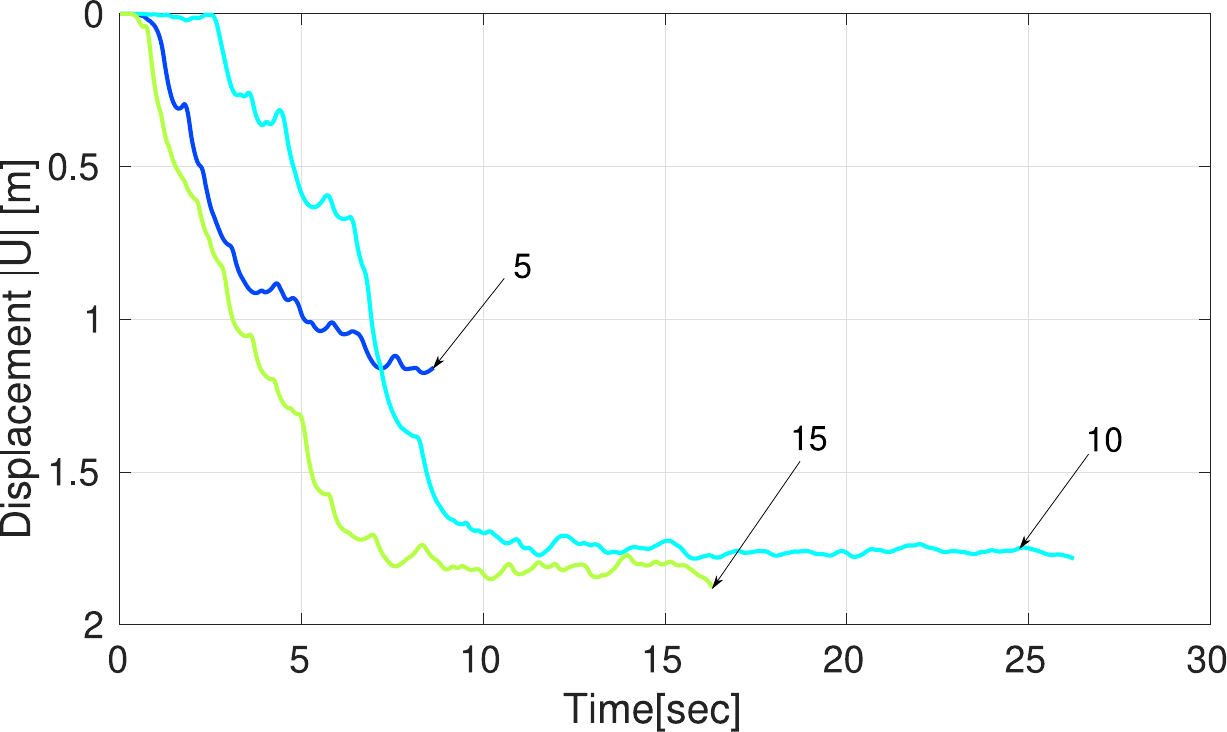}}
\subfigure[Large displacements - Base point]{\includegraphics[height=3.25cm]{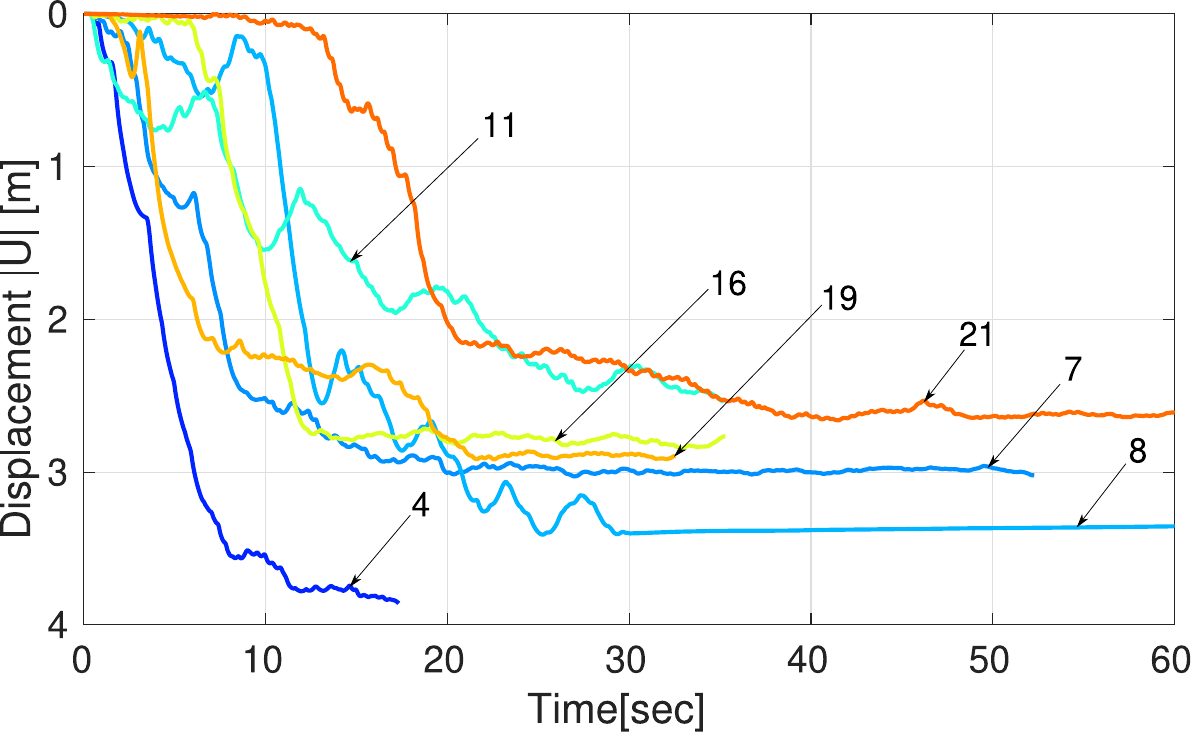}}
%\subfigure[Small displacements  - Crest point]{\includegraphics[height=3cm]{Norm_U_Crest_Green_Gold.pdf}}
%\subfigure[Moderate displacements  - Crest point]{\includegraphics[height=3cm]{Norm_U_Crest_Yellow_Gold.pdf}}
%\subfigure[Large displacements  - Crest point ]{\includegraphics[height=3cm]{Norm_U_Crest_Red_Gold.pdf}}
\subfigure[Small displacements  - Crest point]{\includegraphics[height=3.25cm]{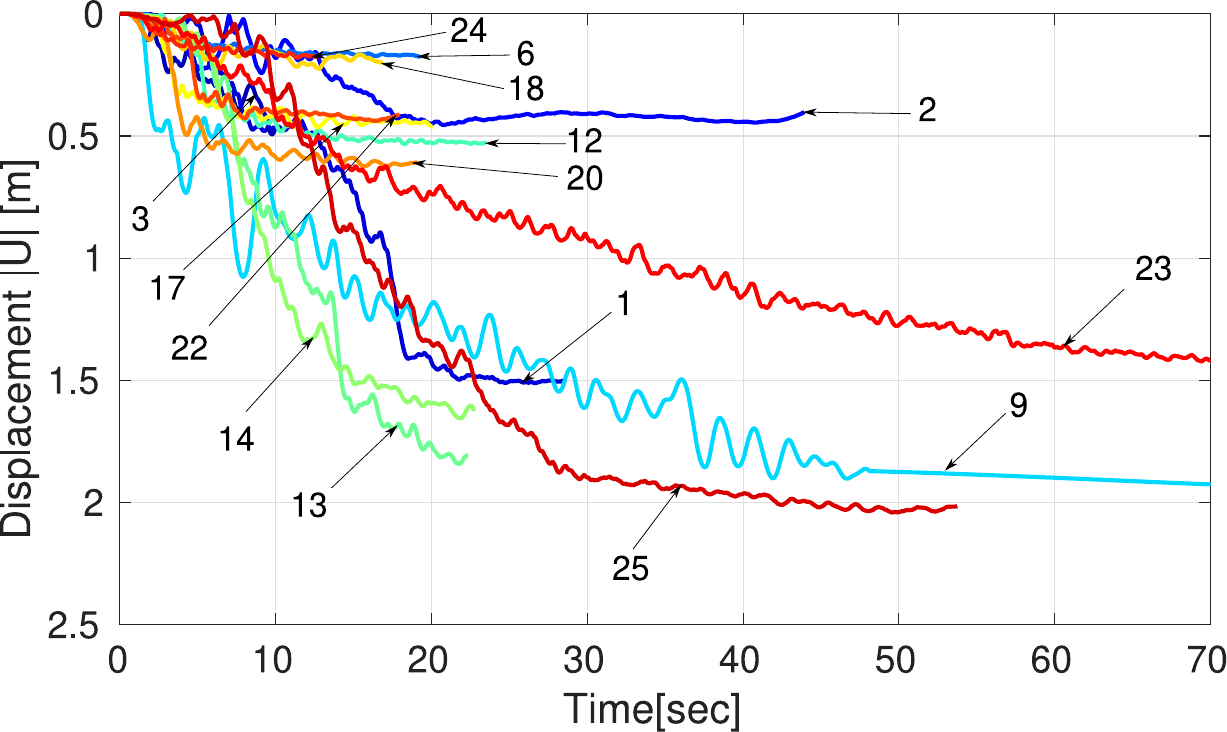}}
\subfigure[Moderate displacements  - Crest point]{\includegraphics[height=3.25cm]{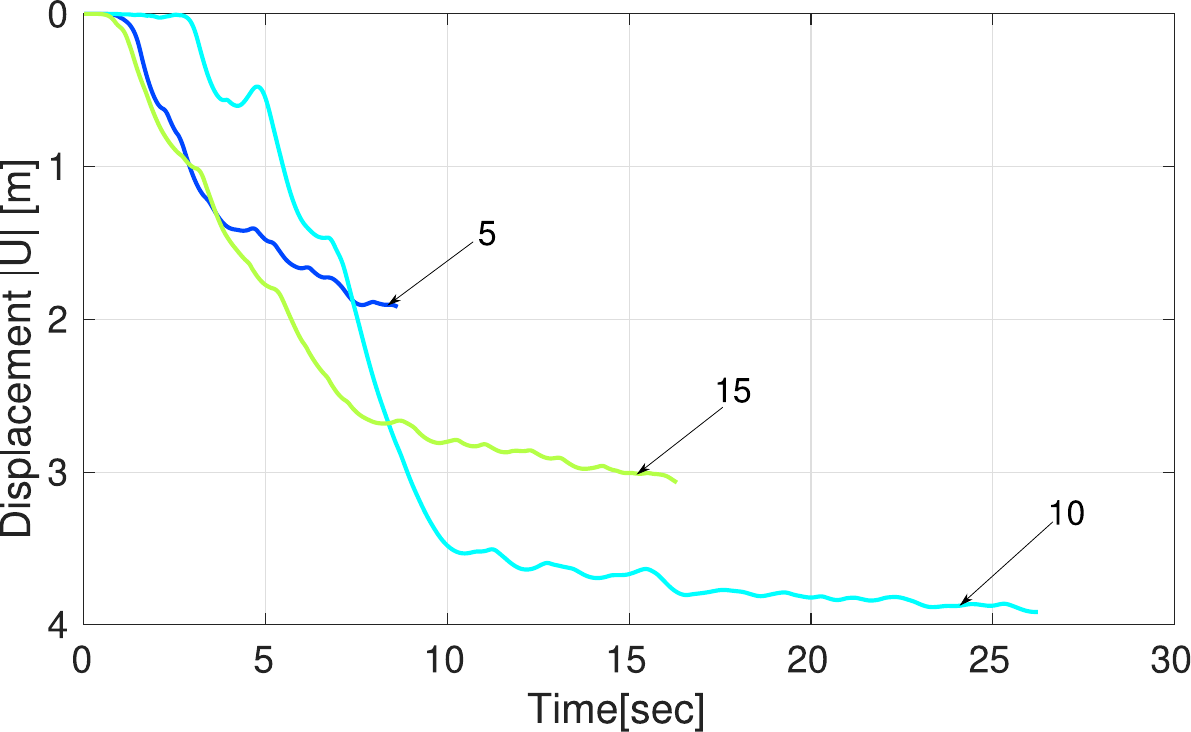}}
\subfigure[Large displacements - Crest point ]{\includegraphics[height=3.25cm]{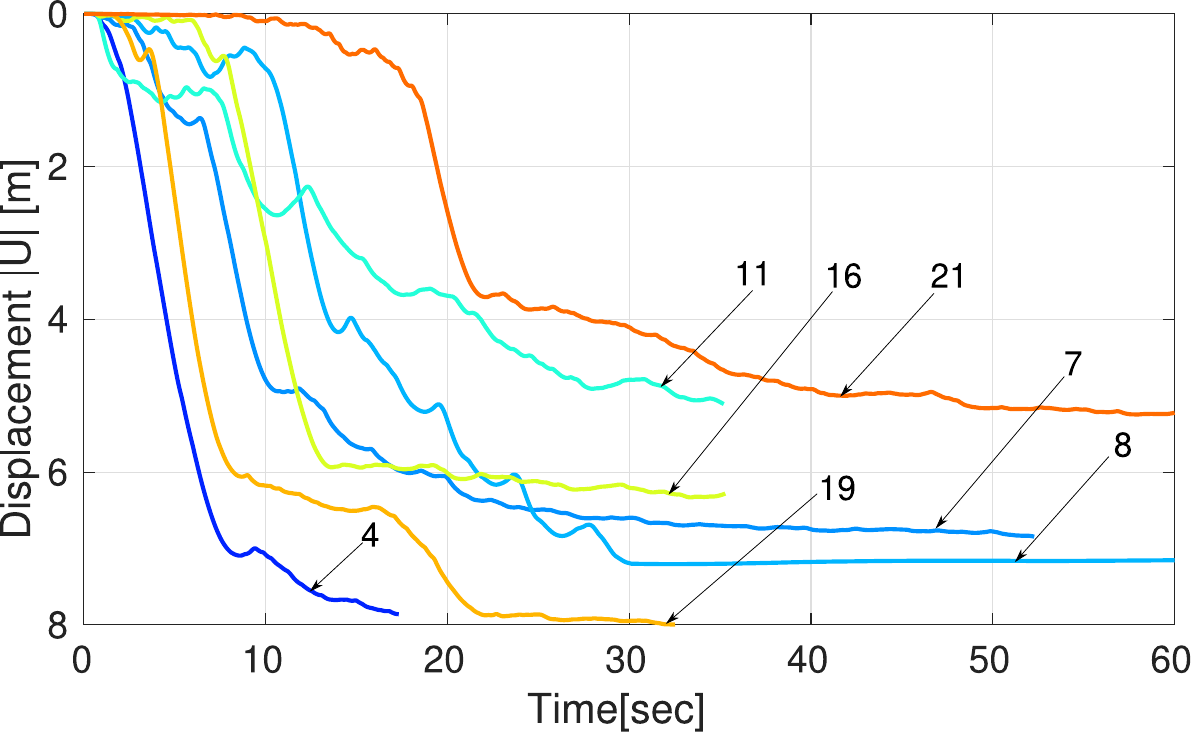}}
\end{center}
\caption{Displacement time-history for base and crest points \NL{for all seismic records using PM4Sand model. Large, moderate and small displacements are referred to \NL{by} colored zones in Figure \ref{fig:correlationsPM4Sand}}.}
\label{fig:DisplacementbasePM4SAND}
\end{figure}

Figures \ref{fig:DisplacementbasePM4SAND} (a) to (c) shows the time history of the base displacements, while Figures \ref{fig:DisplacementbasePM4SAND} (d) to (f) shows the time history at the crest. \sout{Looks like}\NL{Apparently,} there is no correlation, as the scatter is very high for all IMs.  

\begin{figure}
\begin{center}
\subfigure[Record N$^{\circ}9$ with  $AI = 12444$ and $P_{0-2.0 \ Hz}/P_r = 28378.75$.]{\includegraphics[height=2.1cm]{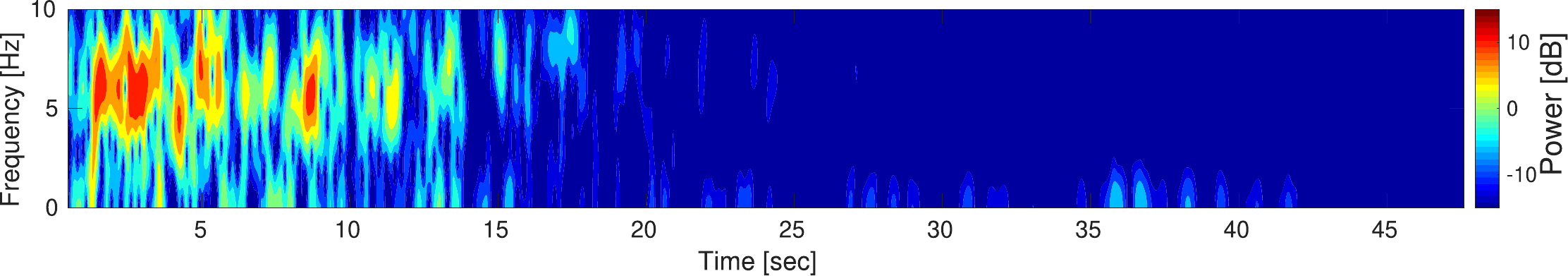}}
\subfigure[Record N$^{\circ}4$ with  $AI = 9057$ and $P_{0-2.0 \ Hz}/P_r = 135265.10$.]{\includegraphics[height=2.1cm]{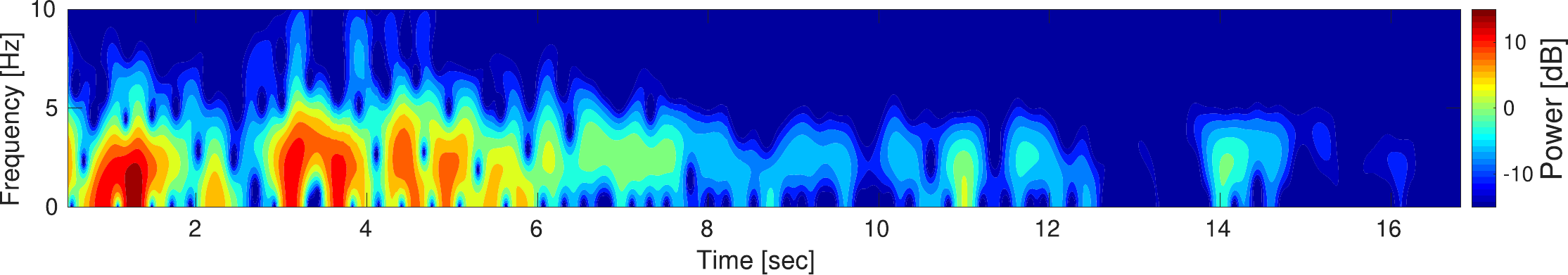}}
\end{center}
\caption{Three seismic signals inducing different demand levels.}
\label{fig:comparisonsspec}
\end{figure}

Attention must be given to the results of seismic records N$^{\circ}9$ and N$^{\circ}4$ to prove the novelty of the method proposed in this paper. Figure \ref{fig:comparisonsspec} shows the spectrograms of the afforementioned signals, where the corresponding to  N$^{\circ}9$ shows that most of its power is concentrated in high frequencies, while the spectrogram for record N$^{\circ}4$ shows exactly the opposite, with most of the energy in the low frequency range. 

\begin{figure} 
\begin{center}
\subfigure[Local deformation for seismic record N$^{\circ}9$.]{\includegraphics[height=8cm]{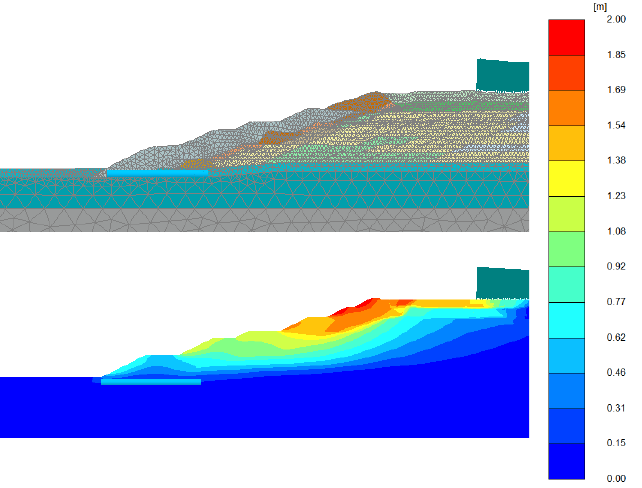}}
\subfigure[Global failure for seismic record N$^{\circ}4$.]{\includegraphics[height=8cm]{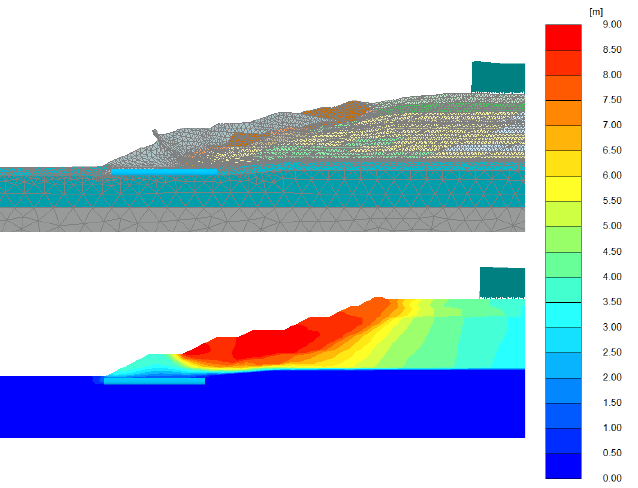}}
\end{center}
\caption{Displacement maps obtained for seismic records presented in Figure \ref{fig:comparisonsspec}, \NL{using PM4Sand model}.}
\label{fig:comparisons}
\end{figure}

If demand were to correlate with Arias Intensity, record N$^{\circ}9$ should be the most demanding but, as shown in Figure \ref{fig:comparisons}, the most demanding is N$^{\circ}4$. This phenomenon is captured by the intensity measure expressed in terms of spectral power, being the power concentrated in low frequencies higher in ground motion \sout{that}\NL{which} produces higher demand. 

Figure \ref{fig:correlationsPM4Sand} shows the maximum time-history displacement obtained for different seismic intensity measures. Similarly to the results obtained for the Newmark and HS-Small exercises, poor results are obtained \NL{with classical IMs}, and the best fit is the one with AI, with a  $R^2=0.07$ in Figure \ref{fig:correlationsPM4Sand} (a), while for CAV and CAV5 in Fig. \ref{fig:correlationsPM4Sand} (b) and (c) respectively, the correlation is even worse. The spectral power intensity measure presented in Fig.  \ref{fig:correlationsPM4Sand} (d)  shows an excellent fit with an $R^2=0.95$ when low frequencies are considered, i.e. $P_{0-2.0 \ Hz}$. When the frequency band becomes wider, the scatter increases obtaining a similar pattern than the one obtained for the AI (Fig.  \ref{fig:correlationsPM4Sand} (e)).

\begin{figure}
\begin{center}
\subfigure[Crest displacements vs AI]{\includegraphics[height=4.5cm]{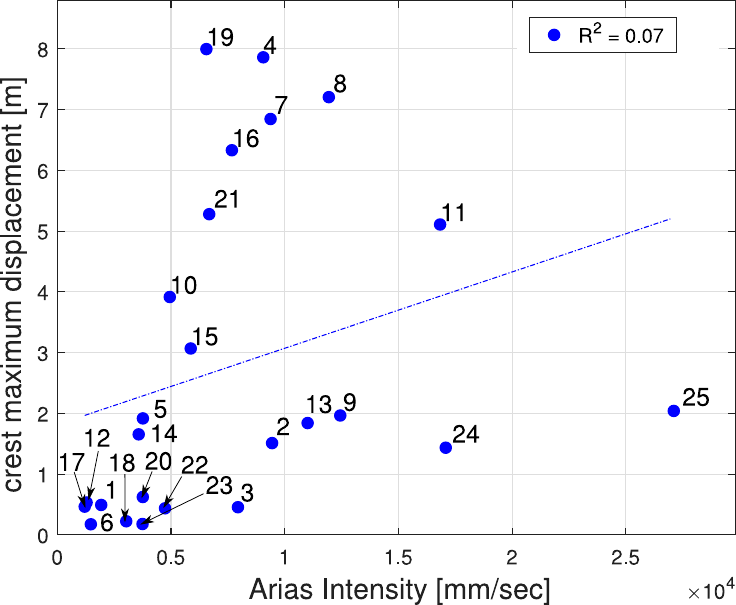}} 
\subfigure[Crest displacements vs CAV.]{\includegraphics[height=4.5cm]{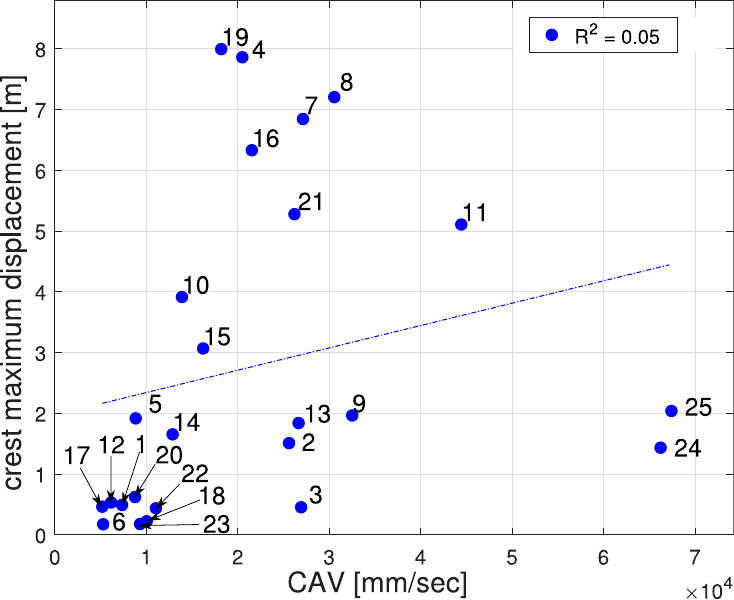}}
\subfigure[Crest displacements vs CAV5.]{\includegraphics[height=4.5cm]{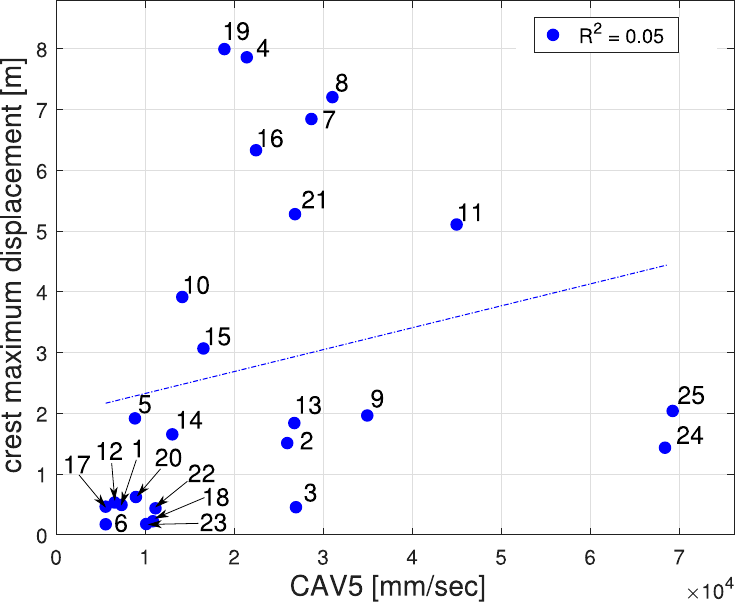}}
%\subfigure[]{\includegraphics[height=3cm]{Energy0a1.png}}
\subfigure[Crest displacements vs $P_{0-2.0 \ Hz}$.]{\includegraphics[height=4.45cm]{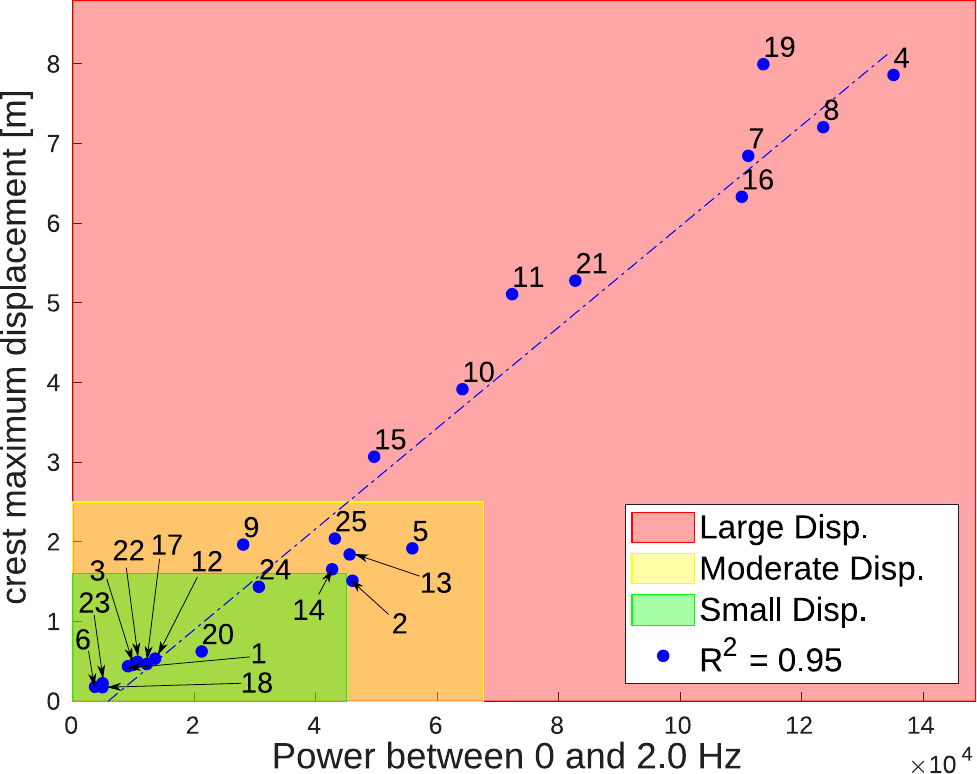}}
%\subfigure[Crest displacements vs $P_{0-5.0Hz}$.]{\includegraphics[trim={1cm 0 1.4cm 0},clip,height=4.75cm]{Energy0a5.png}}
\subfigure[Crest displacements vs $P_{0-15 \ Hz}$.]{\includegraphics[height=4.45cm]{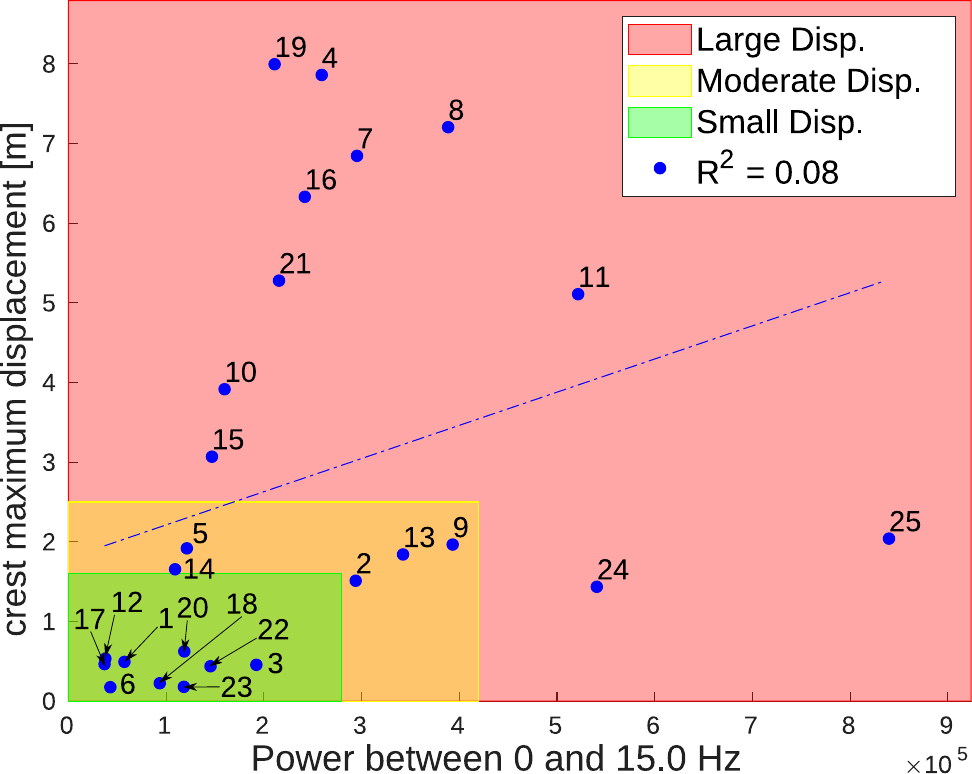}}
\end{center}
\caption{Crest displacements obtained with the PM4Sand  model compared with classical and the proposed intensity measure.}
\label{fig:correlationsPM4Sand}
\end{figure}

The best width of the window of the spectral power IM must of course depend on the problem in hand. In the example shown here, the fundamental frequency of the tailings body was computed in Section \ref{section:tailing} to be $f_1=0.70 \ Hz$, yielding a fundamental period of $T=1.43 \ s$, which suggests that, at least in this case and maybe in similar analyses, a window width of $0-2.0 \ T$ would be a good starting point \sout{for using}\NL{to use} the proposed IM as a tool \sout{for a-priori estimating}\NL{to make a priori estimates of} the seismic demand that a given record would produce to the dam under study.

\section{Conclusions}
\label{section:Conc}

A novel intensity measure (IM) based on the spectral power of seismic records has been presented and applied to the problem of estimating the displacement demand of an upstream-raised tailings storage facility (TSF) subjected to strong earthquakes. The proposal is inspired \sout{in}\NL{by} the evidence that the seismic demand of such TSFs -base and crest displacement- is strongly correlated with the power of the earthquakes at low and medium-low frequency range, close to the first-mode natural period of the dams. 

The numerical model set up for the dynamic deformation analysis of a large TSF \sout{was}\NL{has been} briefly described as an example, including the calibration of \sout{HSsmall}\NL{HS-Small} and PM4Sand constitutive models. The mathematical definition of the proposed intensity measure based on spectral power has been introduced in a discrete domain, detailing the equations required to perform the computations.

It has been shown that the proposed intensity measure based on the spectral power is as generalization of other classical and well-known intensity measures such as Arias Intensity, Cumulative Absolute Velocity and Cumulative Absolute Velocity above $0.05  \ g$, and that it provides a better correlation with seismic demand of the dam for three design procedures used as examples: Newmark-type displacement analysis, time-history deformation analysis using \sout{HSsmall}\NL{HS-Small} and time-history deformation analysis using PM4Sand \NL{model}. It \sout{was}\NL{has been}  demonstrated that the key aspect to be accounted for is the filtering of spectral power in the frequency band relevant to the structure under analysis. 

In the example shown, where the fundamental period is $T=1.43 \ s$, the correlation between the spectral power intensity measure and seismic demand was $R^2 = 0.81$ for the Newmark-type analysis, $R^2 = 0.86$ for the HS-Small deformation \sout{modelling}\NL{modeling} and up to $R^2 = 0.95$ for the PM4Sand deformation \sout{modelling}\NL{modeling}. The best frequency window has a slight dependence on the calculation method, but for all cases frequencies belongs to a low range, being $0-5.0 \ Hz$ to the Newmark case, $0-2.5 \ Hz$ for the HS-small model and $0-2.0 \ Hz$ for the PM4Sand model. It was suggested that a window width of $0-2.0 \ T$ would be a good starting point \sout{for using}\NL{to use} the proposed IM as a tool  \sout{for a-priori estimating}\NL{to make a priori estimates of} the seismic demand that a given record would produce to the dam under study.

\bibliographystyle{unsrt}
\bibliography{Reference}

\appendix
\section{Seismic Records}
\label{section:Appendix}

This appendix includes the seismic records used for all the simulations presented in this paper, \sout{together}\NL{along} with \sout{the}\NL{its} spectrograms\sout{ calculated to be}, used to compute \sout{out}\NL{our} intensity indicator. All of them \sout{has}\NL{have} been downloaded from the PEERS seismic database.

\begin{landscape}
\begin{figure}[!htbp]
\ContinuedFloat
\captionsetup{list=off,format=cont}
\centering
\subfigure[Record N$^{\circ}1$ - Nahanni Canada]{\includegraphics[width=8.5cm]{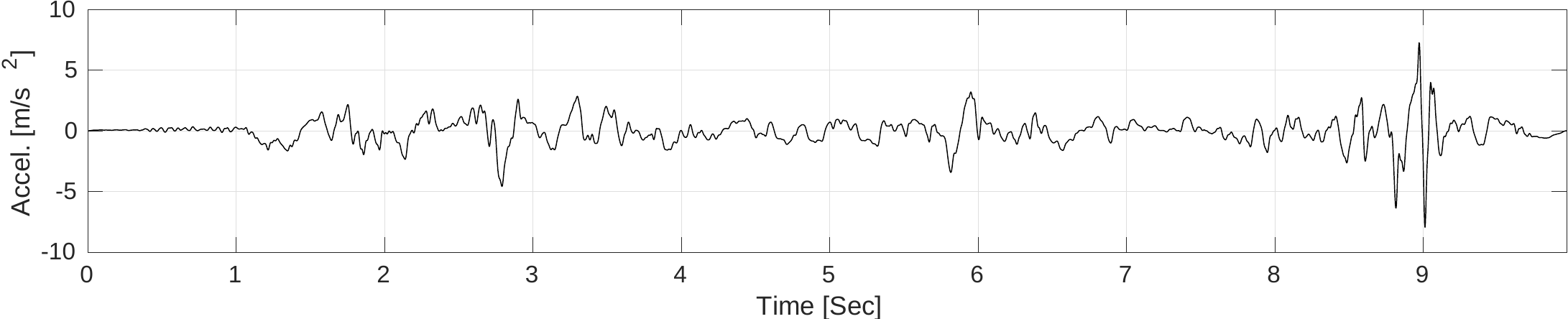}} 
\subfigure[Spectrogram N$^{\circ}1$ - Nahanni Canada]{\includegraphics[width=8.75cm]{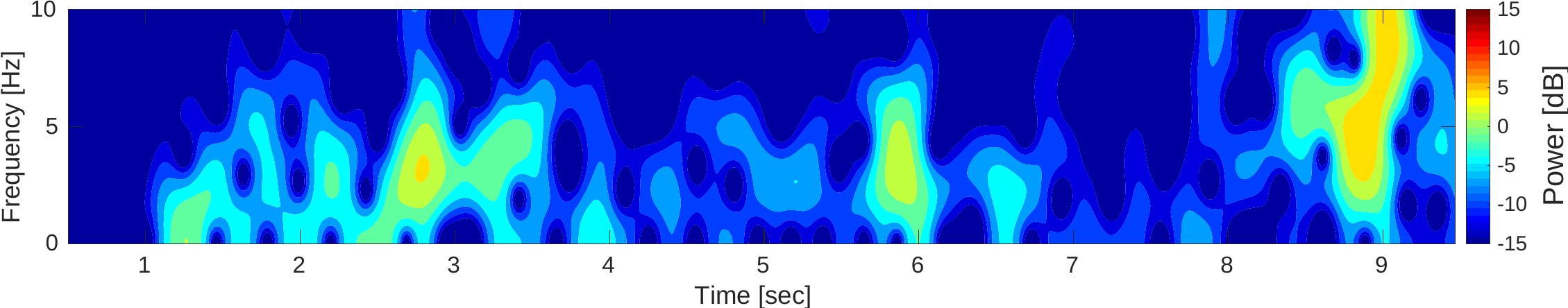}} \\
\subfigure[Record N$^{\circ}2$ - Duzce Turkey]{\includegraphics[width=8.5cm]{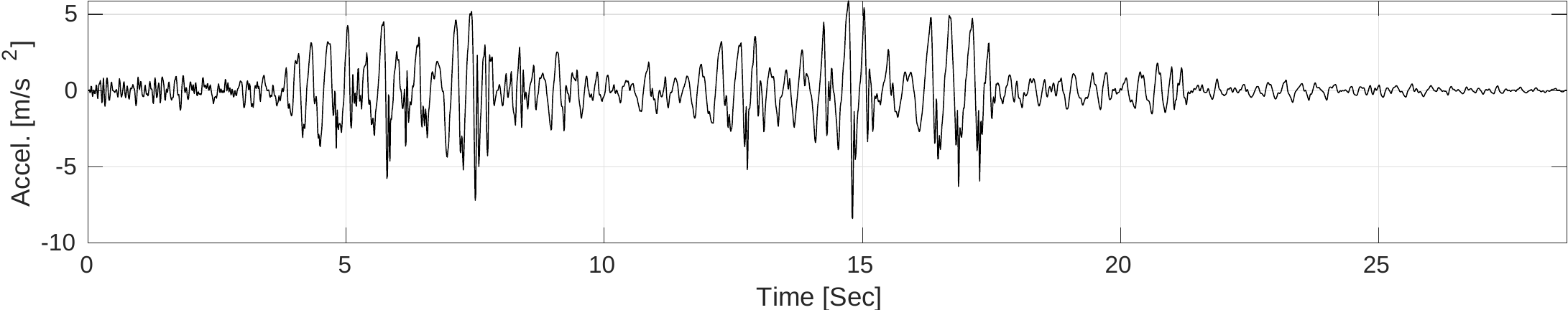}} 
\subfigure[Spectrogram N$^{\circ}2$ - Duzce Turkey]{\includegraphics[width=8.75cm]{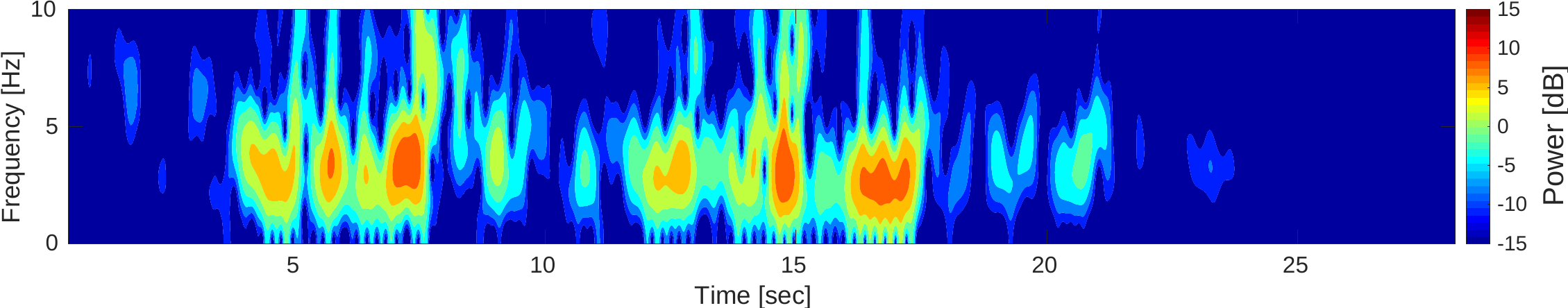}} \\ 
\subfigure[Record N$^{\circ}3$ - Landers]{\includegraphics[width=8.5cm]{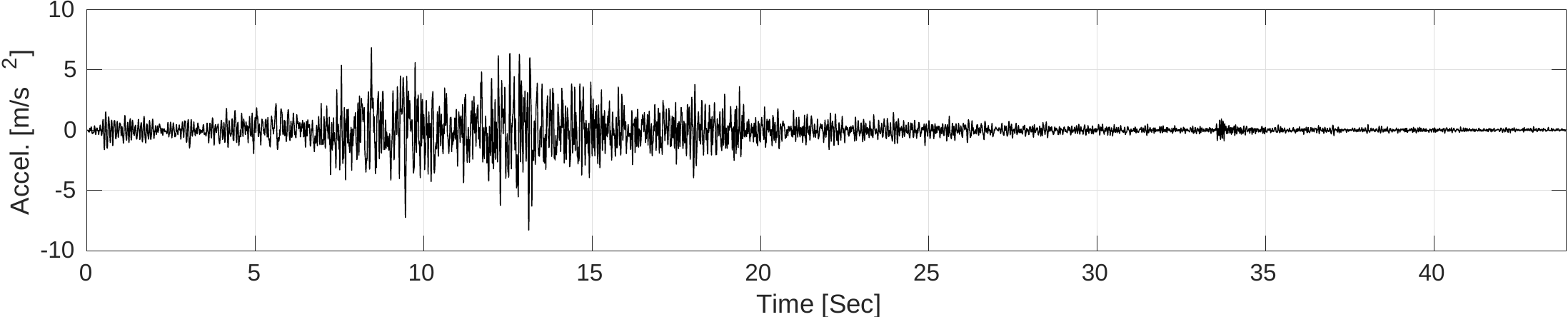}} 
\subfigure[Spectrogram N$^{\circ}3$ - Landers]{\includegraphics[width=8.75cm]{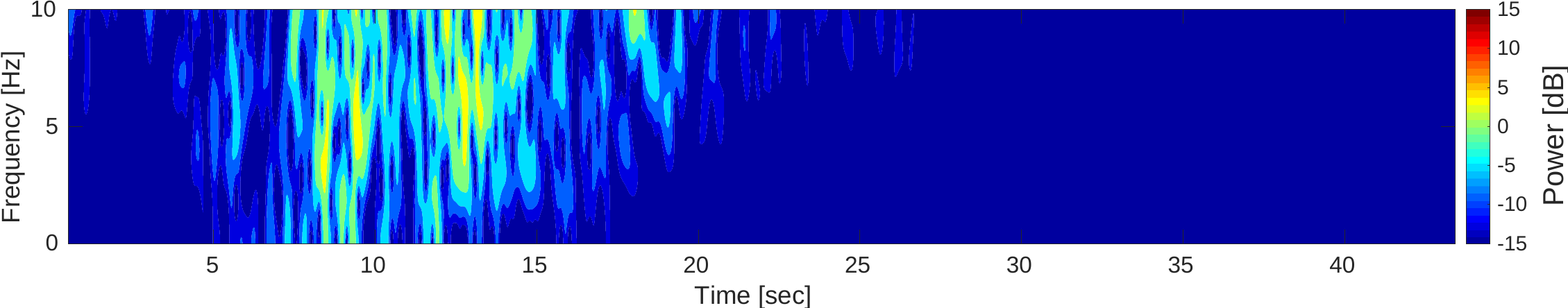}}  \\
\subfigure[Record N$^{\circ}4$ - Kobe Japan]{\includegraphics[width=8.5cm]{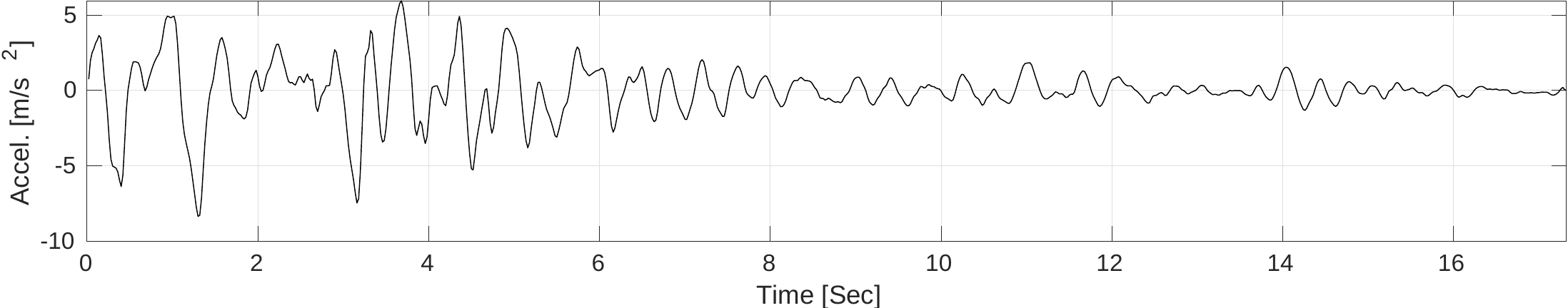}} 
\subfigure[Spectrogram N$^{\circ}4$ - Kobe Japan]{\includegraphics[width=8.75cm]{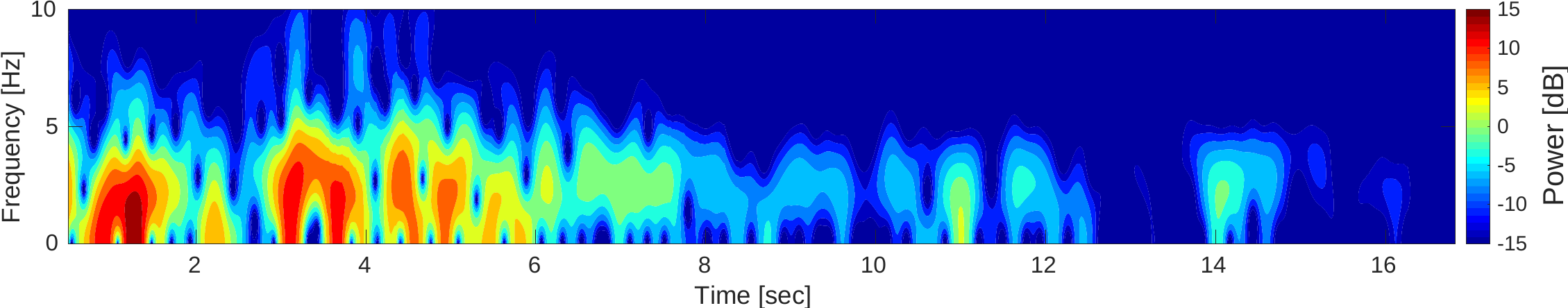}} \\
\subfigure[Record N$^{\circ}5$ - San Salvador]{\includegraphics[width=8.5cm]{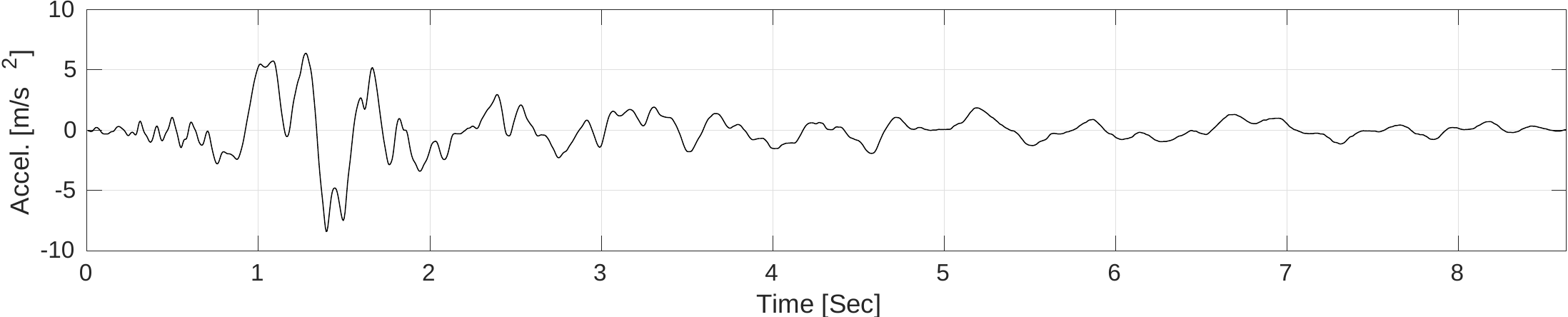}} 
\subfigure[Spectrogram N$^{\circ}5$ - San Salvador]{\includegraphics[width=8.75cm]{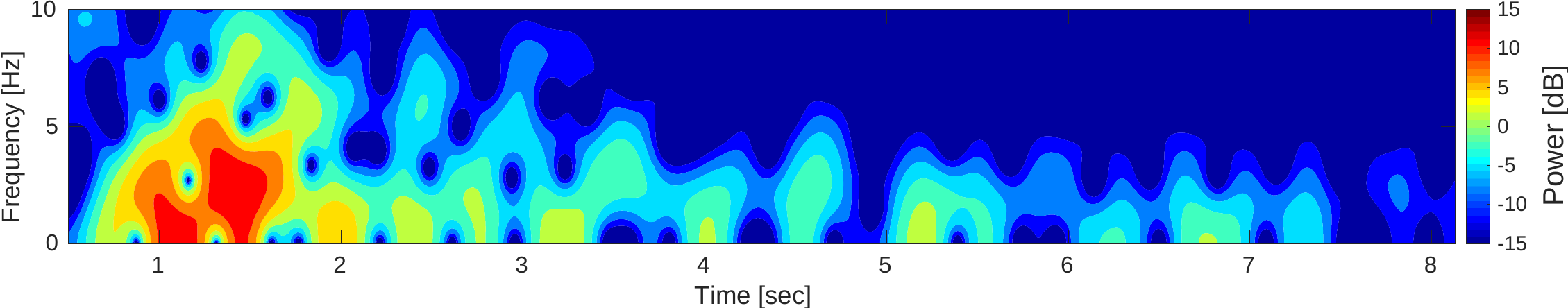}} 
\caption{Time history acceleration and spectrograms for considered seismic events.}
%\label{fig:seismicrecords}
\end{figure}
\begin{figure}[!htbp] 
\ContinuedFloat
\captionsetup{list=off,format=cont}
\centering
\subfigure[Record N$^{\circ}6$ - Parkfield-02 CA]{\includegraphics[width=8.5cm]{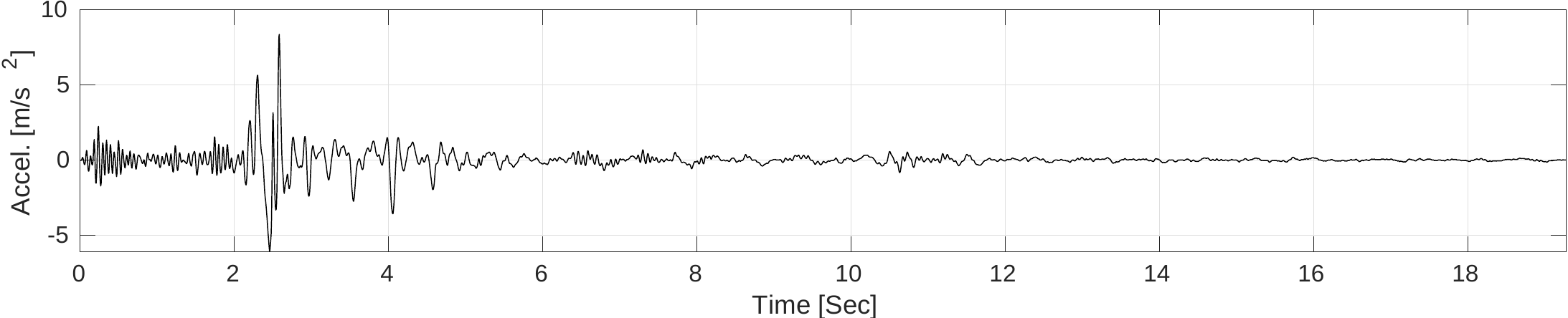}} 
\subfigure[Spectrogram N$^{\circ}6$ - Parkfield-02 CA]{\includegraphics[width=8.75cm]{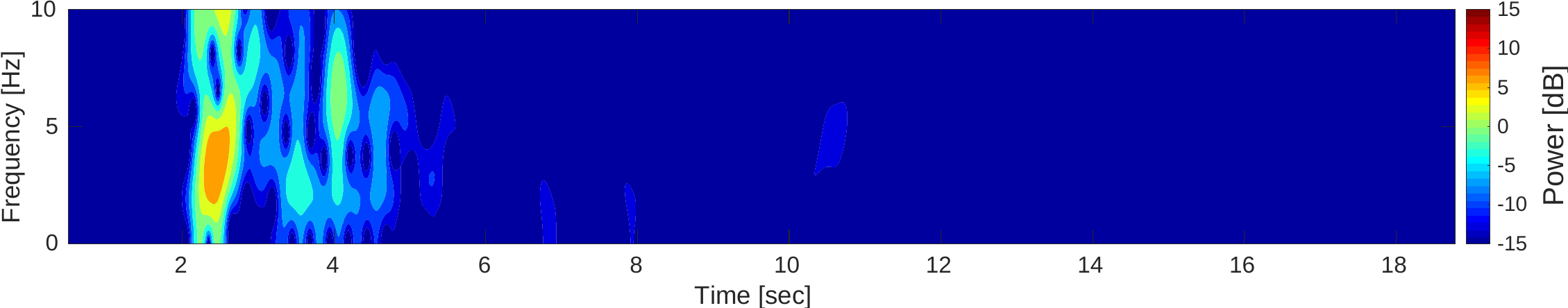}} \\
\subfigure[Record N$^{\circ}7$ - Chuetsu-oki]{\includegraphics[width=8.5cm]{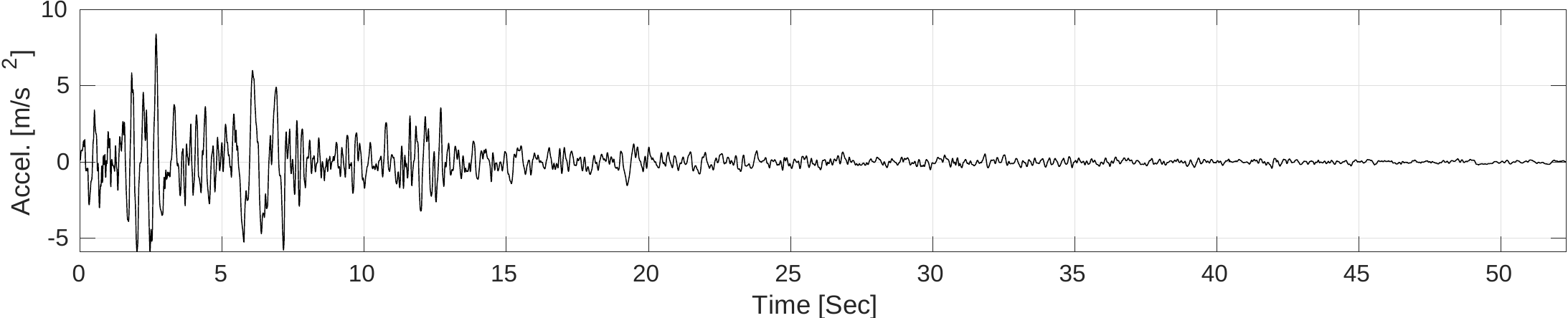}} 
\subfigure[Spectrogram N$^{\circ}7$ - Chuetsu-oki]{\includegraphics[width=8.75cm]{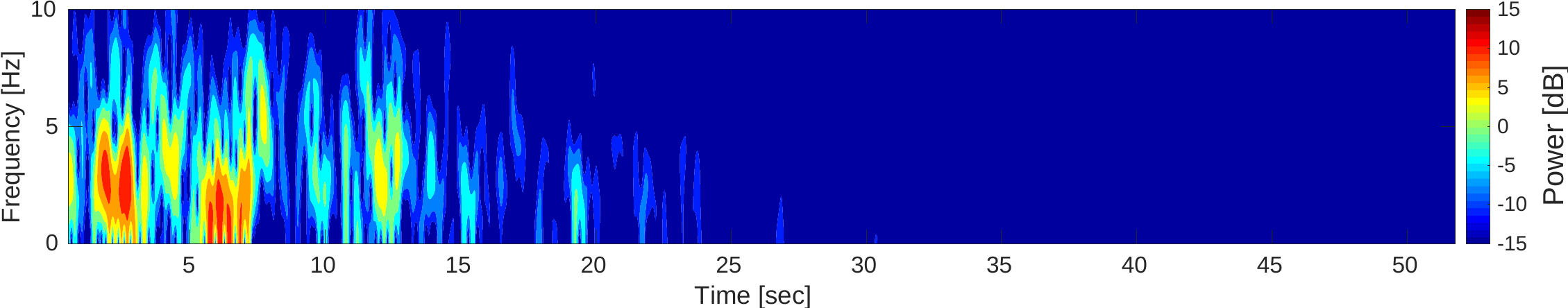}}  \\
\subfigure[Record N$^{\circ}8$ - Tabas Iran]{\includegraphics[width=8.5cm]{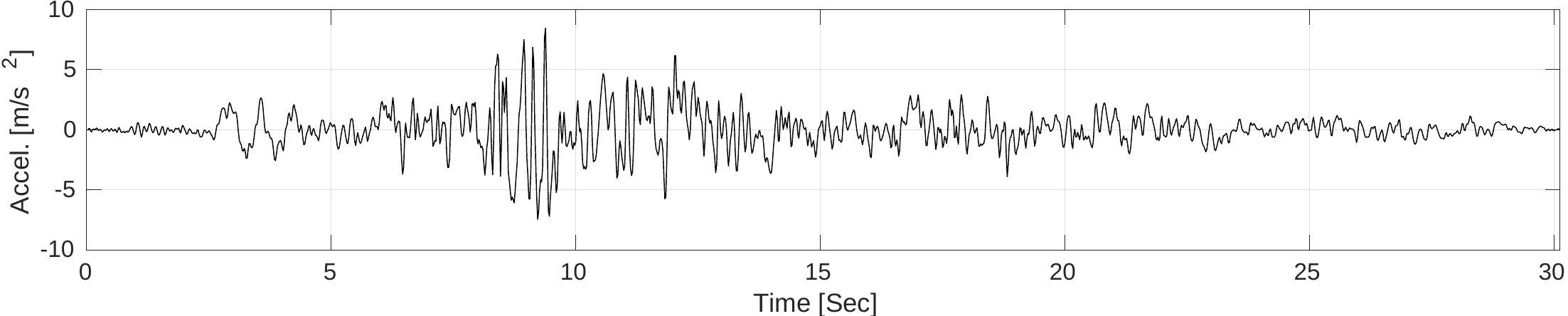}} 
\subfigure[Spectrogram N$^{\circ}8$ - Tabas Iran]{\includegraphics[width=8.75cm]{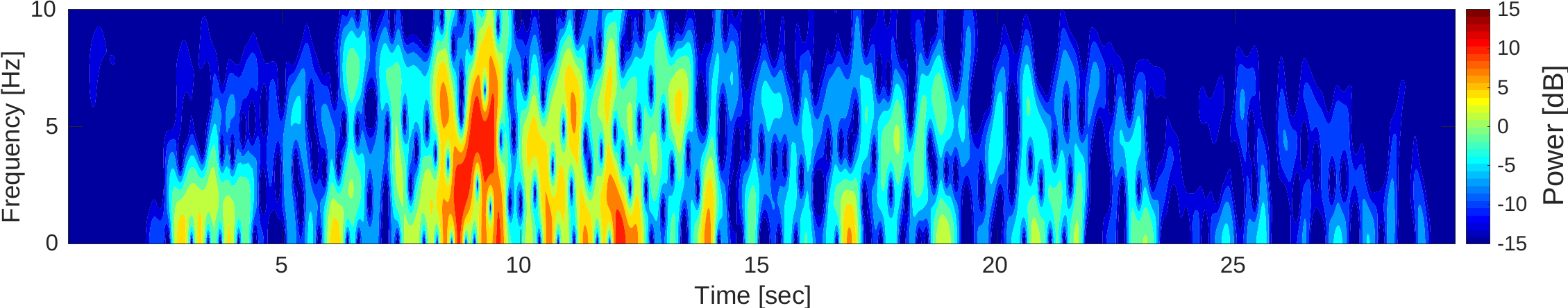}} \\
\subfigure[Record N$^{\circ}9$ - Iwate]{\includegraphics[width=8.5cm]{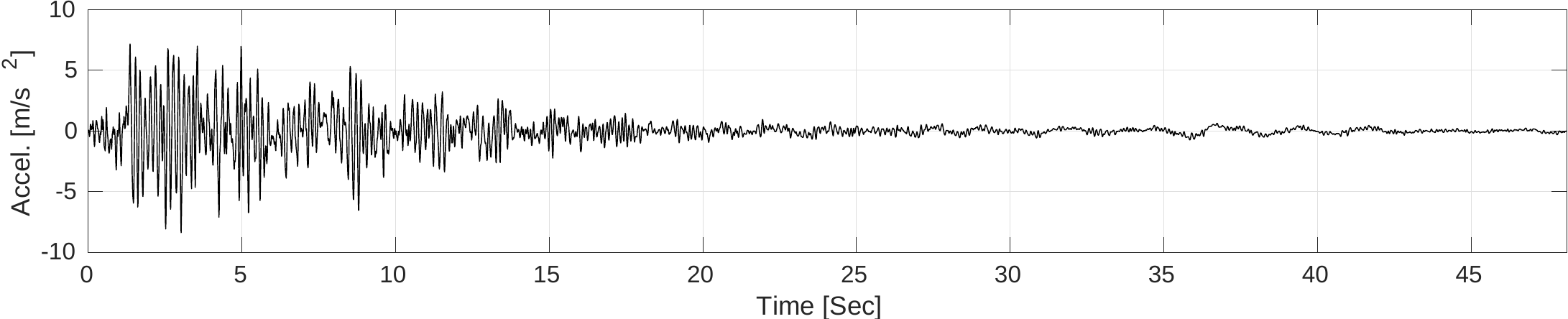}} 
\subfigure[Spectrogram N$^{\circ}9$ - Iwate]{\includegraphics[width=8.75cm]{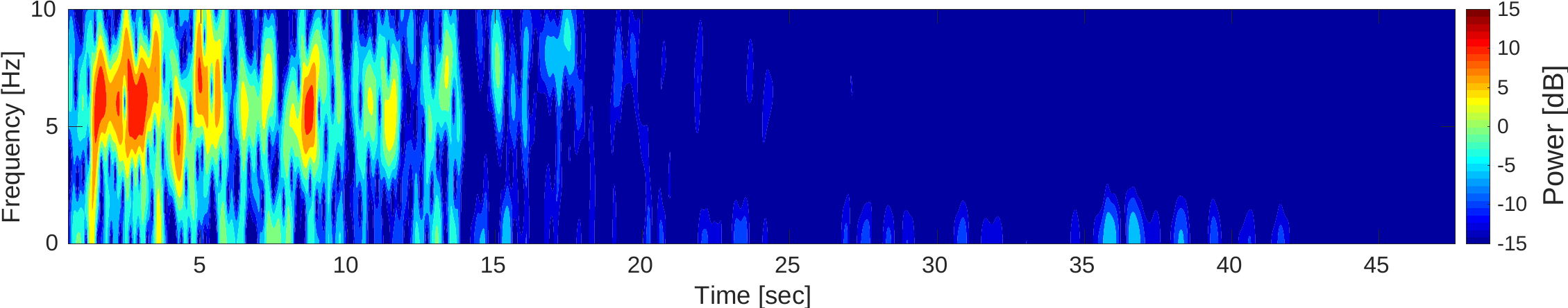}} \\
\subfigure[Record N$^{\circ}10$ - Northridge-01]{\includegraphics[width=8.5cm]{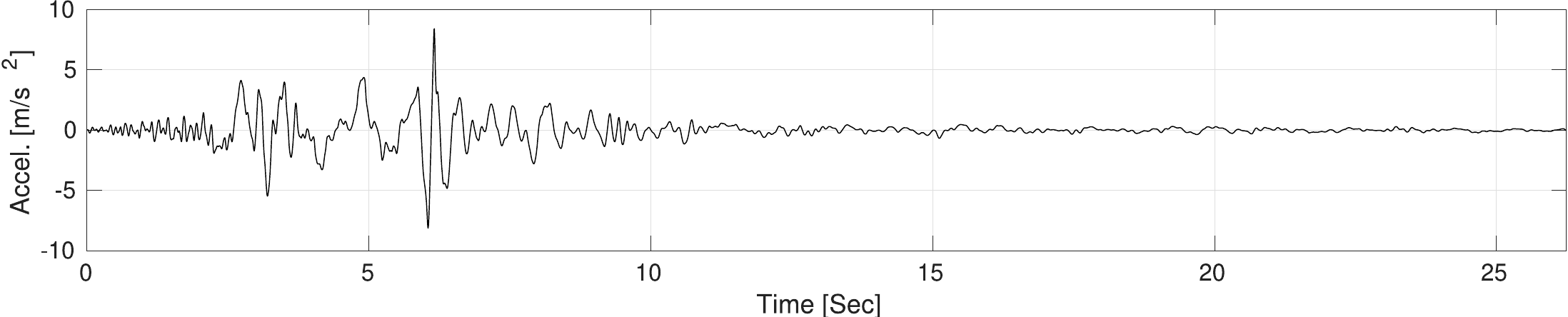}} 
\subfigure[Spectrogram N$^{\circ}10$ - Northridge-01]{\includegraphics[width=8.75cm]{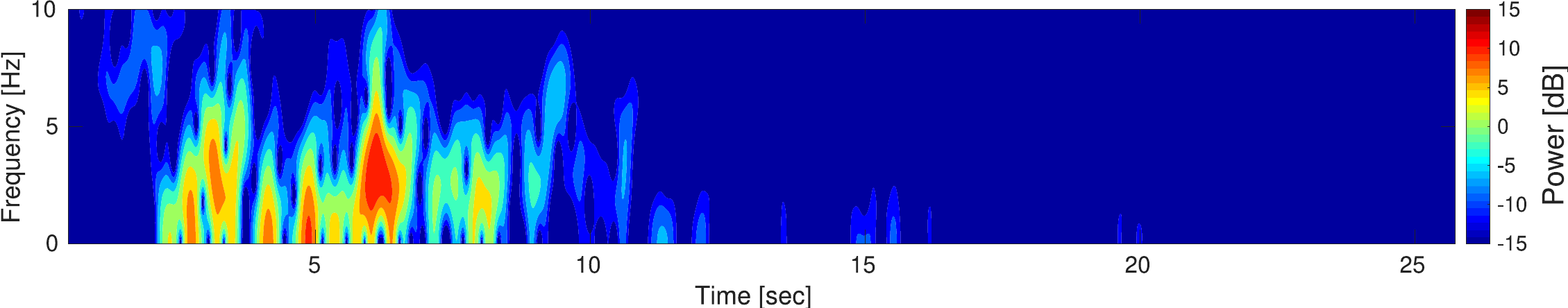}} 
\caption{Time history acceleration and spectrograms for considered seismic events.}
%\label{fig:seismicrecords}
\end{figure}
\begin{figure}[!htbp] 
\ContinuedFloat
\captionsetup{list=off,format=cont}
\centering
\subfigure[Record N$^{\circ}11$ - Chi-Chi Taiwan]{\includegraphics[width=8.5cm]{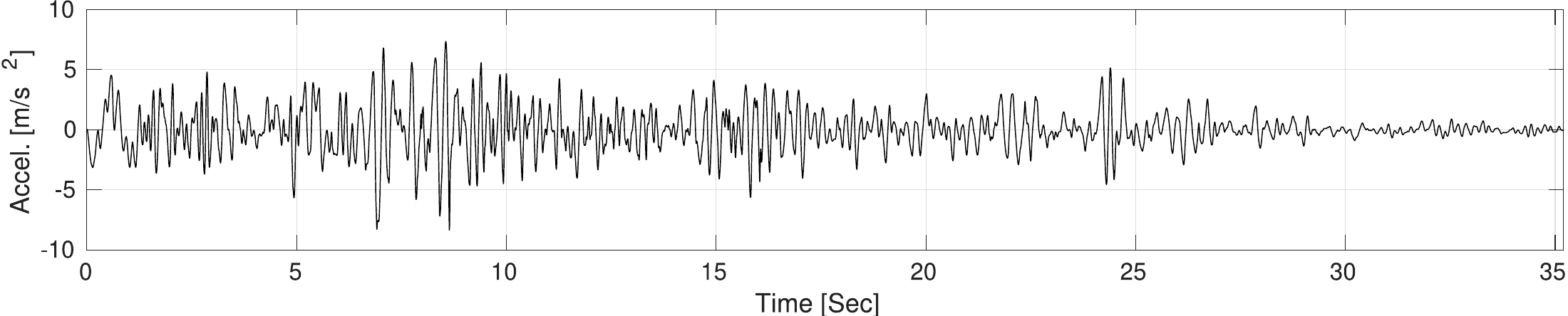}} 
\subfigure[Spectrogram N$^{\circ}11$ - Chi-Chi Taiwan]{\includegraphics[width=8.75cm]{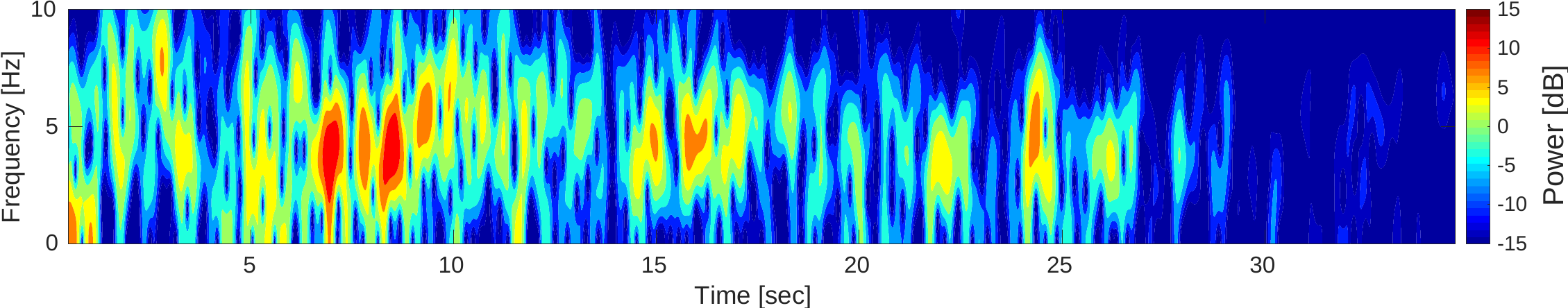}} \\
\subfigure[Record N$^{\circ}12$ - Mammoth Lakes-06]{\includegraphics[width=8.5cm]{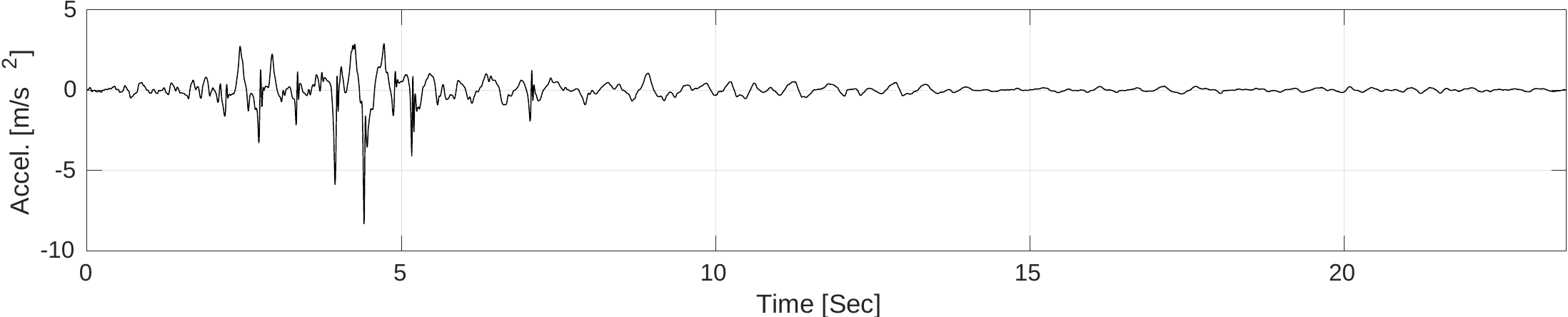}} 
\subfigure[Spectrogram N$^{\circ}12$ - Mammoth Lakes-06]{\includegraphics[width=8.75cm]{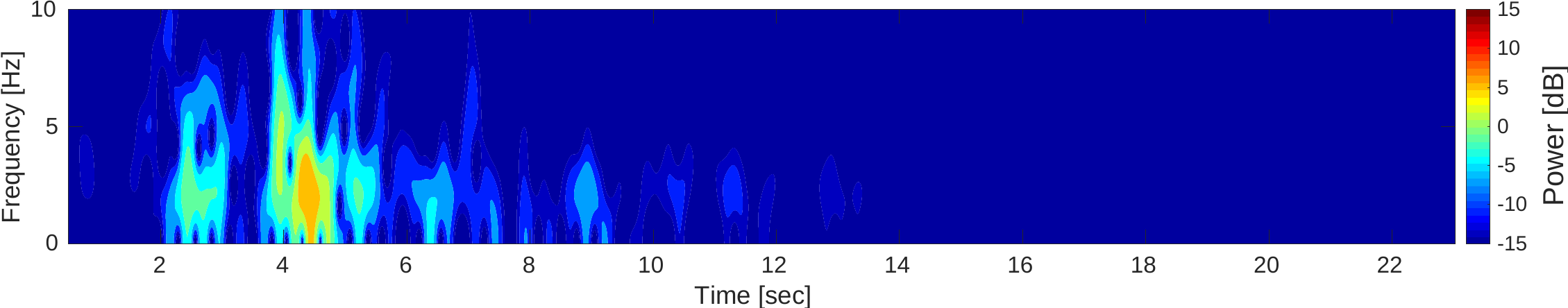}} \\
\subfigure[Record N$^{\circ}13$ - Loma Prieta]{\includegraphics[width=8.5cm]{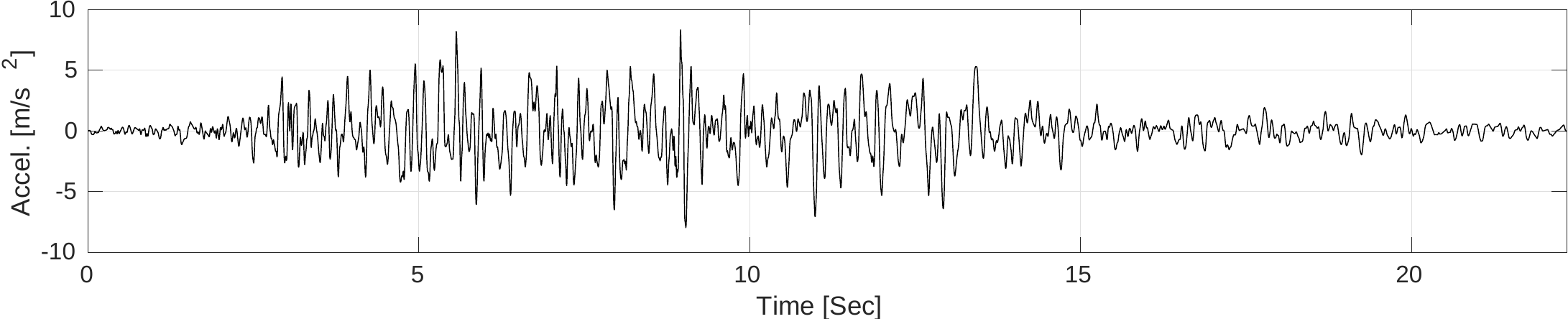}} 
\subfigure[Spectrogram N$^{\circ}13$ - Loma Prieta]{\includegraphics[width=8.75cm]{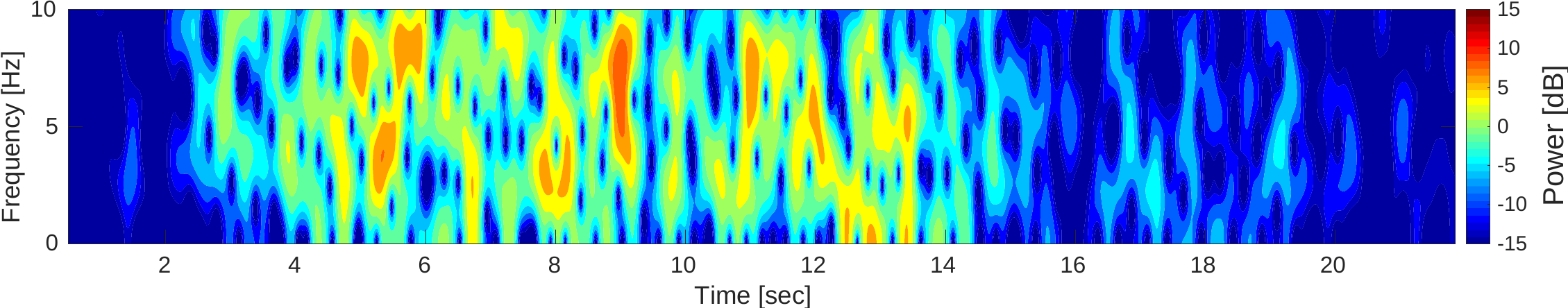}} \\
\subfigure[Record N$^{\circ}14$ - Victoria Mexico]{\includegraphics[width=8.5cm]{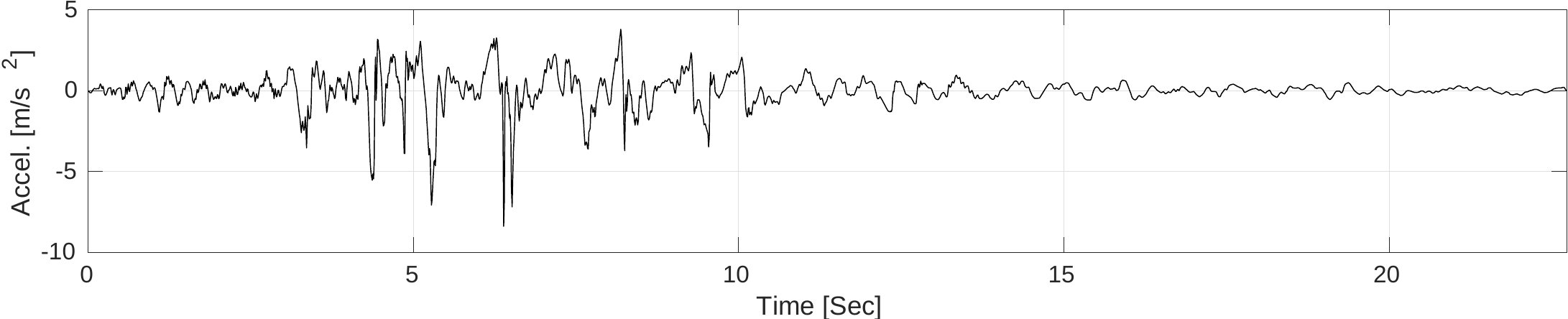}} 
\subfigure[Spectrogram N$^{\circ}14$ - Victoria Mexico]{\includegraphics[width=8.75cm]{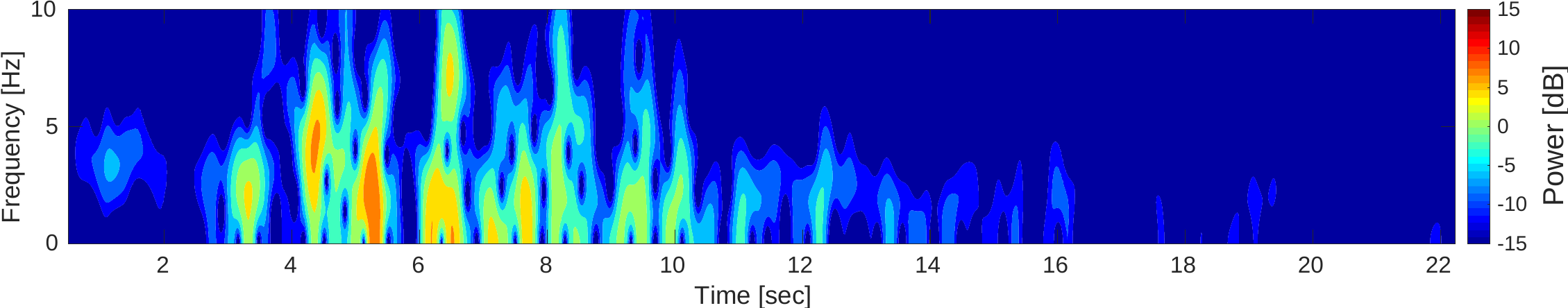}} \\
\subfigure[Record N$^{\circ}15$ - Loma Prieta]{\includegraphics[width=8.5cm]{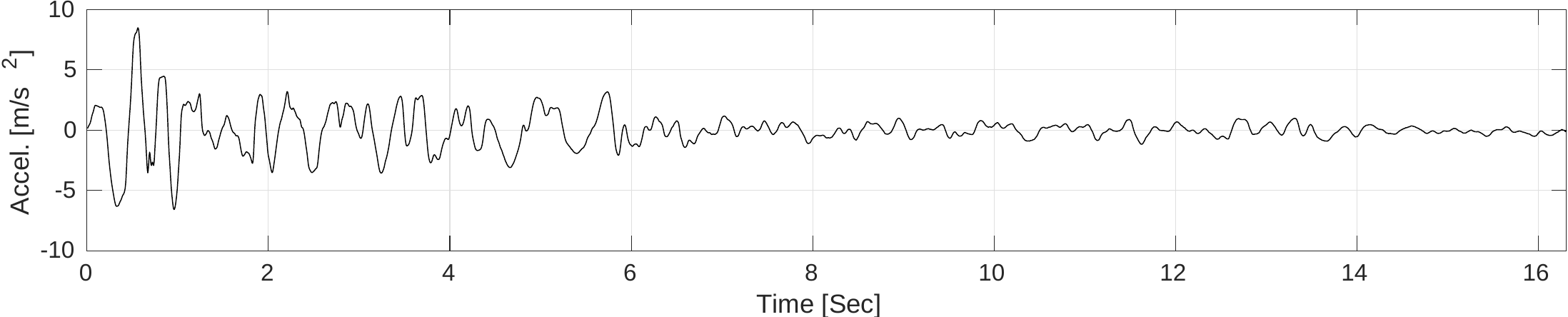}} 
\subfigure[Spectrogram N$^{\circ}15$ - Loma Prieta]{\includegraphics[width=8.75cm]{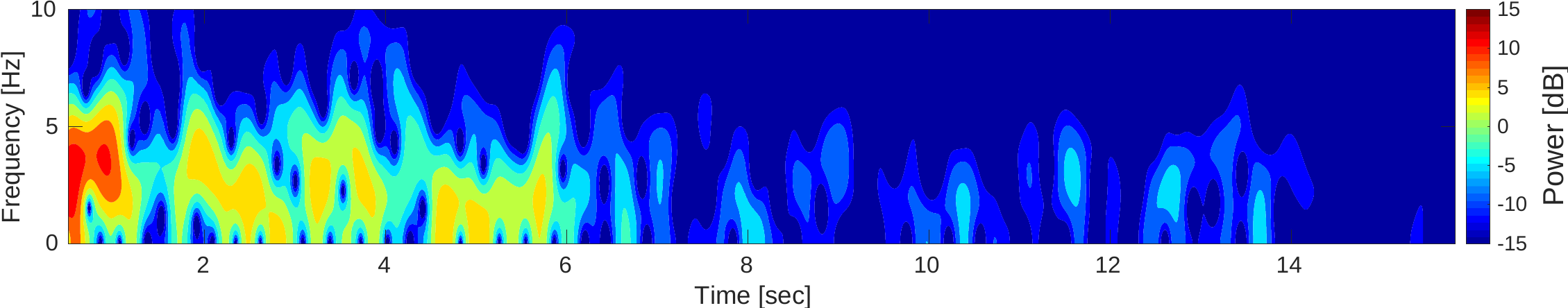}} 
\caption{Time history acceleration and spectrograms for considered seismic events.}
\label{fig:seismicrecords}
\end{figure}
\begin{figure}[!htbp]
\ContinuedFloat
\captionsetup{list=off,format=cont}
\centering
\subfigure[Record N$^{\circ}16$ - Chi-Chi Taiwan (TCU071)]{\includegraphics[width=8.5cm]{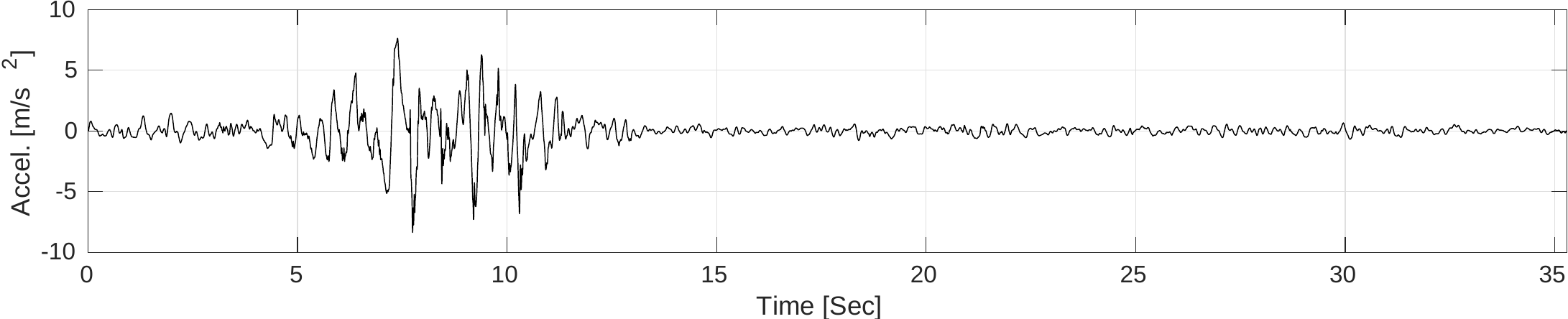}}  
\subfigure[Spectrogram N$^{\circ}16$ - Chi-Chi Taiwan (TCU071)]{\includegraphics[width=8.75cm]{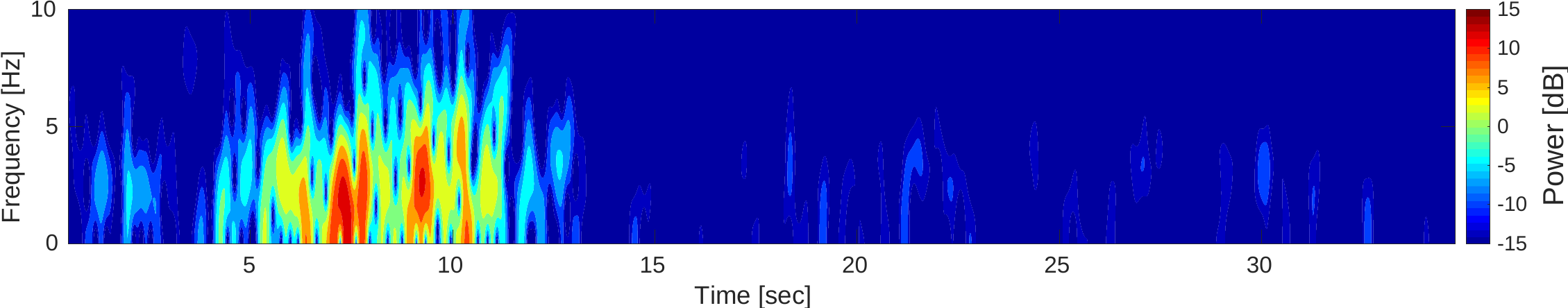}} \\
\subfigure[Record N$^{\circ}17$ - Coalinga-05]{\includegraphics[width=8.5cm]{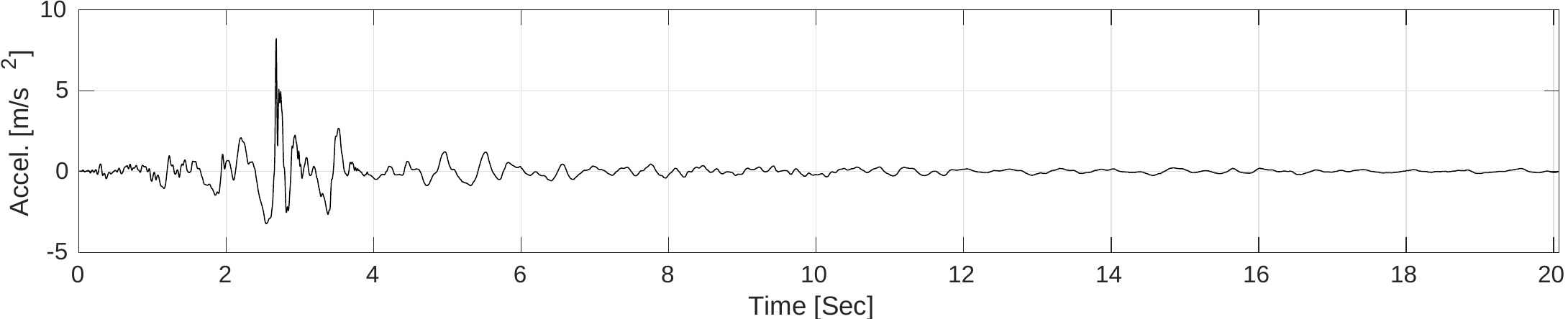}}  
\subfigure[Spectrogram N$^{\circ}17$ - Coalinga-05]{\includegraphics[width=8.75cm]{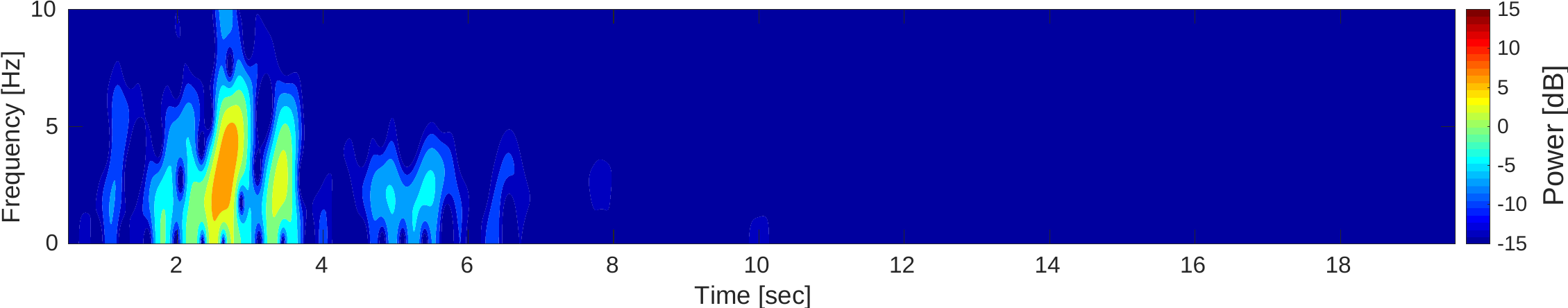}} \\
\subfigure[Record N$^{\circ}18$ - Chuetsu-oki]{\includegraphics[width=8.5cm]{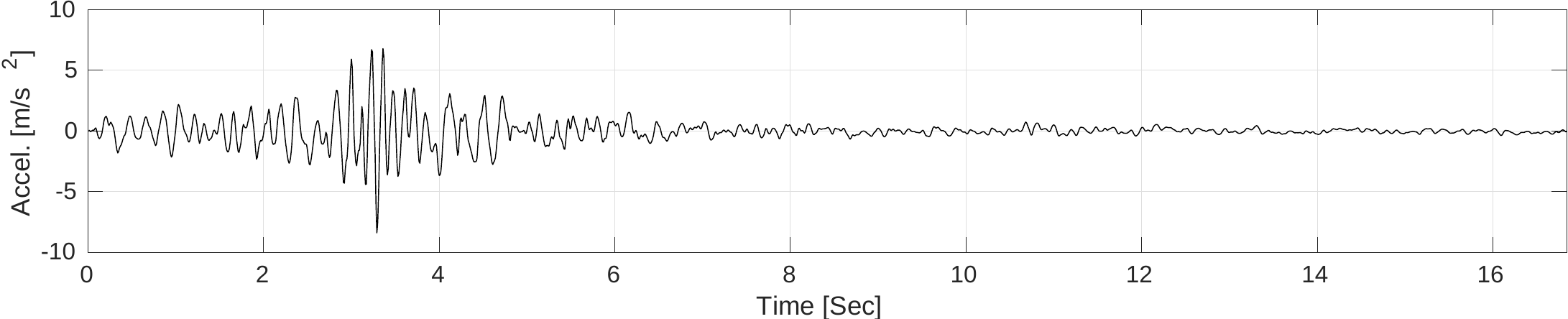}}  
\subfigure[Spectrogram N$^{\circ}18$ - Chuetsu-oki]{\includegraphics[width=8.75cm]{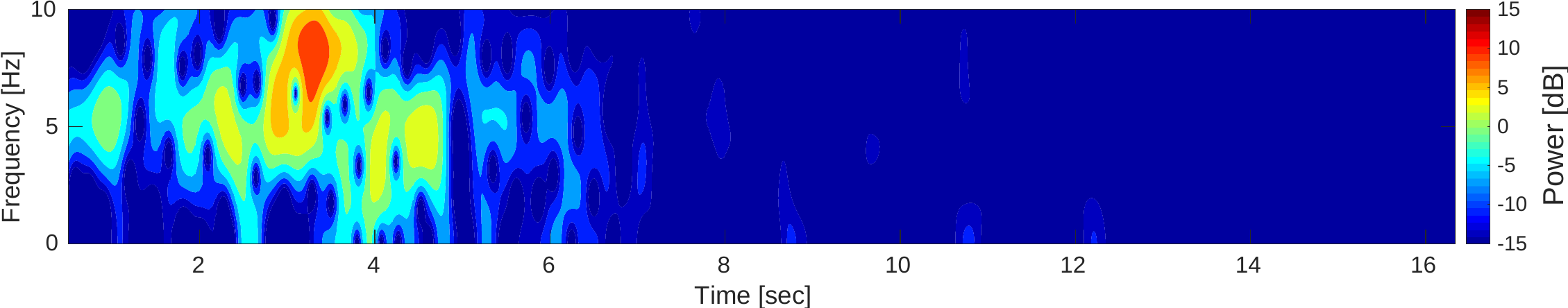}} \\
\subfigure[Record N$^{\circ}19$ - C. Mendocino]{\includegraphics[width=8.5cm]{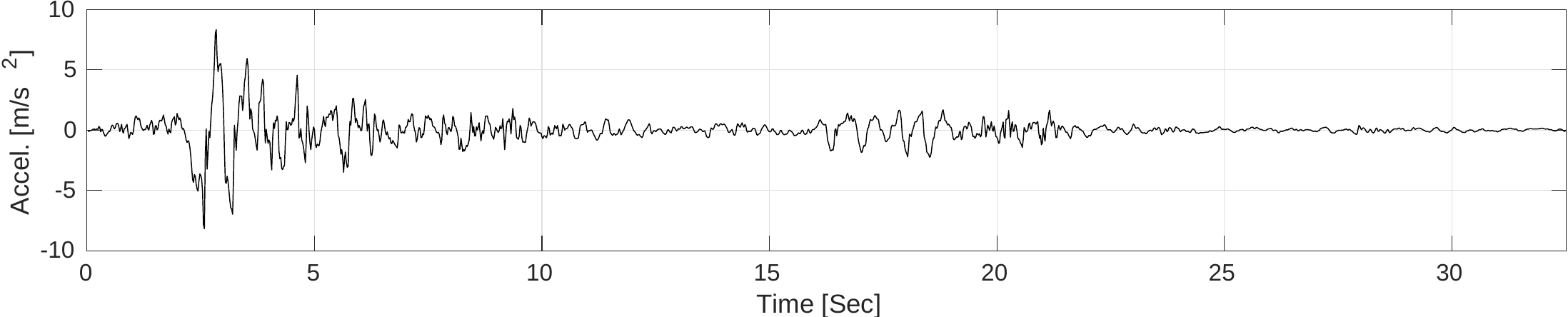}}  
\subfigure[Spectrogram N$^{\circ}19$ - C. Mendocino]{\includegraphics[width=8.75cm]{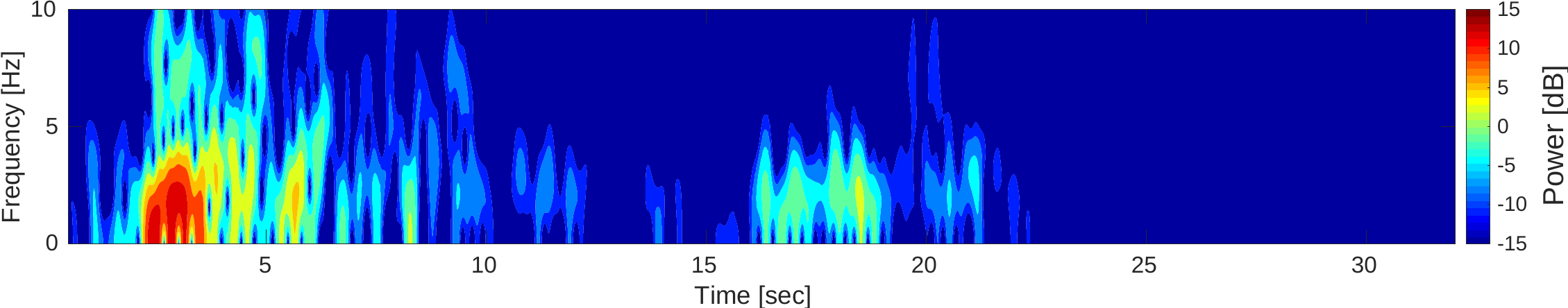}} \\
\subfigure[Record N$^{\circ}20$ - Coalinga-05]{\includegraphics[width=8.5cm]{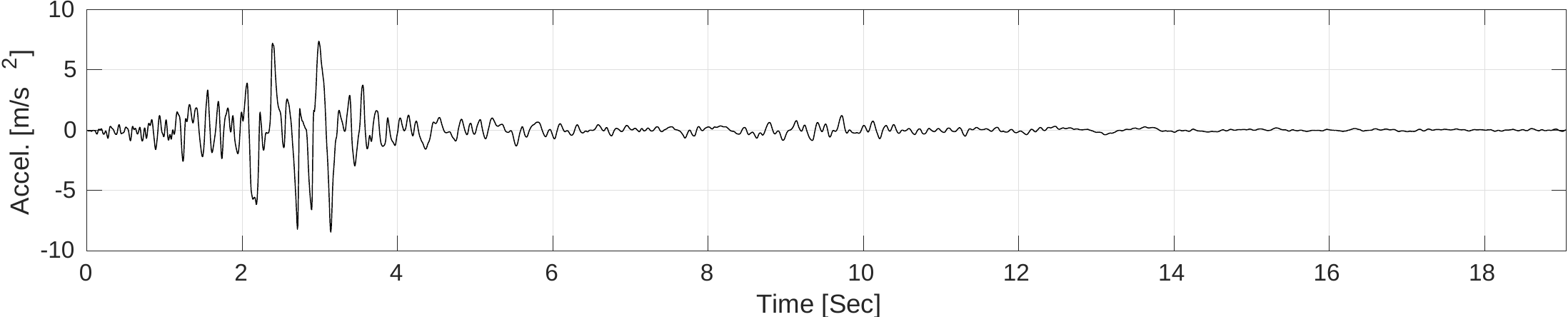}} 
\subfigure[Spectrogram N$^{\circ}20$ - Coalinga-05]{\includegraphics[width=8.75cm]{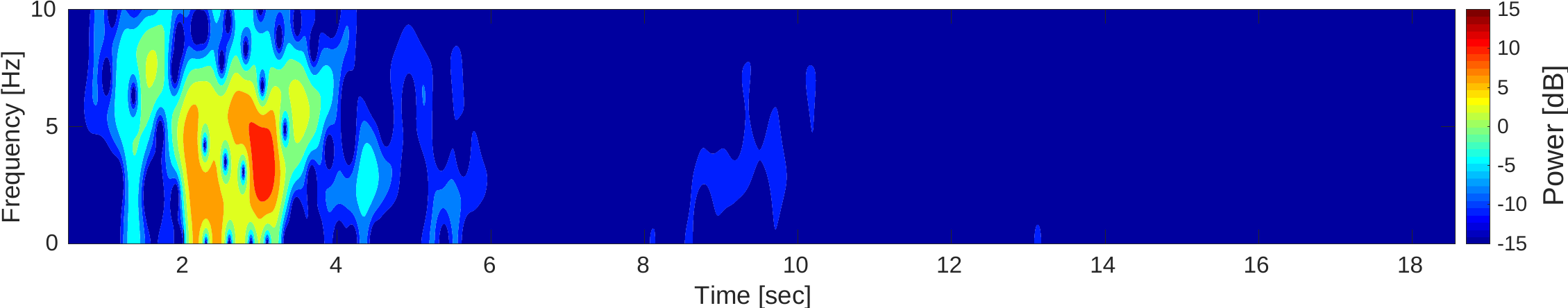}} \\
\caption{Time history acceleration and spectrograms for considered seismic events.}
%\label{fig:spectrogramsrecords}
\end{figure}
\begin{figure}[!htbp]
\ContinuedFloat
\captionsetup{list=off,format=cont}
\centering
\subfigure[Record N$^{\circ}21$ - Chi-Chi Taiwan (CHY041)]{\includegraphics[width=8.5cm]{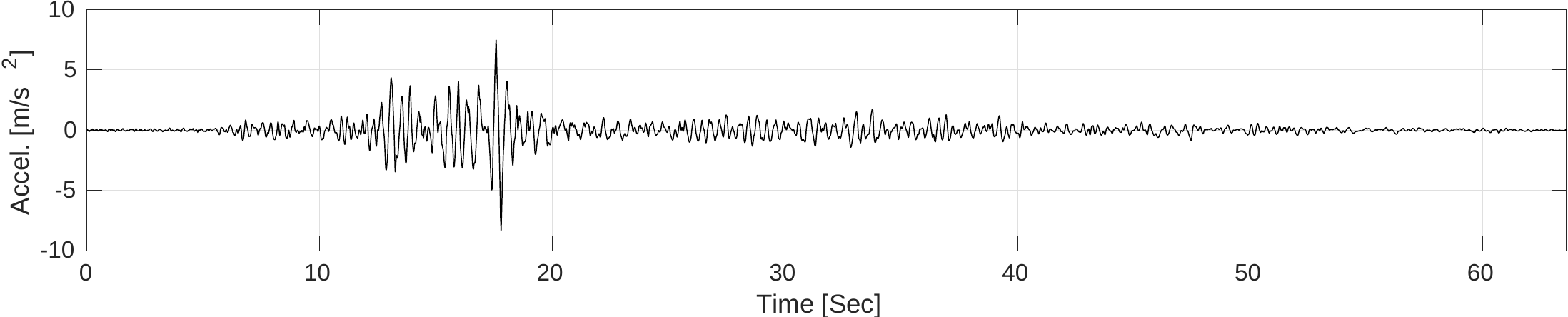}} 
\subfigure[Spectrogram N$^{\circ}21$ - Chi-Chi Taiwan (CHY041)]{\includegraphics[width=8.75cm]{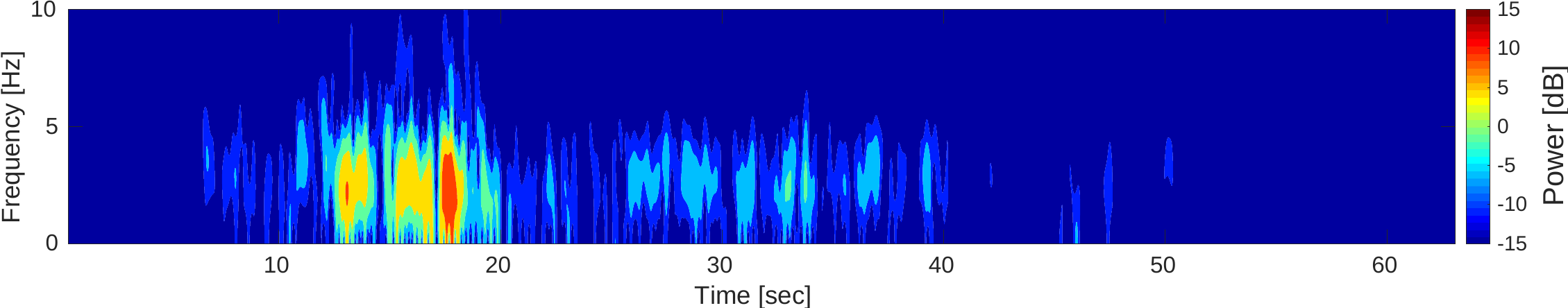}} \\
\subfigure[Record N$^{\circ}22$ - Christchurch NZ]{\includegraphics[width=8.5cm]{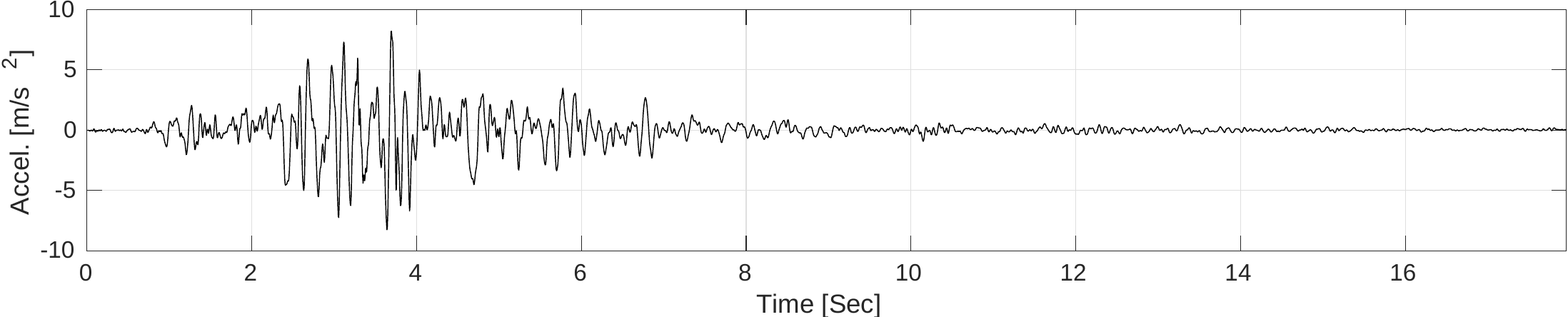}}  
\subfigure[Spectrogram N$^{\circ}22$ - Christchurch NZ]{\includegraphics[width=8.75cm]{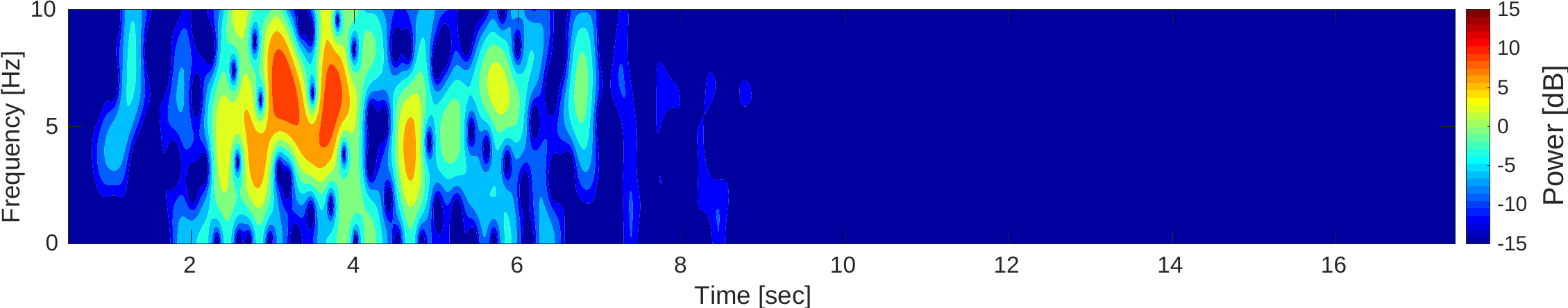}} \\
\subfigure[Record N$^{\circ}23$ - N/A (Ward Fire St)]{\includegraphics[width=8.5cm]{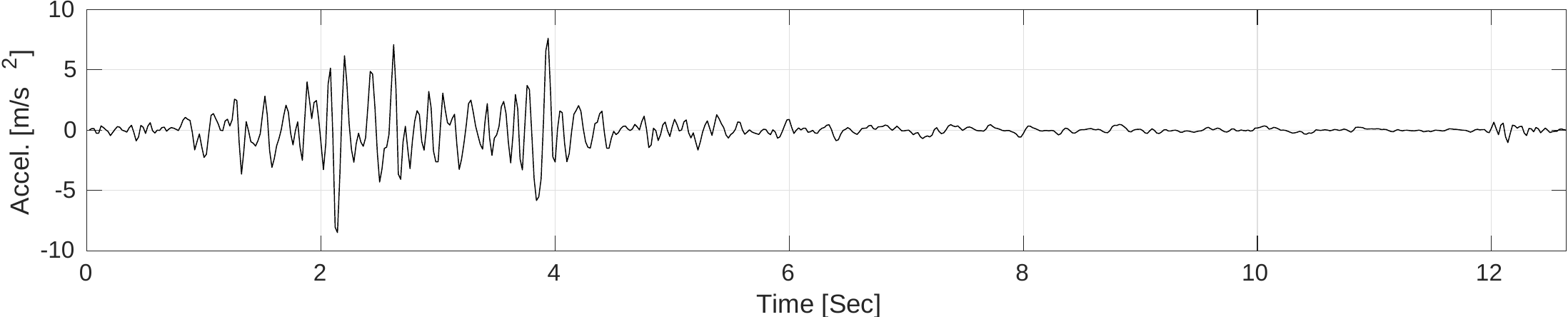}}  
\subfigure[Spectrogram N$^{\circ}23$ - N/A (Ward Fire St)]{\includegraphics[width=8.75cm]{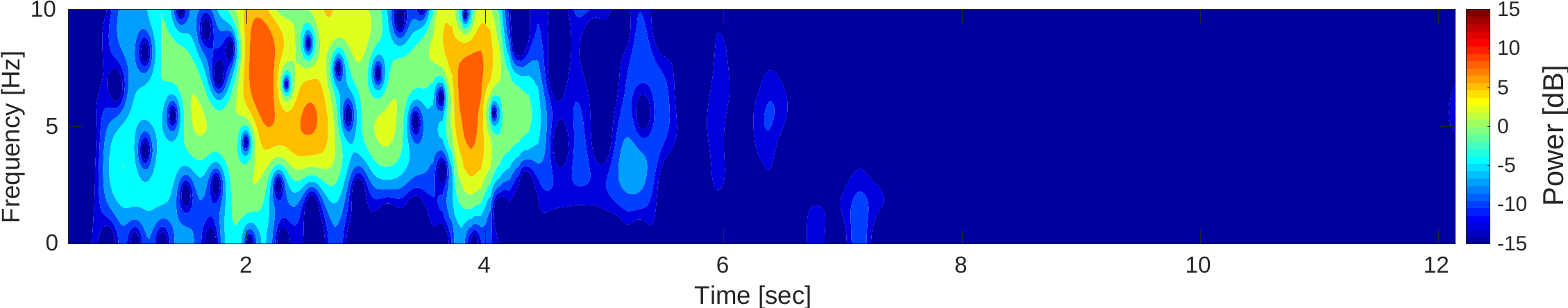}} \\
\subfigure[Record N$^{\circ}24$ - N/A (ANGOL)]{\includegraphics[width=8.5cm]{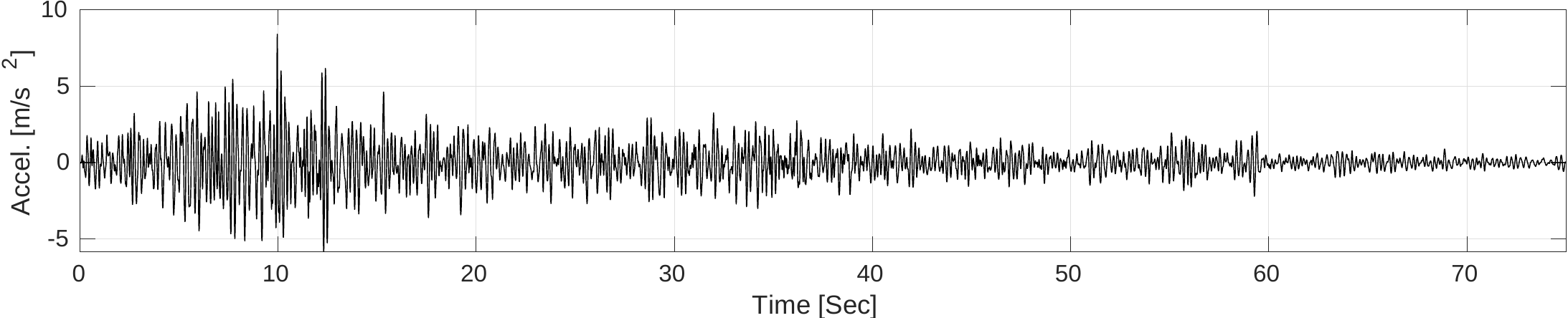}}  
\subfigure[Spectrogram N$^{\circ}24$ - N/A (ANGOL)]{\includegraphics[width=8.75cm]{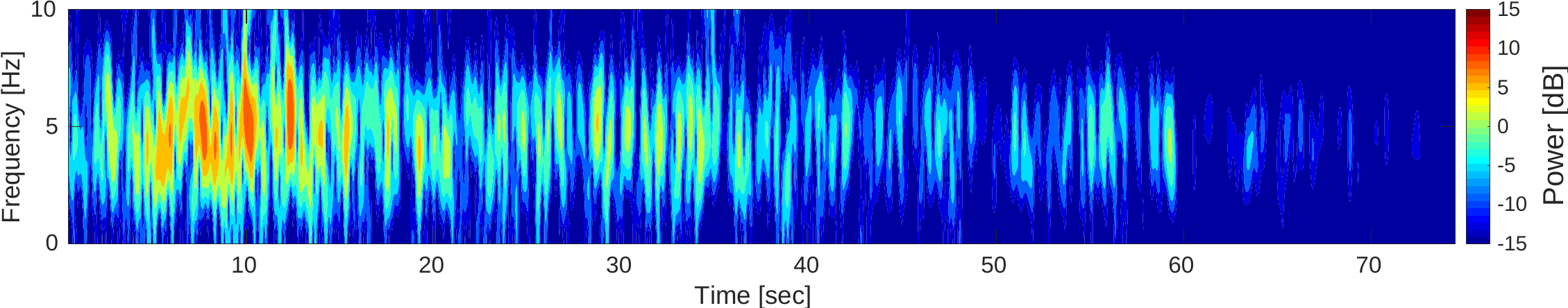}} \\
\subfigure[Record N$^{\circ}25$ - N/A (PICA)]{\includegraphics[width=8.5cm]{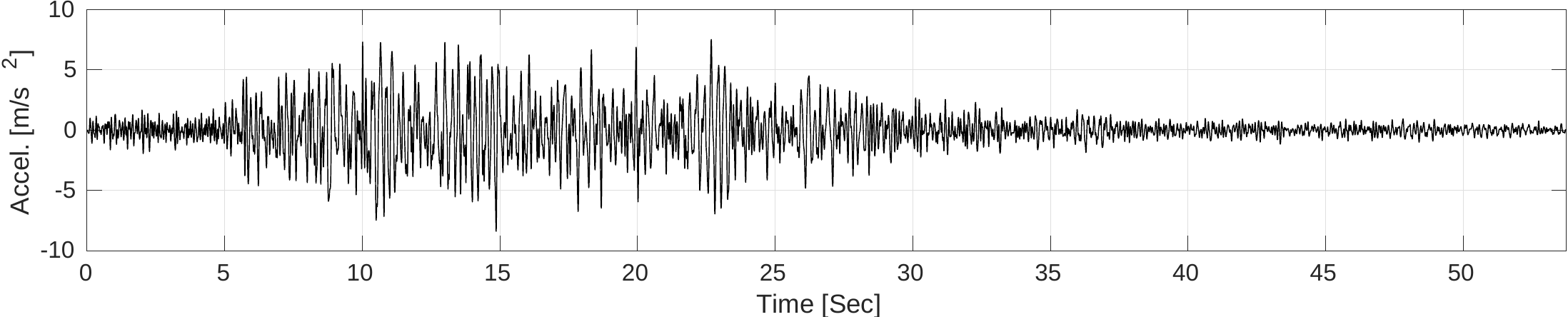}} 
\subfigure[Spectrogram N$^{\circ}25$ - N/A (PICA)]{\includegraphics[width=8.75cm]{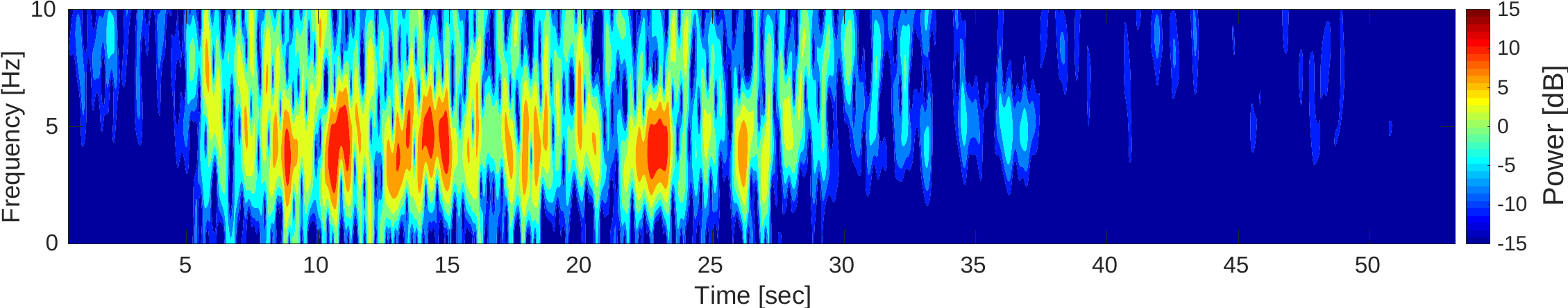}} 
\caption{Time history acceleration and spectrograms for considered seismic events.}
%\label{fig:spectrogramsrecords}
\end{figure}
\end{landscape}

\end{document}